# THE ARAUCARIA PROJECT

Improving the cosmic distance scale

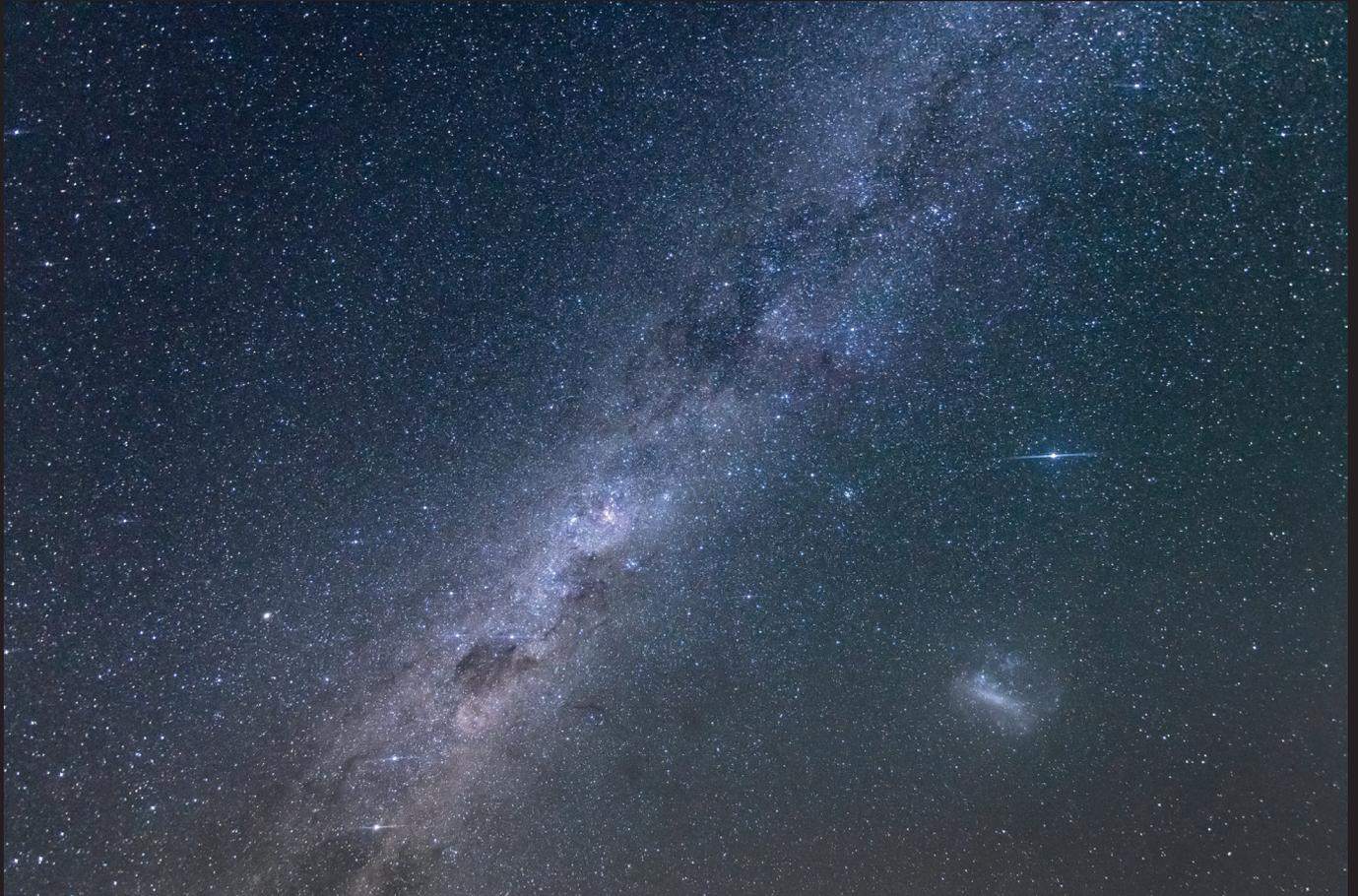

# THE ARAUCARIA PROJECT

## Improving the cosmic distance scale

Warsaw 2021

The Araucaria Project: Improving the cosmic distance scale

# Table of Contents

**Wolfgang Gieren[1] and Grzegorz Pietrzyński[2]**

# A brief history of the Araucaria Project

In early 2000, Grzegorz Pietrzyński started to work with Wolfgang Gieren in Concepción. During this time the Hubble Space Telescope (HST) Key Project on the Extragalactic Distance Scale by Wendy Freedman and collaborators was about to finish, and the topic of distance determination was again, and very strongly, in the focus of the international astronomical community.

Wolfgang and Grzegorz immediately had some ideas about remaining problems in setting up the distance scale. The main systematic uncertainty in the HST Key Project seemed to be the largely unknown effect of environmental properties, mostly metallicity, on the luminosities of classical Cepheid variables which had been extensively used by the Key Project to determine the distances to supernova Ia host galaxies, and to calibrate these SN Ia as standard candles reaching out into the unperturbed Hubble flow where the Hubble constant can be reliably determined. *Our conclusion was that it was absolutely necessary to complement the HST Key Project work with a thorough determination of environmental effects on Cepheids, and a more accurate distance to the Large Magellanic Cloud which served as the anchor galaxy to the HST Key Project SN Ia host galaxy distances.* The perfect laboratories to do this were the galaxies in the nearby Local and Sculptor Groups which offer a wide range of types, and environmental properties. Eventually, we thought that it would be of great interest to not only carry out studies aiming at improving classical Cepheids as standard candles, but also other stellar distance indicators like RR Lyrae stars, red clump stars, the Tip of the Red Giant Branch, Type II Cepheids, eclipsing binaries, and others. Another key consideration was to carry the stellar methods of distance determination, traditionally calibrated in the optical, to the near-infrared regime where extinction was much less a problem than in the optical spectral region, and where some previous theoretical studies had suggested that environmental dependences might be smaller as well. It was good timing for such studies

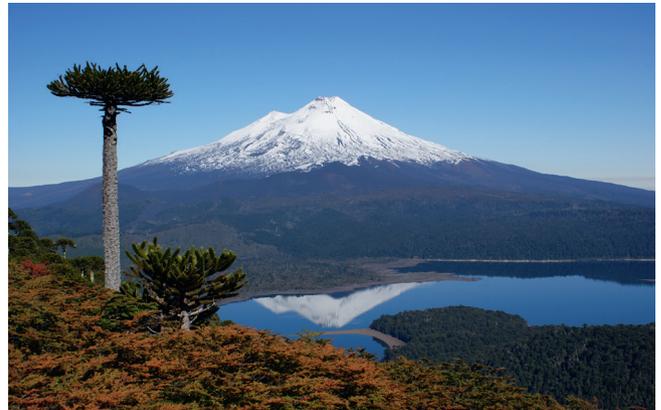

**Fig. 1.** Araucaria tree in Conguillío National Park.

[1] Universidad de Concepción, Departamento de Astronomia, Casilla 160-C, Concepción, Chile
[2] Nicolaus Copernicus Astronomical Center, Polish Academy of Sciences, Bartycka 18, 00-716 Warszawa, Poland

because excellent near-infrared cameras had just become available at several observatories around the beginning of the 21st century. When it became clear to us that we were starting a very ambitious and long-lasting project, we needed a name for it, and decided to call it the "Araucaria Project", making reference to the place the first ideas were born (Araucaria trees are the most beautiful native trees in southern Chile).

In the same year, 2000, Wolfgang and Grzegorz visited Rolf Kudritzki in Munich. Rolf was a leading expert on blue supergiant stars and had some ideas in mind about how to use these extremely luminous objects, visible in very distant galaxies, for distance determination. Wolfgang and Grzegorz were working at this time, together with Pascal Fouqué, who was a visitor at the European Southern Observatory's (ESO) branch in Chile, on a survey for Cepheids in the beautiful spiral galaxy NGC 300 in the Sculptor Group using images obtained with the ESO/MPI Wide-Field Imager at the 2.2-m telescope on La Silla, and it was evident from the blue images that NGC 300 contained a lot of bright blue stars. So we quickly agreed with Rolf that it would be a very nice project to test his spectroscopic method (which eventually became known as the "Flux-Weighted Gravity-Luminosity Relationship") to determine distances using blue supergiants, and check their correctness by comparison with the distance we were about to determine from the NGC 300 Cepheids. This way the Araucaria Project group saw its first expansion, by integrating Rolf Kudritzki and his close collaborators, Fabio Bresolin and Miguel Urbaneja in the project, who were to strongly contribute to the scientific success of the project over the next two decades.

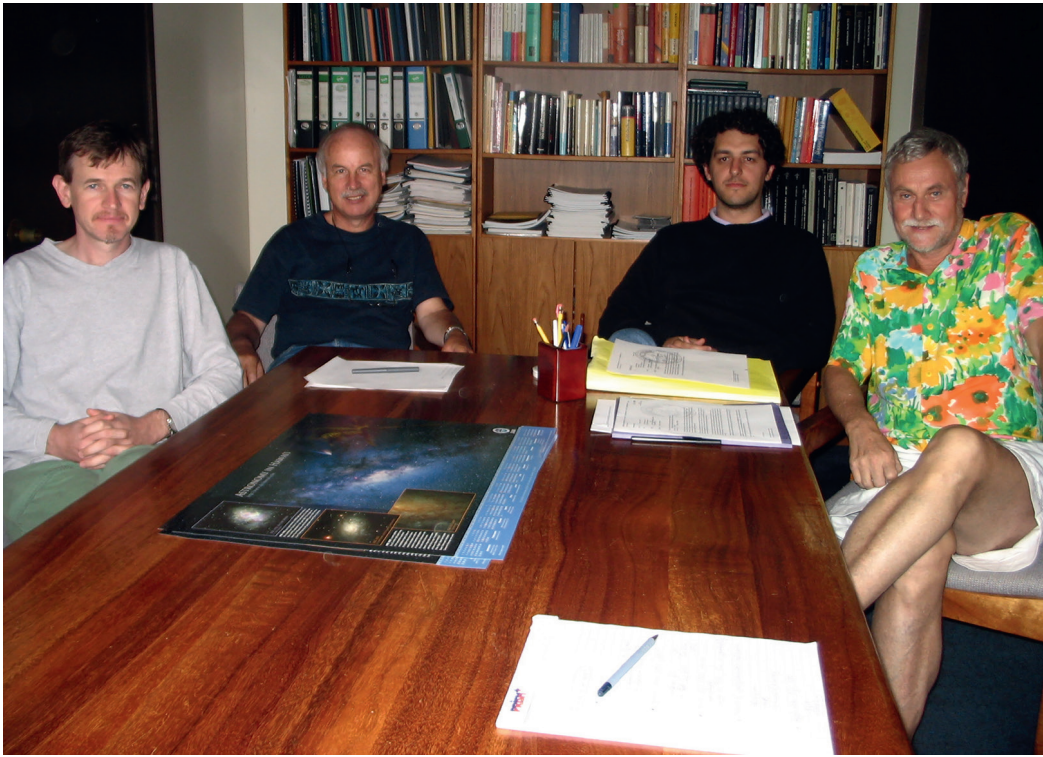

**Fig. 2.** The meeting at the Institute for Astronomy of University of Hawaii in Honolulu, 2006. From left: Fabio, Wolfgang, Miguel, Rolf.

At the same time, Giuseppe Bono from Rome Observatory joined the project which opened up the opportunity to complement the observational work with theoretical model results and predictions, with mutual benefits for both approaches to study and understand pulsating variable stars. Another member of the Araucaria Project who joined our group in its very first years was Jesper Storm from AIP Potsdam. He, Wolfgang and Pascal Fouqué had been working for some years together to develop a near-infrared version of the classical Baade-Wesselink method, the Infrared Surface Brightness (IRSB) Technique, and now went on with this work as part of the Araucaria Project. Another two key members of the Project joined us some years later: Ian Thompson from the Observatories of the Carnegie Institute of Washington, and Darek Graczyk, who joined our team in Concepcion. Ian and Darek became interested in collaborating with the Araucaria Project group when we started to extensively observe special eclipsing binary systems in the Magellanic Clouds which were composed of pairs of red giants. These very rare systems had been discovered by the OGLE Project at Las Campanas Observatory and needed radial velocity follow-up observations to confirm their physical reality, and derive their properties and distances. Ian not only added his experience in eclipsing binary studies to the Project but also, through his special access to the Magellan telescopes at Las Campanas Observatory in Chile, helped to start very ambitious observing campaigns on these faint objects which, after a decade or more of continued data collection, led to the most accurate-ever distance determination to both Magellanic Clouds. Darek was taking care of precision modeling of our binary systems.

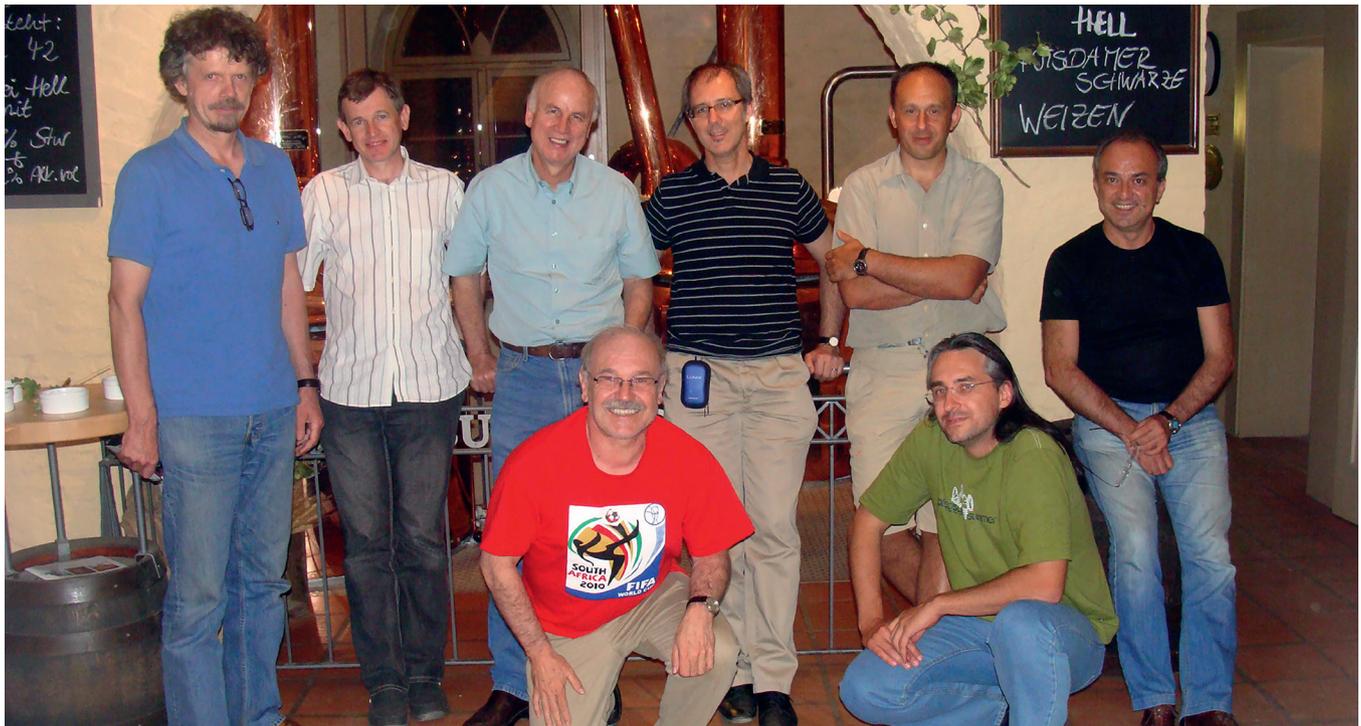

**Fig. 3.** Araucaria Meeting in Potsdam, Germany, 2010.

Through the years, a number of very talented and motivated young scientists joined the Araucaria Project. Igor Soszynski from Warsaw worked for two years in Concepción with Wolfgang and Grzegorz (2004-2006) on classical Cepheids; other postdoctoral fellows working in Concepción were Olaf Szewczyk, Nicolas Nardetto and Marek Górski. Olaf started to work on the improvement of RR Lyrae stars as distance indicators using near-infrared photometry, and eventually obtained good distance determinations to both Magellanic Clouds. Nicolas was working on the modeling of classical Cepheid atmospheres mainly to better understand how observed Cepheid radial velocities could be transformed to the pulsation velocities of these variables, a central problem in the IRSB method to be overcome in order to yield reliable distances from this technique. Marek was working on red clump stars and the TRGB method to improve our understanding of how these stellar methods of distance determination depend on age and metallicity, and also became involved in the IRSB work led by Wolfgang and Jesper.

In 2006 Grzegorz decided to go back to Poland. It was a very difficult decision because he was extremely happy working and living in beautiful southern Chile. However this decision turned out to have a strong positive impact on the Araucaria Project. Thanks to generous support from the Foundation of Polish Science a dedicated working group was established in Warsaw. Many talented researchers joined our team thanks to these funds including Bogumił Pilecki, Paulina Karczmarek, Piotr Konorski, Mónica Taormina and Ksenia Suchomska. Since then the Araucaria team has been able to expand research on several different kinds of eclipsing binaries and even pulsating stars in eclipsing binaries.

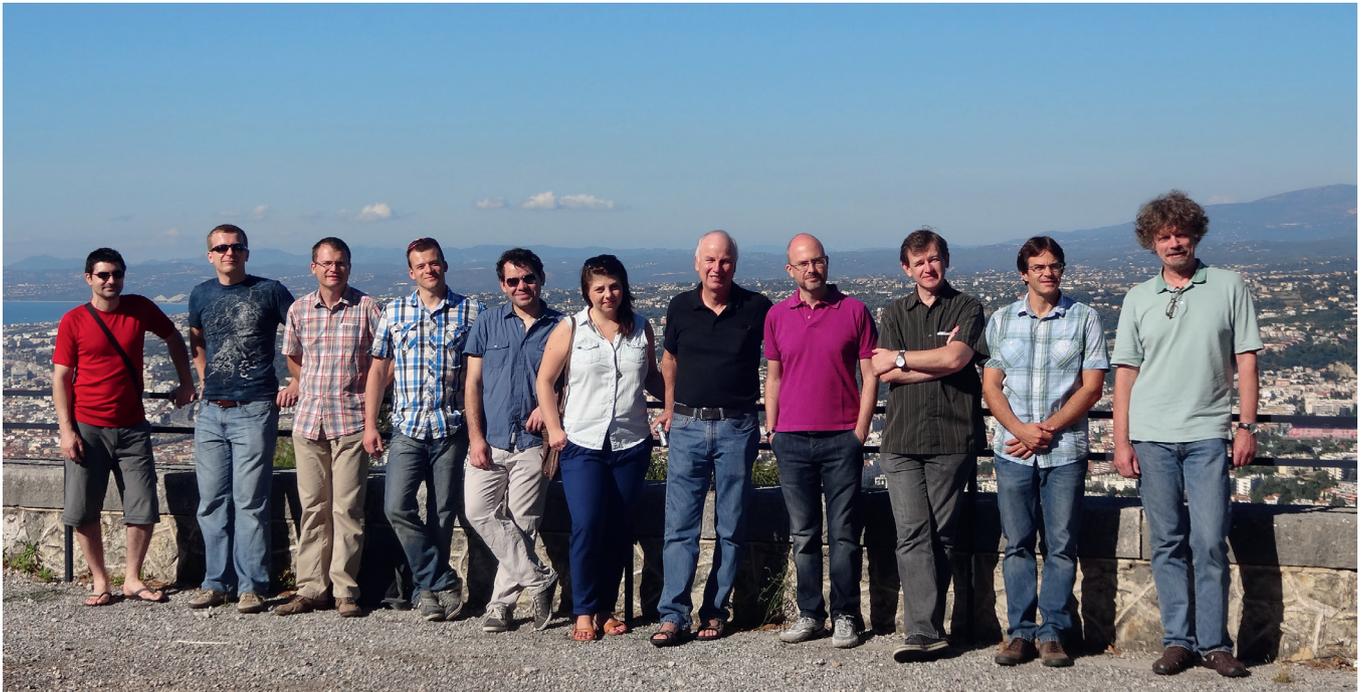

**Fig. 4.** Araucaria Meeting in Nice, France, 2013.

Darek Graczyk has been leading the project on eclipsing binaries in the Solar neighborhood, and also kept working on binaries in the Magellanic Clouds. Bogumił managed to very carefully model several classical and Type II Cepheids in eclipsing binaries leading to the determination of their physical parameters with unprecedented accuracy. Grzegorz and Darek discovered a very special pulsating star which mimics a classical RR Lyrae star but is the product of binary evolution. Mónica obtained very interesting results on very promising eclipsing-binary systems composed of early-type stars. Ksenia dedicated a lot of work to studying Milky Way eclipsing binaries composed of giants. Paulina performed simulations to study in detail the evolution of pulsating stars in binary systems. We also gained another important collaborator, an expert on eclipsing binaries - Pierre Maxted. All these scientific projects opened a new era in the Araucaria project. Most of them have been continued until today (2021) and resulted in several dozen publications, including 4 papers published in the prestigious Nature journal.

Another important expansion of our research line was possible due to the start of a close collaboration with Pierre Kervella and Nicolas Nardetto in 2003, and Alexandre Gallenne in 2008. Thanks to their expertise in interferometry we were now able to conduct several new projects which gave us the opportunity to improve on previous results. By the year 2018, we managed to very precisely measure angular diameters of some 50 red clump stars which, in tandem with high quality near-IR data collected for the same stars at the South African Astronomical Observatory, that allowed us to improve the calibration of the surface brightness-color relationship for stars in the Cepheid color range, an effort which complemented our hectic efforts on eclipsing binaries very well. This is just one very good example of the strong synergy we now had within our group. Alexandre and Pierre worked on several very interesting astrometric binary systems in the Milky Way containing classical Cepheids, work which complements our efforts to study Cepheids in eclipsing binaries in the Large Magellanic Cloud.

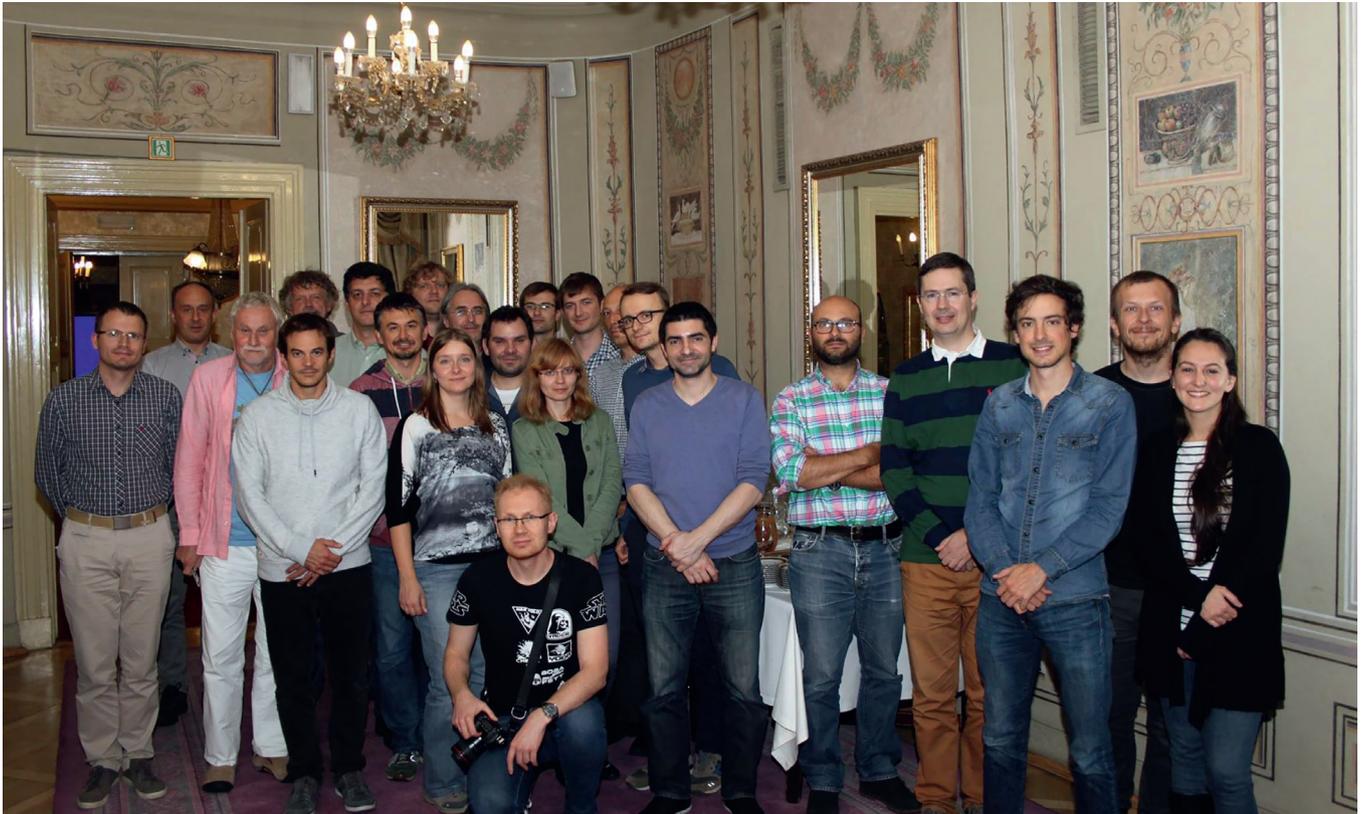

**Fig. 5.** Araucaria Meeting in the Royal Restaurant Wierzynek in Kraków, Poland, 2017.

All these advances allowed us to formulate more ambitious science goals and apply for, and eventually obtain, important new funding from the European Research Council (ERC) in the form of a prestigious ERC advanced grant. This grant, realized from 2016 to 2021, allowed us to build an even stronger science team. Many very valuable collaborators joined our project including Weronika Narloch, Piotr Wielgórski, Bartek Zgirski, Gergely Hajdu, Wojtek Pych, Mikołaj Kałuszyński, Czarek Gałan, Gonzalo Rojas, Megan Lewis, Louise Breuval, Simon Borgniet, Boris Trahin, and Behnam Javanmardi. Louise, Piotr, and Bartek worked mostly on pulsating stars. They produced very precise fiducial period-luminosity relations for classical Cepheids, Type II Cepheids, and RR Lyrae stars in different environments. Gergely studied mostly RR Lyrae stars, and Megan Miras in the Milky Way. Mikołaj and Gonzalo modeled eclipsing binaries. Simon, Boris, and Wojtek worked on spectroscopic data, while Behnam used archival Hubble Space Telescope data to independently determine the Cepheid distance to one of the key SN host galaxies, using new algorithms which allowed a valuable check on the results and methods applied by previous research teams.

Meanwhile we realized that, in order to take full advantage of the huge amount of precision data from satellite missions like Gaia, TESS, or Kepler to improve the extragalactic distance scale, one needs to complement these satellite data with very precise ground-based data. The only way to do that given the extremely large amount of data needed, is to collect them with dedicated telescopes at a very good site. A breakthrough in our project occurred around 2016 when we started to closely collaborate with Rolf Chini's group who operated their own telescopes at Cerro Armazones in northern Chile. Our team members not only had a unique opportunity to obtain the required observations but also to work and collaborate with many colleagues from Bochum including Martin Hass, Michael Ramolla, Francisco Pozo, Moritz Heckstein, Christian Westhues, Catalina Sobrino Figuaredo, Sadegh Noorozi, and others. This particularly successful collaboration allowed us to start a few instrumental projects funded by the Polish Ministry of Science and Higher Education, including building a few new modern telescopes.

On the basis of all our previous work we successfully applied for an ERC Synergy grant on the continuation and extension of the Araucaria Project. Within this grant which started in late 2021, we will build a 2.5m telescope which will extend enormously our observing capabilities, and together with our smaller telescopes, will enable us to collect an enormous amount of high quality photometric and spectroscopic data for eclipsing binaries, pulsating stars, and also for Active Galactic Nuclei which offer a completely different and independent route to build the extragalactic distance scale and extend it out to enormous distances, observing and measuring galaxies in an era when the Universe was only half as old as today. Therefore, our scientific panorama for the next 6 years looks extremely promising, paving the way to a solid measurement of Hubble constant, reaching a 1% accuracy, and even tracing the possible variation of its value over cosmic time.

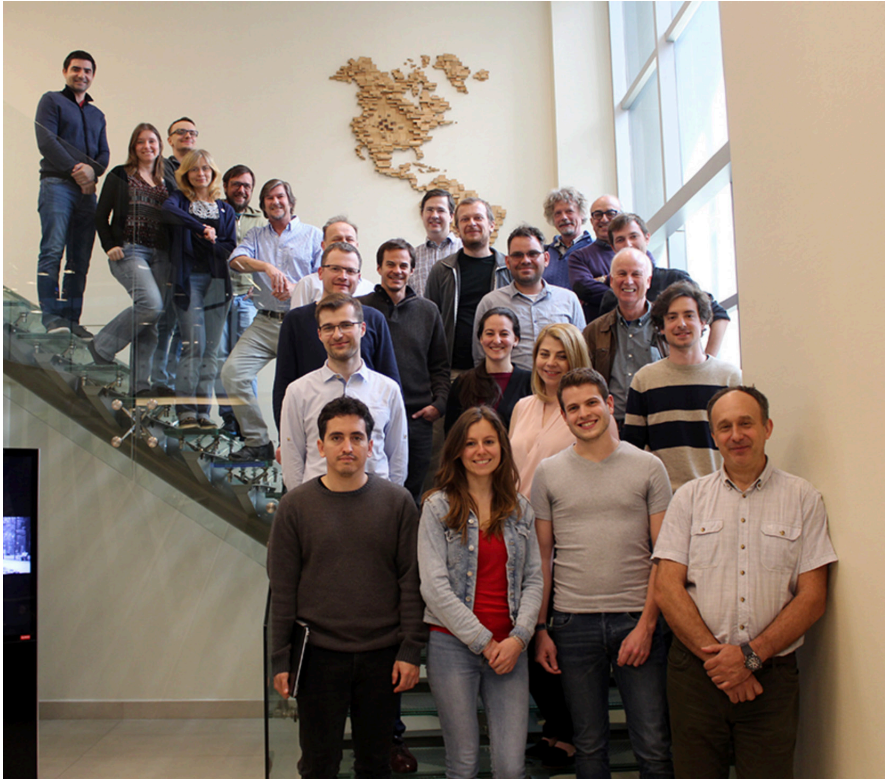

**Fig. 6.** Araucaria Meeting in Concepción, Chile, 2019.

The Araucaria Project has so far produced about 200 papers in the most prestigious astronomical journals, and it is far beyond the scope of this article to mention all the science results it has produced over the two decades since its start in 2000. A few highlights however should be mentioned here:

- in 2001, the first successful survey for classical Cepheids in NGC 300 was finished, pushing the number of known Cepheids in this galaxy from 18 to 117, and demonstrating that our discovery strategy was successful (Pietrzyński et al. 2002, AJ, 123, 789)

- in the same year, we discovered and obtained spectra for a large number of blue supergiants in NGC 300 which were later used to calibrate blue supergiants as distance indicators (Bresolin et al. 2002, ApJ, 567, 277)

- in 2003, a first study on environmental effects on red clump stars in optical and near-infrared bands was published (Pietrzyński et al. 2003, AJ, 125, 2494)

- in 2005, the most accurate-ever (at the time) distance to a galaxy was derived for NGC 300 by applying the multiband optical/near-infrared photometric method developed by the Araucaria Project to its Cepheid variables (Gieren et al. 2005, ApJ, 628, 695)

- also in 2005, an in-depth pulsation modeling study of classical Cepheids yielded a period-age relation for these stars which up to now has become the standard reference in the field (Bono et al. 2005, ApJ, 621, 966)

- in 2006 we published the first paper reporting results from an improved TRGB application to red giants in NGC 300, with data obtained with the Advanced Camera for Surveys onboard the Hubble Space Telescope (Rizzi et al. 2006, ApJ, 638, 766)

- in 2008, we obtained an accurate distance to the Large Magellanic Cloud using RR Lyrae stars in the near-infrared (Szewczyk et al. 2008, AJ, 136, 272)

- also in 2008, the Araucaria Project published a definitive calibration of the FGLR spectroscopic method for blue supergiants which was later used to determine the distances to most of the Araucaria Project target galaxies, with results mostly in excellent agreement with the Cepheid distances (Kudritzki et al. 2008, ApJ, 681, 269)

- in 2009, a first paper analyzing one of the rare eclipsing binary systems in the LMC consisting of a pair of red giants was published, demonstrating the extraordinary potential these systems have for a super-accurate distance determination (Pietrzyński et al. 2009, ApJ, 697, 862)

- also in 2009, a study was published which compared the metallicities and metallicity gradients in NGC 300 as derived from young stars to those derived from H II regions, showing that within very small uncertainties the two methods yield identical results. This is a very important result in the context of distance determination, which needs information about metallicities in the target galaxies whose distances are to be measured (Bresolin et al. 2009, ApJ, 700, 309)

- in the same year, 2009, we published an accurate Cepheid-based distance determination to the so-far most distant galaxy in our project, NGC 247 in the Sculptor Group (Gieren et al. 2009, ApJ, 700, 1141)

- in 2010, our group finished the analysis of the first classical Cepheid discovered in an eclipsing binary system in the LMC. The paper reported the first dynamical mass determination of a classical Cepheid ever, accurate to one percent, and in this way resolved the famous mass discrepancy problem for classical Cepheids showing that the masses predicted by pulsation theory were correct (Pietrzyński et al. 2010, Nature, 468, 542)

- in 2011, we published a re-calibration of the IRSB Baade-Wesselink method using for the first time a large sample of Large Magellanic Cloud Cepheids which set additional constraints on the calibration not available from Galactic Cepheids alone. This new calibration represented a very significant improvement over the previous one and provided much improved distances (Storm et al. 2011, A&A, 534, A94 and A95)

-in 2012, we published precision near-IR photometry of some 200 nearby red clump stars. This opened the road to improve the surface brightness-color relation for late-type stars and was a necessary condition to measure a 1% distance to the LMC in the future. (Laney et al. 2012, MNRAS, 419, 103)

- also in 2012, we announced the discovery of a new class of pulsating stars, called Binary Evolution Pulsators. (Pietrzyński et al. Nature, 484, 75). Pulsating models for such stars were then developed by Araucaria scientist Radek Smolec et al. (2013, MNRAS, 428, 3034) while Paulina Karczmarek studied in detail possible evolutionary channels for such stars.

- in 2013 we presented a 2% distance determination to the LMC based on eclipsing binaries (Pietrzyński et al. 2013, Nature, 495, 56) which was the most precise-ever distance measured to the LMC at the time. The distance to the SMC based on the same technique could be measured by our team with a precision of 3% (Graczyk et al. 2014, ApJ, 780, 59). These were the first very solid results from our decade-long (at that time!) observing efforts on this project and received outstanding attention from the international astronomical community.

- in 2014 we presented the results of the analysis of a very exotic eclipsing binary system which revealed that it is composed of two classical Cepheids (Gieren et al. 2014, 786, 80). It was (and still is) the first-ever firmly established binary system with two Cepheid stars orbiting each other.

- in 2016 Gallenne et al. (A&A, 586, 35), presented a very precise and accurate distance measurement to the binary system TZ For, based on a combination of its astrometric and spectroscopic orbits.

- also in 2016 another important study on chemical abundances in the high-metallicity spiral galaxy M83 was published by Bresolin et al. (ApJ, 830, 64), while Kudritzki et al. (ApJ, 829, 70) reported another distance determination to the galaxy NGC 55 using the FLGR method.

- in 2017 Nardetto et al. published an extensive theoretical study on the p-factor problem in the Baade-Wesselink technique. The important conclusions reached in this paper were based on high-resolution spectroscopic data for the Cepheid prototype star Delta Cephei (A&A, 597, 73).

- also in 2017 a new distance determination to the Fornax dwarf galaxy from near-IR photometry of RR Lyrae stars was announced by Paulina Karczmarek et al. (AJ, 154, 263). This work concluded our long-term effort on determining improved distances to nearby galaxies from this method.

- also in 2017 a new and improved calibration of the FLGR method based on the detailed study of blue supergiants in the LMC, was published by Urbaneja et al. (AJ 154, 102).

- in 2018, an extremely precise determination of the physical parameters for six classical Cepheids members in eclipsing binary systems in the LMC was presented (Pilecki et al. ApJ, 862, 43). This work has yielded the most accurate masses and radii known to date for this important class of pulsating stars. It also concluded our long term efforts on studying classical Cepheids in eclipsing binaries.

- also in 2018 the effect of metallicity on the brightness of classical Cepheids was precisely estimated from a Baade-Wesselink analysis of Cepheids in the Milky Way, LMC and SMC (Gieren et al. A&A, 620, A99). For the first time, a very sizable sample of metal-poor Cepheids in the SMC could be included in the analysis thanks to data which were obtained over five years as part of an ESO Large Programme.

- also in 2018 a new multi-band calibration of the absolute magnitude of the TRGB was published by our group (Górski et al. AJ, 156, 278). This work demonstrated that the TRGB absolute magnitude can be precisely and fully empirically calibrated with multiband photometric data.

- in 2019 several results regarding the binarity fraction of Cepheids (Kervella et al. A&A, 623, 110, Gallenne et al. 2019, A&A, 622, A164) and RR Lyrae stars (Kervella et al. A&A, 623, A116) were published which strongly expanded our previous efforts on this topic.

- also in 2019 we published our most precise LMC distance from eclipsing binaries (Pietrzyński et al. Nature, 567, 200). We managed to achieve a 1% precision, which makes the LMC the best anchor for the extragalactic distance scale at the present time. This paper has been enthusiastically received by the community and has become the standard reference in both cosmological studies and studies on astrophysical objects in the Large Magellanic Cloud.

- also in 2019 and based on Strömgren photometry, Narloch et al. (MNRAS, 489, 3285) detected many candidates for non-pulsating stars located in the main Cepheid instability strip. The physical nature of these objects remains unclear so far but is potentially very interesting.

- also in 2019 the first paper on eclipsing binaries consisting of early type components was published by Taormina et al. (ApJ, 886, 111). Apart from interesting results on the physical parameters of massive stars in the LMC, these results point towards the possibility to use such systems to extend the calibration of the surface brightness-color relation to very early type, hot stars, opening a window for using bright, hot stars in galaxies to determine their distances rather precisely.

- in 2020 we published extended reddening maps of the LMC and SMC which are very important and useful for most of the studies of these galaxies (Górski et al. ApJ, 889, 179).

- also in 2020 an important advance on the empirical calibration of the SBCR based on interferometric data was presented by Nardetto et al. (2020, A&A, 639, A67).

- also in 2020 our group published a 2% geometric distance to the SMC from late type eclipsing binary systems (Graczyk et al. 2020, ApJ, 904, 13). This work improved on our 2014 result and concluded our 15-year-long project, which was one of the most important ones in the Araucaria Project. It has already become the standard reference for the SMC distance in the literature.

- in 2021 we continued studies on binarity of Cepheids and RR Lyrae stars. Pilecki et al. 2021 (ApJ, 910, 118)

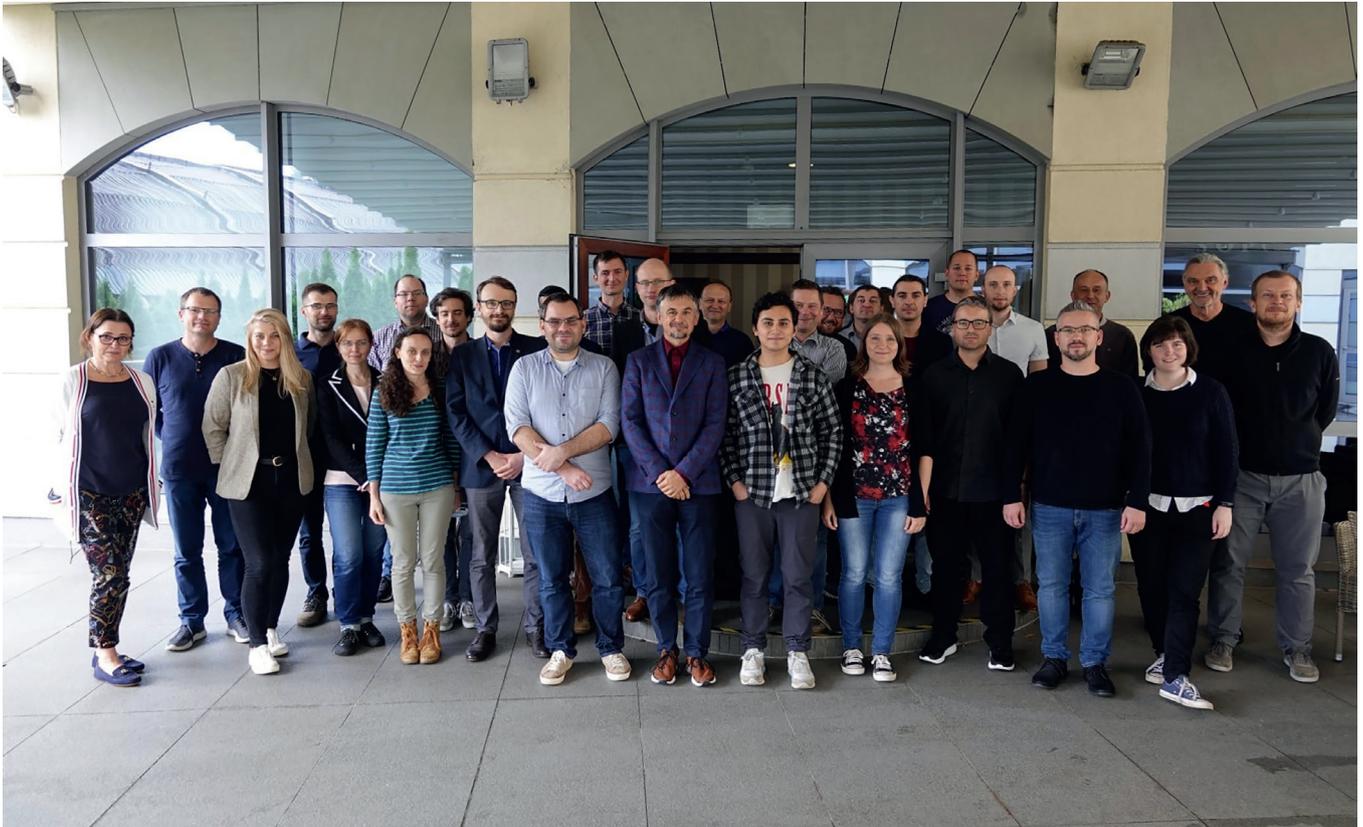

**Fig. 7.** Araucaria Meeting in Sopot, Poland, in 2021, twenty-one years after founding.

detected a numerous population of double-lined binaries in the LMC hosting a Cepheid component; Hajdu et al. (ApJ, 915, 50) presented a very detailed study of binaries containing RR Lyrae stars as detected from a study of their period variability.

- also in 2021, thanks to our geometrical distances and Gaia DR3 parallaxes another precise determination of the Cepheid metallicity effect was derived and published (Breuval et al. ApJ, 913, 38). The results are in excellent agreement with those found in a completely independent way by our group before (Gieren et al. 2018, A&A, 620, A99).

- also in 2021 the first calibration of the SBCR from eclipsing binaries and Gaia DR3 parallaxes was published (Graczyk et al. A&A, 649, A109). The results show the great potential of this approach.

The Araucaria Project team members held several internal workshops over the years (sometimes enriched by external collaborators) which proved to provide excellent opportunities for brainstorming, and social events which helped to strengthen the spirit of friendship among our group, including the ever-increasing number of graduate students who became part of the project over the years.

**Paulina Karczmarek[1], Wolfgang Gieren[1], Grzegorz Pietrzyński[2]**

# Optical survey of classical Cepheids

The Araucaria Project began from, and gained momentum with, a study of a graceful galaxy in the Sculptor Group, NGC 300. It is a spiral, near face-on galaxy, with a metallicity gradient ranging from solar-like in the center to SMC-like on the peripheries.

NGC 300 resembles the Milky Way in almost every way, and hosts a plethora of standard candles: classical Cepheids, blue supergiants, tip of the red giant branch stars, eclipsing binaries, and more. It is a generously equipped cosmic laboratory, where one can study a lot: the metallicity gradient in the disc from the spectroscopic analysis of blue supergiants (Bresolin et al. 2002), the effect of metallicity on the intrinsic brightness of classical Cepheids, but above all, the distances determined independently from different standard candles, especially from classical Cepheids (hereafter just Cepheids). The only problem before the year 2002 was that NGC 300 was virtually uncharted territory, and the reported population of classical Cepheids was only 18 (Graham 1984). Why has NGC 300 not been studied thoroughly? A few reasons stacked up. NGC 300 is a large galaxy in terms of angular size (19′ x 13′), and covering such an area on the sky required a large field camera, which before the year 1999 simply hadn't existed. In 1999 the Wide Field Imager (WFI, FoV: 34′ x 33′) on MPG/ESO 2.2-m telescope at the La Silla Observatory, Chile, was commissioned, making the observations of spatially extended objects feasible and very time-efficient. The enormous images produced by the WFI required dedicated time and human resources to be properly processed and analyzed, which meant that a new postdoc position at the Astronomy Department of the Universidad de Concepción was announced. A resourceful and ingenious Polish astronomer, Grzegorz Pietrzyński, joined the team of Wolfgang Gieren, and they embarked on a scientific journey to find more Cepheids in NGC 300.

Prompted by the fact that NGC 300 shows signs of recent, massive star formation, and therefore should host a significant number of Cepheids, Wolfgang applied for observing time at ESO and was awarded the generous number of 29 nights between July 1999 and January 2000. During this time around 150 images of NGC 300 in each of B, V, I bands were collected. From around 32,000 observed stars Grzegorz selected targets of specific color index (0.4 mag < (B-V) < 1.5 mag), that showed the asymmetrical sawtooth light curves typical for classical Cepheids, and that had larger pulsation amplitudes in the B-band than in the V-band, with the B-band amplitude exceeding 0.4 mag. Crawling through a database of tens of thousands objects took some time but every now and then, a Cepheid would pop out! Soon not only had we recovered all previously known 18 Cepheids, but we increased the known sample by a factor of six, resulting in a collection of 117 Cepheids (Pietrzyński et al. 2002).

[1] Universidad de Concepción, Departamento de Astronomía, Casilla 160-C, Concepción, Chile
[2] Nicolaus Copernicus Astronomical Center, Polish Academy of Sciences, Bartycka 18, 00-716 Warszawa, Poland

The discovered Cepheids span a broad range of pulsation periods, from 115 to 5.4 days, with a period uncertainty of only 0.005-0.01 days, thanks to greatly extended time baseline since the first epoch observations (Graham 1984), which now corresponds to about 76-1000 pulsation cycles. The high accuracy of our data is reflected by the uncertainties of the photometric zero points, which in both B- and V-bands are within 0.03 mag, and agree very well with those present-ed by Freedman et al. (2002). The completeness of our sample drops suddenly for objects fainter than about 22.5 mag in the V-band, corresponding to a pulsation period of around 10 days. In other words, our survey is virtually complete for Cepheids with periods longer than 10 days. The spatial distribution of Cepheids, that shows how they trace the spiral arms of the galaxy, can be appreciated in Figure 1, together with finding charts and light curves of five exemplary Cepheids.

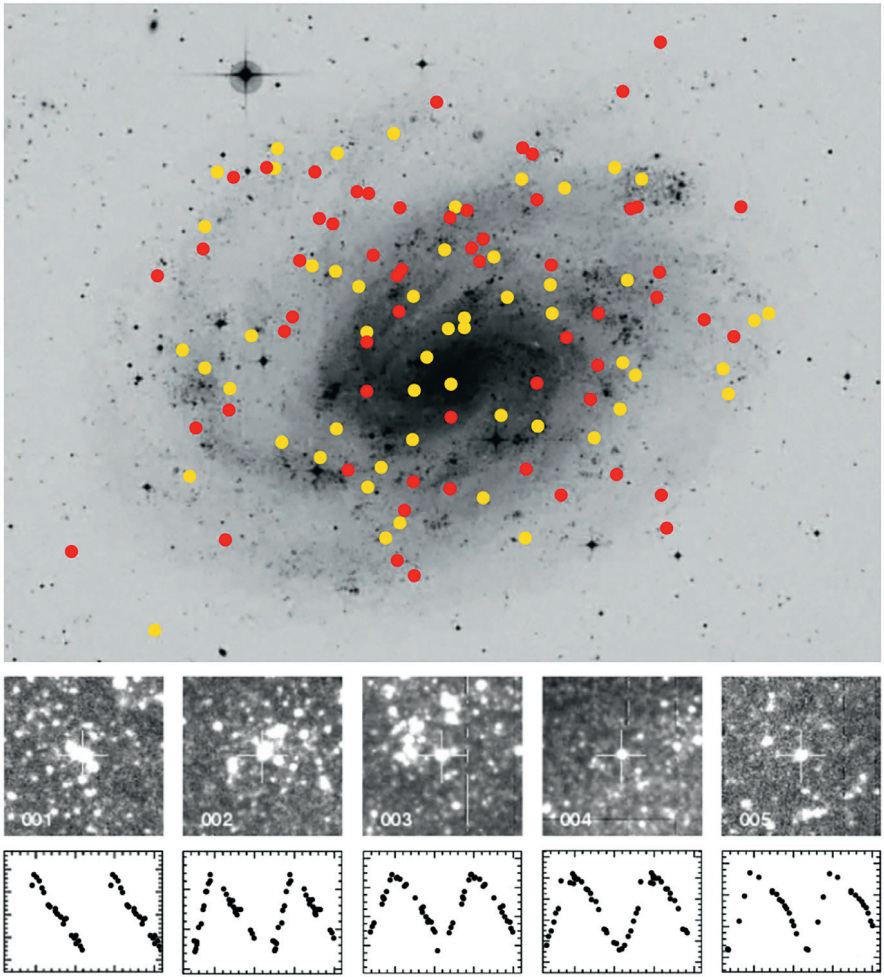

**Fig. 1.** NGC 300 with the positions of de-tected Cepheids with periods shorter (red) and longer (yellow) than 15 days; north is up, and east is to the left. Below are finding charts and corresponding B-band light curves of five exemplary Cepheids. Figure assem-bled from figures 1, 3, 5 of Pietrzyński et al. (2002).

Figure 2 shows that all 117 Cepheids follow the period-luminosity relation in the B- and V-band, however the spread is too large to accurately determine the distance to NGC 300 from the visual passbands alone. A much more accurate distance determination can be achieved in the near-infrared (NIR) domain, where the Cepheids follow the period-luminosity relation more tightly and are less affected by the reddening. The idea of carrying out NIR observations has been integrated into a Large Programme that we applied for at ESO in 2003. A survey in the visual passbands was necessary to find Cepheids by their large amplitude, asymmetrical light curves, and pick the best sample for a follow-up NIR photometry. With the visual and NIR period-luminosity relations one can determine the distance and reddening simultaneously, and derive a true, dereddened distance of superior accuracy (Gieren et al. 2005 and in this volume).

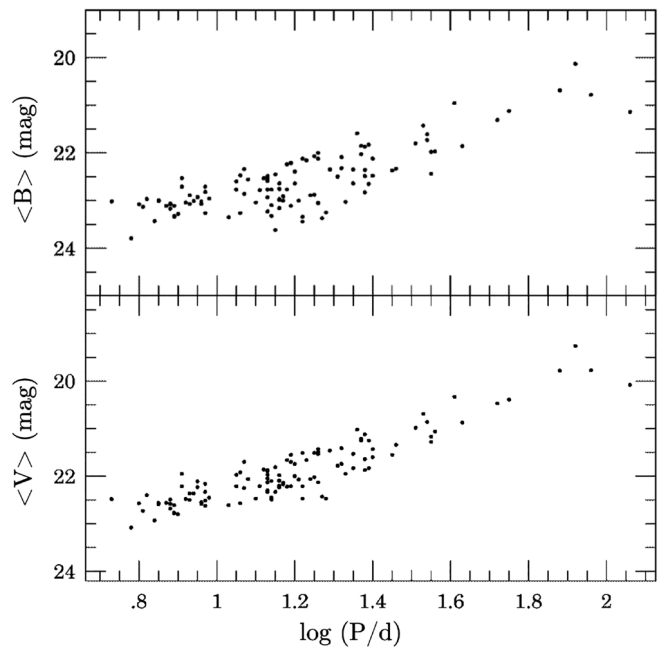

**Fig. 2.** Period-luminosity relations in the B- and V-band for the NGC 300 Cepheids detected in our survey. Adapted from figure 10 of Pietrzyński et al. (2002).

Wolfgang Gieren[1]

# Near-infrared follow-up of extragalactic classical Cepheids previously discovered in optical photometric surveys

Classical Cepheids (hereafter just Cepheids) are the most accurate stellar distance indicators in the range from about 1-30 Mpc. The James Webb Space Telescope will extend this range to perhaps 50-100 Mpc. It is therefore of paramount importance to calibrate the Cepheid method with ever-increasing precision and accuracy.

In a Large Programme approved by the European Southern Observatory to the Araucaria Project members in 2003, many nights became available to our group to perform follow-up photometric observations at the ESO Paranal and La Silla telescopes in the near-infrared J and K bands (at 1.25 and 2.2 microns, respectively) of subsamples of the Cepheids we had previously discovered in optical bands in a number of spiral and irregular galaxies in the Local and Sculptor Groups. Photometric observations of Cepheids in the near-infrared (NIR) spectral region offer three important advantages for distance determination with period-luminosity (PL) relations over work restricted to the optical spectral region. First, absorption corrections in the NIR are much smaller than in optical passbands which is important because Cepheids, as young stars, are usually found in dusty regions in their host galaxies. Second, the intrinsic, or cosmic dispersion of the PL relation due to the finite width of the Cepheid instability strip on the Hertzsprung-Russell diagram decreases with increasing wavelength and is in K only about half of the dispersion in the optical V band, tightening the NIR PL relations and making them a more precise tool for distance determination. Third, and very importantly, even a very few (1-2) random-phase Cepheid magnitudes in a NIR band can already produce a distance determination that can compete in accuracy with the one coming from optical photometry using full, well-sampled light curves – this was demonstrated by our group in Soszyński et al. (2005). One prime interest of the Araucaria Project related to Cepheids was to investigate the effect of metallicity on the PL relation, particularly in NIR bands which are, as mentioned above, the most useful ones for distance determination to galaxies. At the time of starting our project in 2000, there were wildly different results on the "metallicity effect" in the literature; even the sign of the effect was disputed. With the intention of measuring the local Hubble constant, $H_0$, with an accuracy approaching one percent, as pursued for example by the SH0ES Project (Riess et al. 2021), it was clear that a very accurate determination of the metallicity effect was an indispensable element in achieving such an extremely high accuracy for $H_0$.

Our first target galaxy was the southern spiral NGC 300 located in the Sculptor Group. This beautiful galaxy is shown in Fig. 1. Our previous optical survey had discovered 117 Cepheids in NGC 300; a subsample of these objects was observed with the VLT and the ISAAC near-infrared camera at ESO/Paranal in the J and K bands producing the PL relations shown in Fig. 2. Comparison with the corresponding PL relations in the Large Magellanic Cloud obtained by Persson et al. (2004), demonstrated that the slopes of the PL relations in NIR bands were exactly the same, providing the first convincing empirical evidence that the slope of the Cepheid PL relation in NIR bands is not sensitive to me-tallicity. This conclusion was later strengthened by our work on other spirals and irregulars in the Local and Sculptor Group.

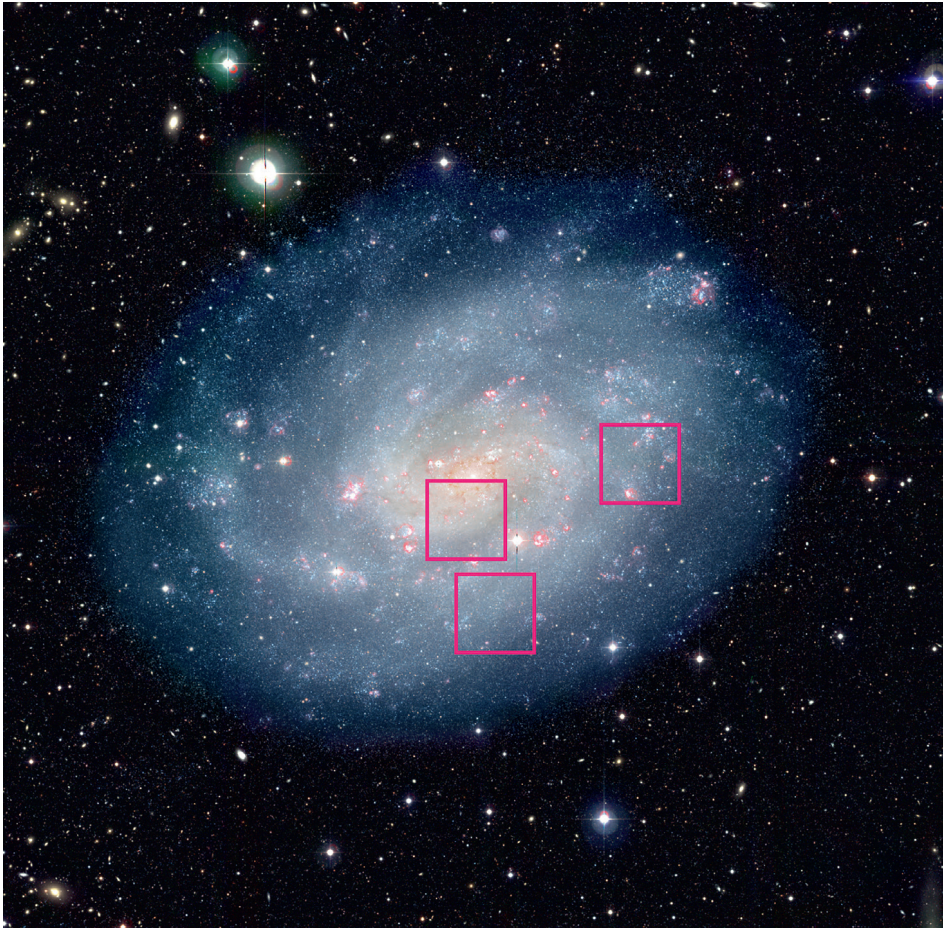

**Fig. 1.** Location of the observed VLT ISAAC fields in NGC 300, excerpted from Gieren et al. (2005).

The best way to determine the true distance modulus of a target galaxy was found to combine the results for the apparent, reddened distance moduli in optical and NIR bands (VIJK) and plot them against the ratio of selective to total absorption $R_\lambda$, assuming a reddening law (we used Schlegel et al. 1998). The corresponding plot for NGC 300 is shown in Fig. 3. A linear fit to the data points in this diagram reveals both, the total reddening affecting the Cepheids of the target galaxy, and the true distance modulus of the target galaxy from the intersect of the fit line with the $(m-M)_\lambda$ axis. The case of our pilot galaxy NGC 300 produced the most accurate distance for a galaxy located outside the Local Group at the time, reaching a distance accuracy of 3% (Gieren et al. 2005), proving that our strategy of combining optical and near-infrared photometry of Cepheids in the described way was indeed capable of producing distance results of exquisite accuracy.

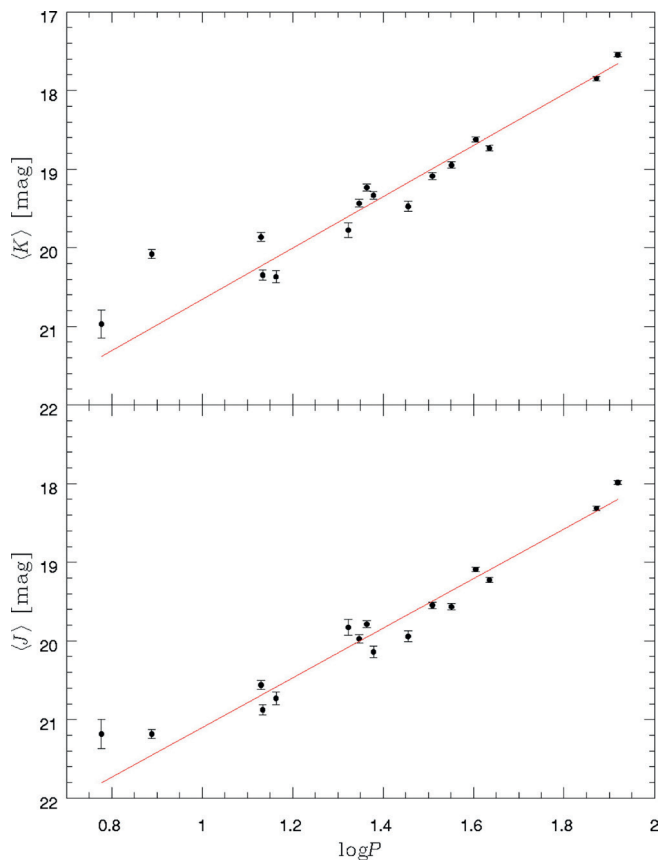

**Fig. 2.** Final adopted intensity mean magnitudes of the Cepheids in NGC 300, plotted against logP (in days). Overplotted are the best-fitting lines, for which we have adopted the slopes measured by Persson et al. (2004) on a large sample of LMC Cepheids. Figure excerpted from Gieren et al. (2005).

Taking advantage of the great amount of observing time allotted to us at ESO, and also at Las Campanas Observatory in Chile to collect both, optical multi-epoch photometry for Cepheid discovery and NIR follow-up photometry of extragalactic Cepheid samples, we measured very accurate Cepheid-based distances to five Local Group galaxies (IC 1613: Pietrzyński et al. 2006; NGC 6822: Gieren et al. 2006; NGC 3109: Soszyński et al. 2006; WLM: Gieren et al. 2008; M33: Gieren et al. 2013) and four spirals in the nearby Sculptor Group (NGC 300: Gieren et al. 2005; NGC 55: Gieren et al. 2008; NGC 247: Gieren et al. 2009; NGC 7793: Zgirski et al. 2017). This set of precision distances, together with the near-geometrical 1-2% eclipsing binary distances to the Magellanic Clouds measured by the Araucaria Project scientists, forms a unique sample of galaxies of very different environmental properties with extremely well-known distances allowing us to improve, among other things, our understanding of stellar astrophysics and stellar evolution. A number of such studies have been carried out by our occasional collaborators (e.g. Prada Moroni 2012; Marconi et al. 2013).

Last but not least, the decade-long work on Cepheid variables of the Araucaria Project has finally succeeded in determining the metallicity effect on Cepheid NIR PL relations with high precision, paving the way for an extremely accurate determination of $H_0$. From two very different techniques, we now know that there *is* a significant effect of metallicity on the J, H and K band PL relations of size 0.2 mag/dex, in the sense that more metal-rich Cepheids are intrinsically more luminous than their more metal-poor counterparts of the same pulsation period. One of these studies compared the Cepheid PL relations for Milky Way, LMC and SMC Cepheids derived by using the Infrared Surface Brightness Technique (Gieren et al. 2018) whereas the other study used Gaia EDR3 parallaxes of Milky Way Cepheids in tandem with the very accurate eclipsing binary distances of LMC and SMC measured by the Araucaria Project (Breuval et al. 2021, and this volume). The agreement between these independent and very precise determinations makes us believe that the long-standing question of the metallicity sensitivity of Cepheid PL relations is now definitively settled.

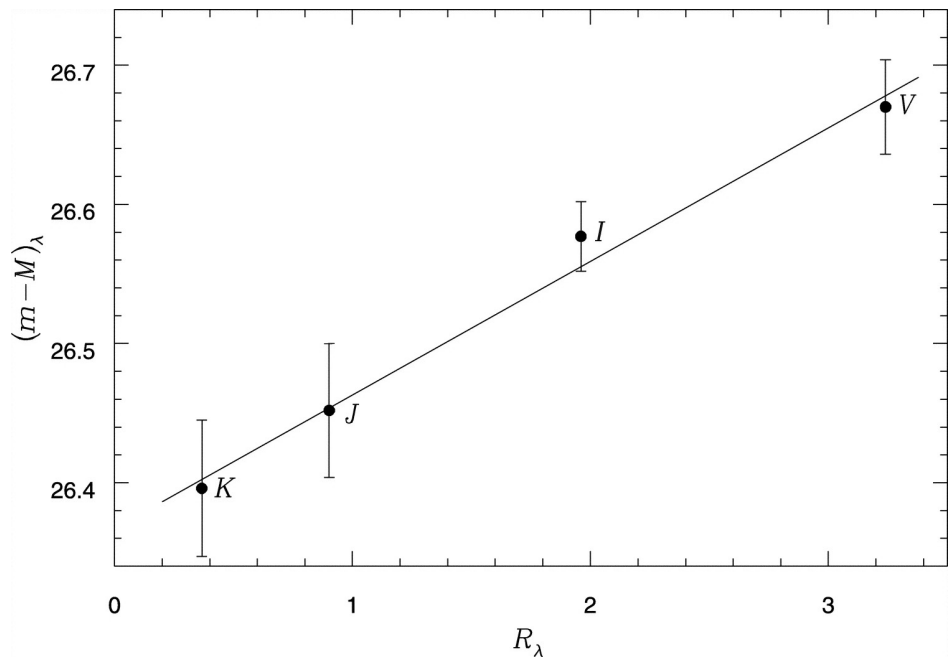

**Fig. 3.** Distance moduli for NGC 300 measured in different photometric bands, against the ratio of selective to total absorption for these bands. The slope of the best-fitting line gives the total color excess, and the intersect gives the true distance modulus of the galaxy. Figure excerpted from Gieren et al. (2005).

Marek Górski[1]

# Tip of the Red Giant Branch stars as distance indicators

The tip of the red giant branch (TRGB) is a sharp cut-off of the bright part of the red giant branch visible on the color-magnitude diagrams of globular clusters and galaxies. It marks the final stage of evolution for low-mass stars during the red giant branch (RGB) phase terminated by a helium flash. The immediate reason the TRGB can be used as a standard candle is that all stars with masses smaller than 2 solar masses have a similar brightness just before the helium flash. This brightness is, however, modified by the metallicity of the stellar atmosphere for particular bands (bolometric correction). While in the optical I-band the metallicity effect is small, the magnitude of the TRGB in the near-infrared K-band is changed by more than 0.5 mag with 1 dex change of the metallicity.

The first attempt to use the tip stars to measure distance was made by Baade in 1944, when he observed the central region of the Andromeda galaxy and its two companion galaxies, M32 and NGC 205 with red-sensitive photographic plates. He noticed that the brightest red giants in all three galaxies have the same magnitude and color, which led him to the conclusion that all three galaxies are at the same distance. In 1971 Sandage found that the tip stars in the IC 1613 galaxy have the same absolute magnitude as the brightest red giants in the M31 and M33 galaxies. During the next two decades more sophisticated observing techniques were developed, and with the arrival of CCD measurements, the TRGB I-band magnitude was used to determine the distances to almost all Local Group galaxies. The improvement of this technique was also achieved with the introduction of the quanti-tative edge detection technique – the Sobel filter (Lee, Freedman & Madore 1993). Comparison of the distances obtained with the TRGB I-band brightness with distances obtained with the Cepheid period-luminosity relation and RR Lyrae stars for 10 Local Group galaxies showed that all three distance-measurement methods yield distances that agree to within 5% (Lee, Freedman & Madore 1993). To date, the I-band TRGB method has been applied to determine the distances to more than 300 galaxies up to 17 Mpc (Jacobs et al. 2009; Tully et al. 2016; Hatt et al. 2018).

The TRGB is one of a very few distance-measurement methods that can be used to calibrate the absolute brightness of Supernovae (SN) Ia, which are used to measure the Hubble constant. Based on the TRGB I-band absolute magnitude calibration in the Large

[1] Nicolaus Copernicus Astronomical Center, Polish Academy of Sciences, Bartycka 18, 00-716 Warszawa, Poland

Magellanic Cloud, Freedman et al. (2019) and Yuan et al. (2019) measured distances to 18 SN Ia type host galaxies and calculated the value of the Hubble constant. While Yuan et al. (2019) obtained value $H_0 = 72.4 \pm 1.9$ km/s/Mpc, consistent with the value obtained with the Cepheid distance ladder, Freedman et al. (2019) obtained value of $H_0 = 69.8 \pm 1.3$ km/s/Mpc which sits in the middle of the currently defined Hubble tension. This discrepancy illustrates how important it is to properly calibrate the absolute magnitude of the TRGB, taking into account reliable measurement of the apparent magnitude of the TRGB, transformation of the ground-based to Hubble Space Telescope photo-metric system and applying reddening correction. Additional errors of this distance-measurement method can occur when the bright part of the RGB is not well populated or when AGB stars are contaminating the red giant branch. Finally, the most important systematic error of distance measurements with the I-band TRGB is the reddening and extinction, caused by dust located in the Milky Way and target galaxies.

To improve the TRGB method to obtain distances with precision better than a few percent, it is necessary not only to precisely calibrate the absolute magnitude of the TRGB, but also to utilise other bands and calibrate the metallicity dependence of the absolute magnitude – especially in the near-infrared bands. In 2004 Valenti, Ferraro and Orgilla calibrated the J-, H- and K-band in terms of the metallicity based on the measurements of the TRGB brightness in the galactic globular clusters. This calibration was used by Pietrzyński et al. (2009) and Górski et al. (2011) to obtain distances to seven Local Group galaxies. Unfortunately, the uncertainty of the zero-point of the calibration based on the globular clusters, metallicity error propagation and sparse possibilities to measure the red giant branch metallicities limit the usage of this calibration. Furthermore, Górski et al. (2016), showed that population effects can lead to 5% errors of the distances when the calibrations of Valenti, Ferraro and Orgilla (2004) are used. This effect was also studied theoretically and predicted by Salaris & Girardi (2005) and Serenelli et al. (2017).

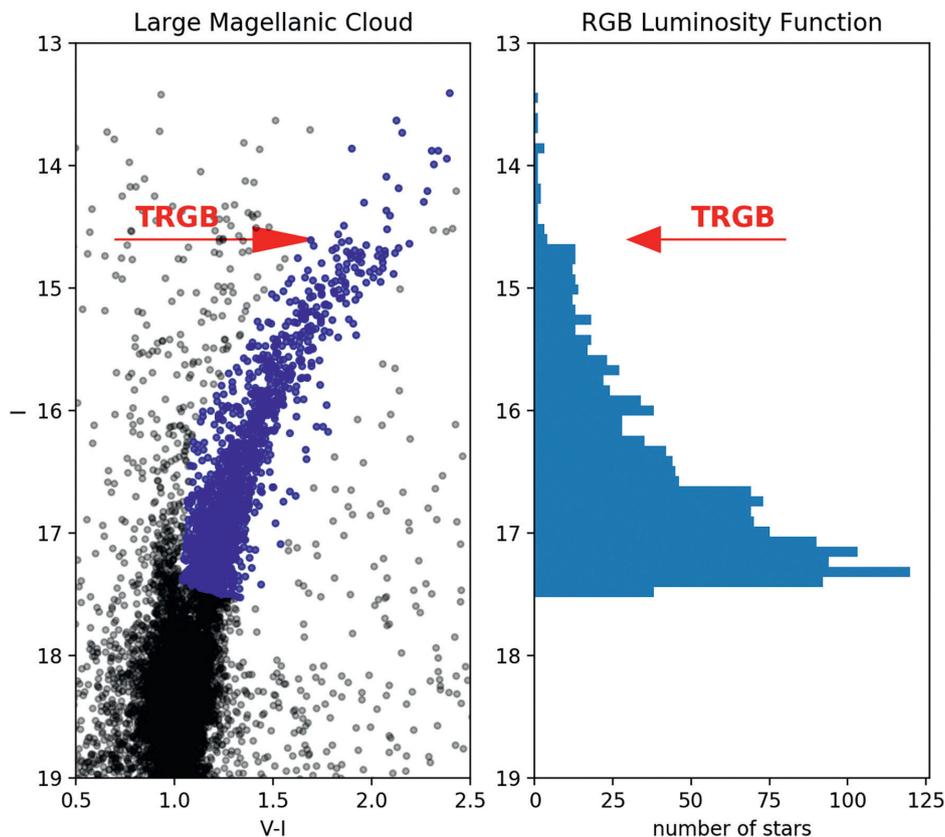

**Fig. 1.** The color-magnitude diagram (left panel) of stars in the central part of the Large Magellanic Cloud. The corresponding luminosity function (right panel) is created only with selected red giant branch stars (blue points). The red arrows mark detected TRGB.

The most promising approach is based on the calibrations of the TRGB absolute magnitude in terms of the color of the red giant branch. This technique is not only more feasible, but also population effects have much less impact on derived distances. Madore et al.

(2018) and Górski et al. (2018) prepared preliminary calibrations of the absolute magnitude of the TRGB in the near infrared J-, H- and K-band in terms of the (V-I), (J-K) and (V-K) colors. Both calibrations are, however, limited to a narrow range of colors, and have to be verified with scrutiny before they can be applied to measure distances with required precision.

**Marek Górski[1]**

# Red Clump stars as distance indicators

The red clump (RC) stars are core-helium-burning, low-mass stars, forming a cluster located on the red giant branch, visible on the color-magnitude diagram. They are the metal-rich equivalent of the horizontal-branch stars. The average brightness of the RC seems to be a very attractive distance indicator, because even in dwarf galaxies there are a lot of clump stars, and therefore the statistical uncertainty of the average brightness and color of the RC is usually very small (below 0.01 mag). Therefore, RC measurements have been an important base for investigations of stellar structures, distances and reddenings, especially in the Milky Way Bulge and Magellanic Clouds.

In the mid-1990's, based on the I-band brightness of the RC, a series of distance measurements to the galactic center, Magellanic Clouds and M31 galaxy were published. Those measurements were tied to the RC absolute brightness calibrated with a sample of roughly 600 stars in the solar neighborhood with parallaxes known to better than 10% in the Hipparcos catalog (Paczyński & Stanek 1998; Stanek et al. 1998; Stanek & Garnavich 1998, Udalski et al. 1998). Those measurements supported the "short distance" to the Magellanic Clouds, however they were assuming negligible effects of age and metallicity on the I-band brightness of the RC.

Soon after, the near-infrared absolute magnitudes of the RC were calibrated with the Hipparcos parallaxes by Alves et al. (2002). In contrast to the optical measurements, the distance modulus to the Large Magellanic Cloud (LMC) based on the near-infrared K-band using the Alves et al (2002) calibration turned out to be consistent with the classical 18.5 mag value (Pietrzyński and Gieren 2002; Laney et al. 2012). Theoretical studies by Cole (1998) and Girardi (1998) suggested that this obvious disagreement of the LMC distances is caused by population effects affecting the I-band brightness of the RC. Surprisingly, the first empirical investigations indicated that while there is a dependence on the metallicity, the dependence is linear and fairly weak (Sarajedini 1999, Udalski 2000). In 2002, Salaris and Girardi pointed out that the brightness of the RC is affected by numerous population effects in a complex way, which makes it impossible to calibrate RC absolute magnitude with simple linear relations. Employing the star formation history, chemical evolution history and initial mass function, they provided theoretical corrections for RC brightness in the Magellanic Clouds.

[1] Nicolaus Copernicus Astronomical Center, Polish Academy of Sciences, Bartycka 18, 00-716 Warszawa, Poland

While those corrections bring the distance modulus of the Magellanic Clouds close to the expected value of 18.5 mag, they are laden with significant uncertainties of population synthesis models.

The final empirical confirmation that population effects strongly affect the V- and I- band magnitude of RC stars in a complex way was done by Pietrzyński et al. (2010). They compared the brightness of the TRGB and RC in 23 nearby galaxies based on precise V- and I-band photometry available from the Hubble Space Telescope. This is a direct and superb approach, because the absolute I-band magnitude of the TRGB depends very weakly on metallicity and age (variations are smaller than 0.1 mag), and the reddening affecting apparent magnitudes of stars in principle should have very similar values for the TRGB and RC stars. As the outcome, Pietrzyński et al. (2010) showed that the V- and I-band RC stars are not an accurate method for the determination of distances to nearby galaxies, because population effects are significant and reach almost 0.5 mag.

Despite the large variation of the absolute brightness of the I-band, the (V-I) color of the RC is much more stable (Girardi et al 1998, Cole 1998, Pietrzynski et al. 2010), which makes the RC color excess measurement an important tool to map the reddening in nearby stellar systems, especially the Galactic Bulge and Magellanic Clouds.

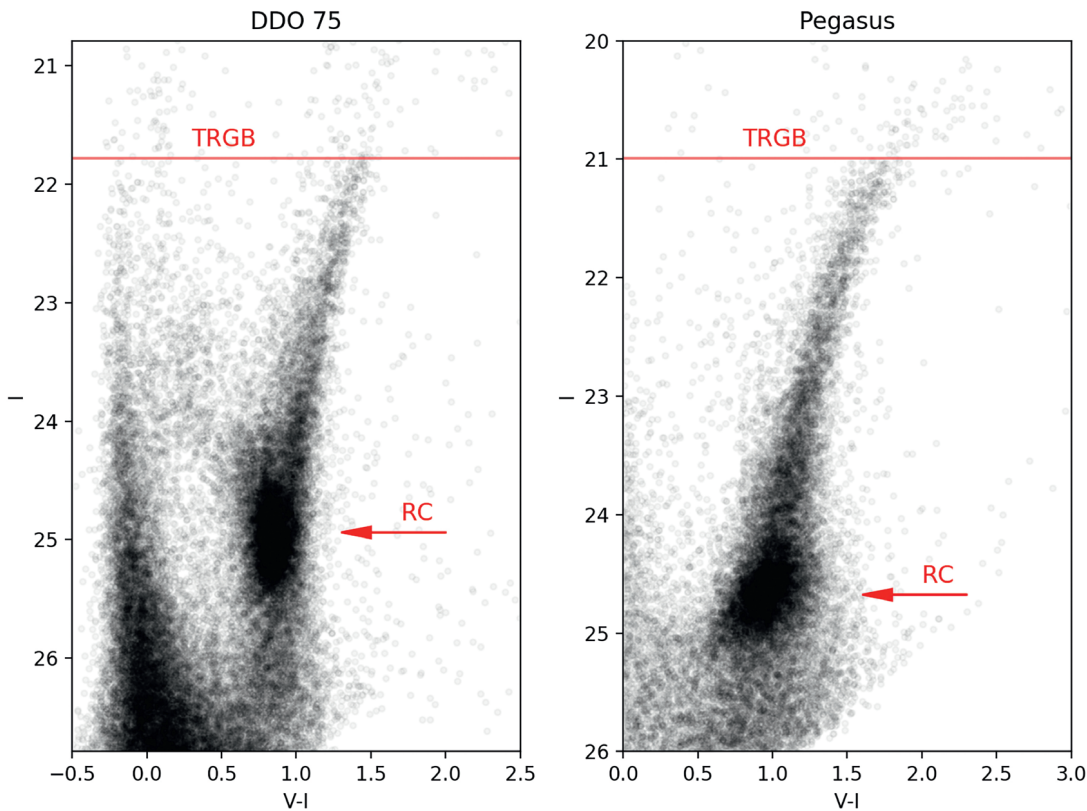

**Fig. 1.** The color-magnitude diagrams of the red giant branches of the DDO 75 galaxy (left panel) and Pegasus dwarf irregular galaxy (right panel). Red arrow marks the average brightness of the red clump. The red horizontal line marks the TRGB. It can be easily seen that the magnitude difference between the TRGB and the RC is higher by roughly 0.5 mag. for the Pegasus dwarf compared to DDO 75 galaxy.

The most recent reddening maps of the Magellanic Clouds were prepared by Górski et al. (2020) and Skowron et al. (2021). In order to obtain RC (V-I) color excess, Skowron et al. (2021) calculated the fiducial (V-I)$_0$ RC color by correcting apparent color for the foreground dust of the Milky Way in the outer areas of the LMC and SMC. To obtain the intrinsic color of the red clump, Górski et al. (2020) used reddenings obtained from late-type eclipsing binary systems, measurements for blue supergiants and reddenings derived from Strömgren photometry of B-type stars. The reddening values are however, again, affected by population effects (metallicity and age gradients in galaxies), fortunately, to much lesser extent that the RC I-band brightness.

Although the RC is an important tool to investigate reddening and structure of different stellar systems, its application for distance measurements requires much better understanding of the influence of numerous population effects and detailed analysis of the star formation history and chemical evolution of target galaxies.

**Bartłomiej Zgirski[1]**

# Distance determinations based on carbon stars*

Atmospheres of thermally-pulsating, old (100 Myr - 3 Gyr) asymptotic giant branch (AGB) carbon stars have a surplus of carbon relative to oxygen, which makes them significantly redder (with temperatures between 2500K and 3900K) than their oxygen-rich evolutionary progenitors. The convective envelopes of carbon stars transport carbon to the surface from the helium-burning shell during the third dredge-up. This alters the molecular opacity of the stellar photosphere, which decreases its effective temperature (Marigo et al. 2008).

Luminosities of carbon stars are subject to the *funneling effect* of confinement, which makes these stars interesting candidates for distance indicators. Carbon-rich photospheres are obtainable only for stars having masses from a relatively compact interval, and this directly restricts their magnitudes. The hot-bottom burning that occurs in massive AGB stars results in conversion of carbon into nitrogen at the base of their convective envelopes. Even though the exact mass value needed for ignition of such a process depends on metallicity of a star, it typically occurs for $M > 3.5\,M_{\odot}$, but may appear even for $M > 2\,M_{\odot}$ for stars having low metallicity. On the other hand, the convective envelopes of low-mass ($M < 1.3\,M_{\odot}$) AGB stars are not extensive enough to bring carbon to their surface.

Richer et al. (1984) used the mean value of the luminosity function of carbon stars in the I-band to determine the distance to a nearby galaxy, NGC 205, for the first time. That work and the several subsequent studies were based on optical VRI photometry. Narrow-band color (8100 - 7800)Å was used to distinguish M-type stars, having strong TiO absorption lines, from carbon stars.

Nikolaev & Weinberg (2000) used 2MASS near-infrared photometry to divide the CMD of the Large Magellanic Cloud into different regions. One of them, the well-separated J-region, is the habitat of carbon stars. In their second work, Weinberg & Nikolaev 2001 showed that J-region stars yield a well-defined, linear luminosity-color relation $K_s$ vs. $(J-K_s)$. They showed that central values of luminosity distributions in *JHK* are standard candles when founded on narrow 0.1-mag $(J-K_s)$ color intervals. Authors were able to trace the spatial structure of the LMC using this fact. Moreover, from the results of their work, it was clear that the central J-magnitude of carbon stars is constant for the whole color interval of the J-region stars defined as $(J-K_s) \in (1.4, 2.0)$ mag.

Revival of distance determinations using central, typical values of carbon stars' luminosity functions in near-infrared has come with the works of Ripoche et al. (2020) and Madore & Freedman (2020). Authors used median and mean luminosity functions of carbon stars, respectively, in order to derive distances to nearby galaxies. First results showed a very good agreement with TRGB distances (Freedman & Madore 2020).

The Araucaria team decided to take advantage of our vast database containing near-infrared photometry of nearby galaxies, gathered originally for the purpose of distance determinations using multiband Cepheid period-luminosity relations. In order to take into account contamination of J-band luminosity functions of carbon stars, we applied a profile that was a superposition of a Gaussian and of a quadratic function, proposed originally to model luminosity functions of Red Clump stars by Paczyński & Stanek (1998). The central value of the Gaussian component serves as a standard candle. We have calibrated the method in the LMC based on the IRSF photometry of Kato et al. (2007) and the very accurate distance to the LMC of Pietrzyński et al. (2019) and reddening maps of Górski et al. (2020).

We obtained excellent agreement with previously determined distances from Cepheids for eight neighbors of the Milky Way outside of the Magellanic System. Our work shows that the method allows for precision comparable to that obtained using Cepheids.

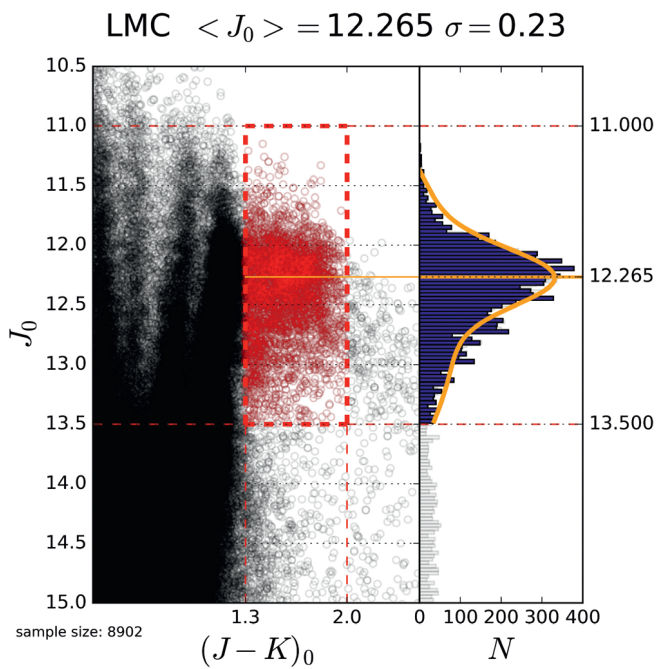

**Fig 1.** [fig1.png] Carbon stars in the LMC and their corresponding luminosity function with the fit of the custom profile.

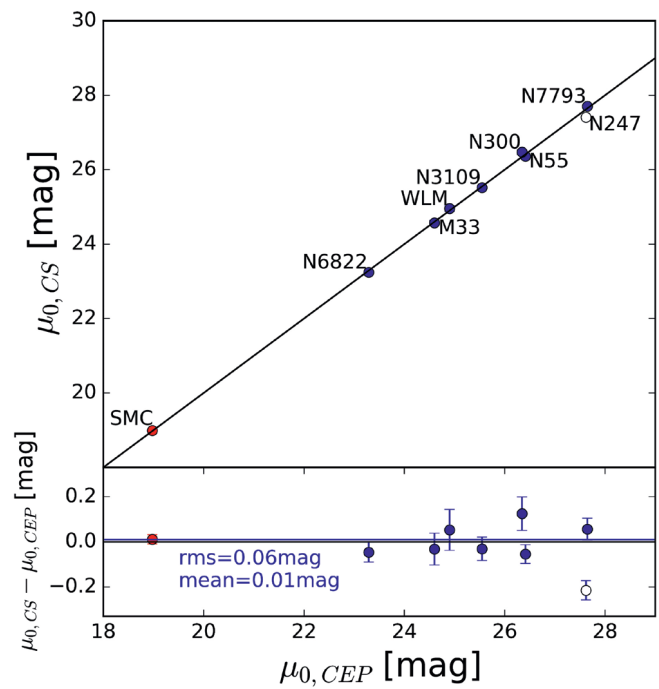

**Fig 2.** [fig2.png] Comparison between distances obtained from carbon stars and classical Cepheids – the agreement is very good with a mean difference of 0.01 mag, and a corresponding spread of 0.06 mag for 7 out of 8 galaxies.

Even though it is not yet entirely clear what, among contamination by foreground stars and background galaxies, causes differences in the shapes of luminosity functions of carbon stars from different populations, our method allows for more precise determinations of their central magnitudes and thus serves as a precise tool of distance determinations.

Central magnitudes of carbon stars in J-band correspond to those of classical Cepheids having periods of around 20 days. With the advent of the new generation of telescopes this method may allow us to establish distances to objects 50-60 Mpc away.

**Paulina Karczmarek[1]**

# RR Lyrae stars as near-infrared distance indicators to Local Group galaxies

Complementary to classical Cepheids, RR Lyrae (RRL) variables offer a way to measure distances to clusters and galaxies that host old (10-12 Gyr) and metal-poor ($-2.5$ dex < [Fe/H] < $-0.5$ dex) stars.

In such environments classical Cepheids (belonging to the young and metal-rich population) are often non-existent, which gives RRL stars some leverage over classical Cepheids. Another advantage of RRLs is that they closely obey a period-luminosity-metallicity (PLZ) relation in the near-infrared (NIR) J and K bands, and are considerably less affected by metallicity than in the optical domain (Bono et al. 2003, Catelan et al. 2004, Sollima et al. 2008). Characteristic light curves of RRLs make them easy to identify in optical photometric surveys of globular clusters and nearby galaxies, while the NIR peak-to-peak amplitudes are typically 0.3 mag (Storm et al. 1992, Marconi et al. 2003), 2-3 smaller than in the optical bands. That means, that instead of covering the entire phased light curve with data points in order to calculate the mean magnitude, only one random-phase data point can approximate the mean magnitude in the NIR. As a result, NIR observations of RRLs are very time efficient, since just one deep exposure of a large field of view suffices to extract single-phase magnitudes of dozens of RRLs in that field. The reduced effect of interstellar extinction in the NIR domain, known from the previous chapters, applies to RRLs as well. All of these features make RRL stars in the NIR domain a reliable tool for determining distances to nearby globular clusters and galaxies.

In the course of the Araucaria Project, we determined the distances to five Local Group galaxies in the southern hemisphere, using the PLZ relation for RRL stars in the NIR domain: the LMC, the SMC, Sculptor, Carina, and Fornax, as shown in Table 1. All distance moduli were calculated with the same method, which secured homogeneity and high quality results (Karczmarek et al. 2021). The main points of this method are: (i) collection of data using world-class infrared cameras SOFI/NTT and HAWK-I/VLT in the ESO Observatories in La Silla and Paranal, respectively; (ii) reduction, photometry, and calibration with a pipeline tailored for the purpose of the Araucaria Project; (iii) calculation of absolute magnitudes of RRLs from the literature PLZ relation; (iv) plot of random-phase RRL magnitudes on the period-luminosity plane and linear least squares fit, which yields the zero point of the calibration, while the slope is intentionally fixed on the literature value to minimize the systematic error (Figure 1); (v) correction of the zero point for the effect of reddening.

The final result from the method described above is the true distance modulus. About 10 different NIR PLZ relations are available in the literature up to date (e.g. Bono et al. 2003, Catelan et al. 2004, Dékány et al. 2013, Sollima et al. 2008) and can be employed simultaneously to determine the averaged distance to a target galaxy with the relative error at the level of 5%. This error originates mainly from the zero point of the PLZ calibrations, and will be reduced to about 3% once more accurate Gaia parallaxes for thousands of Galactic RRL stars are available (Gould & Kollmeier 2017).

In the near future we plan to determine distances to two more dwarf galaxies: Draco and Ursa Minor, which are located in the northern sky. This will require expanding the inventory of NIR instruments to include for example WFCAM/UKIRT in the Mauna Kea Obser-vatory. Draco and Ursa Minor are among the most faint and distant, and the least studied galaxies in the Local Group. Dozens of RRL stars in Draco and Ursa Minor have been reported in the literature (Bellazzini et al. 2002; Bonanos et al. 2004; Kinemuchi et al. 2008, 2016) but all the studies have been carried out in the optical domain. This means that the accuracy of the distance determinations from the optical data of RRLs could be impaired by unresolved or underestimated errors of reddening and/or metallicity.

The effort put in reducing the statistical error and controlling the systematic error of the distance determination using RRLs in the NIR domain may contribute to significant improvement in the accuracy of secondary distance indicators. Thus, the NIR RRL distance determinations to the Local Group galaxies are not only vital but also very timely, because RRLs coupled with the Tip of the Red Giant branch stars have been proposed as an alternative and independent method towards determination of the Hubble constant (Hatt et al. 2017), which might shed more light on the nature of the Hubble tension.

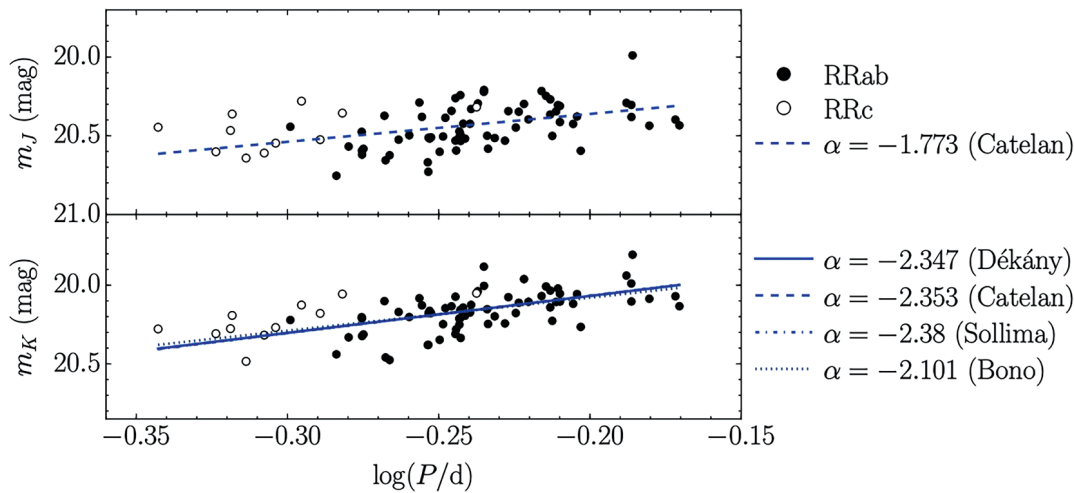

**Fig. 1.** Apparent magnitudes of RRL stars in the Fornax galaxy as a function of their log pulsational period. Lines denote different slopes of PLZ relations, α, extracted from the literature. The figure excerpted from Karczmarek et al. (2017b).

**Tab. 1.** Distance moduli to five Local Group galaxies determined within the Araucaria Project

| Galaxy | $n_{RRL}$ | $(m - M)_0 \pm \sigma_{syst.} \pm \sigma_{stat.}$ | Reference |
|--------|-----------|---------------------------------------------------|-----------|
| Sculptor | 78 | $19.67 \pm 0.02 \pm 0.12$ | Pietrzyński et al. (2008) |
| LMC | 65 | $18.58 \pm 0.03 \pm 0.11$ | Szewczyk et al. (2008) |
| SMC | 34 | $18.97 \pm 0.03 \pm 0.12$ | Szewczyk et al. (2009) |
| Carina | 33 | $20.118 \pm 0.017 \pm 0.110$ | Karczmarek et al. (2015) |
| Fornax | 77 | $20.818 \pm 0.015 \pm 0.116$ | Karczmarek et al. (2017a) |

**Piotr Wielgórski[1]**

# Type II Cepheids as distance indicators

Pulsating stars are extremely important in astrophysics. They are unique laboratories to study stellar interiors using asteroseismology and to improve our understanding of stellar evolution. Radially pulsating stars from the main Instability Strip are of vital interest within the Araucaria Project as they serve as very precise distance indicators. In this article we will summarize our research related to old members of the Cepheid family – Type II Cepheids.

Evolutionary scenarios leading to formation of Type II Cepheids are still not fully understood, but it is believed that they are AGB and post-AGB stars with masses of about $0.5\ M_\odot$ and metallicity in the range from –2.5 to 0 dex. They are divided into four subtypes: short period BL Herculis stars, medium period W Virginis and peculiar W Virginis stars, and long period RV Tauri stars. As they are population II stars, they are observed in the Galactic bulge, old Galactic disc and globular clusters. Several hundred Type II Cepheids are known in the Magellanic Clouds, mainly from observations of the Optical Gravitational Lensing Experiment (OGLE, Soszynski et al. 2018), and they are also observed in other Local Group galaxies (Majaes et al. 2009).

Achieving high-accuracy distance measurements is not possible without a deep understanding of the distance indicators used, thus Araucaria research focuses also on precision measurements of physical parameters of stars. Pilecki et al. (2017) and Pilecki et al. (2018) performed a very detailed analysis of two peculiar W Virginis stars, OGLE-LMC-T2CEP-098 and OGLE-LMC-T2CEP-211, located in the Large Magellanic Cloud. As a result, they obtained a precision determination of physical parameters of the stars including the first measurements of dynamical masses of Type II Cepheids, which are $0.64 \pm 0.02$ and $1.51 \pm 0.09\ M_\odot$ for OGLE-LMC-T2CEP-098 and OGLE-LMC-T2CEP-211, respectively.

Period-luminosity relations of Type II Cepheids have been determined in the optical and near-infrared bands in Galactic globular clusters (Matsunaga et al. 2006), Galactic bulge (e.g. Bhardwaj et al. 2017) and in the Magellanic Clouds (e.g. Soszynski et al. 2018; Ripepi et al. 2015). Ciechanowska et al. (2010) used the near-infrared photometry of a sample of Type II Cepheids in the SMC, obtained as part of the Araucaria Project with the SOFI camera (ESO La Silla Observatory), to measure the distance to this galaxy. They obtained an SMC distance modulus of $18.85 \pm 0.07$ (statistical) $\pm\ 0.07$ (systematic) mag. Before Gaia, the zero-point of the distance scale of Type II Cepheids was based on Hipparcos parallaxes of two nearby stars and its accuracy was worse than 10% (Feast et al. 2008). Moreover, one of these stars seems to be a peculiar W Virginis type star. Currently, the Gaia space mission is delivering accurate parallaxes of stars within 5 kpc from the Sun, which gives us the opportunity to calibrate period-luminosity relations using stars of the general field. Wielgórski et al. (2022, in prep.) selected a sample of nearby Type II Cepheids which can be used to calibrate the period-luminosity relation and collected precision photometry of these relatively bright stars in the near-infrared and in the visual passbands in the Observatorio Cerro Armazones with the IRIS and VYSOS 16 telescopes. This unique dataset and Gaia Early Data Release 3 parallaxes were used to calibrate for the first time the period-luminosity relations for Type II Cepheids in the solar neighborhood in J, H and $K_S$ bands and WJK wesenheit. They improved the accuracy of the Type II Cepheid distance scale zero-point to about 5%. The obtained period-luminosity relation in $K_S$ band is presented in Figure 1.

[1] Nicolaus Copernicus Astronomical Center, Polish Academy of Sciences, Bartycka 18, 00-716 Warszawa, Poland

As Type II Cepheids are radially pulsating stars, the Baade-Wesselink (BW) technique can be applied to measure their distances and radii. The main source of uncertainty in the BW method is the projection factor (p-factor), which translates the observed radial velocity into the velocity of the pulsating surface. The first empirical measurement of the p-factor of a Type II Cepheid was done by Breitfelder et al. (2015). They found the p-factor of κ Pavonis to be $1.26 \pm 0.07$. The aforementioned studies of Pilecki et al. (2017, 2018) resulted in very similar values of the p-factor for the analyzed stars. Using the photometric data used in Wielgórski et al. (2022, in prep.) from OCA and high resolution spectra from state-of-the-art spectrographs we will perform a BW analysis of Type II Cepheids in order to calibrate their p-factors. As a by-product, we will determine precision radii of the stars, which will set constraints on models of Type II Cepheids. An example BW analysis of a nearby Type II Cepheid, V971 Aql, is presented in Figure 2.

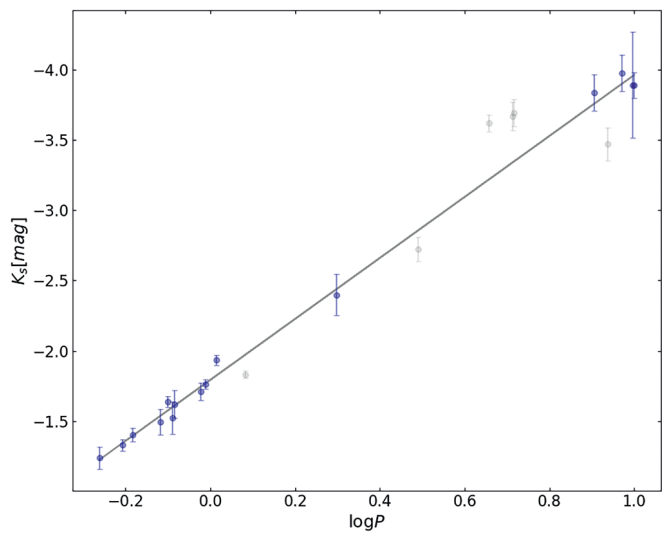

**Fig. 1.** Period-luminosity relation in $K_S$ band for Type II Cepheids in the solar neighbourhood based on Gaia EDR3 parallaxes (Wielgórski et al. 2022, in prep.).

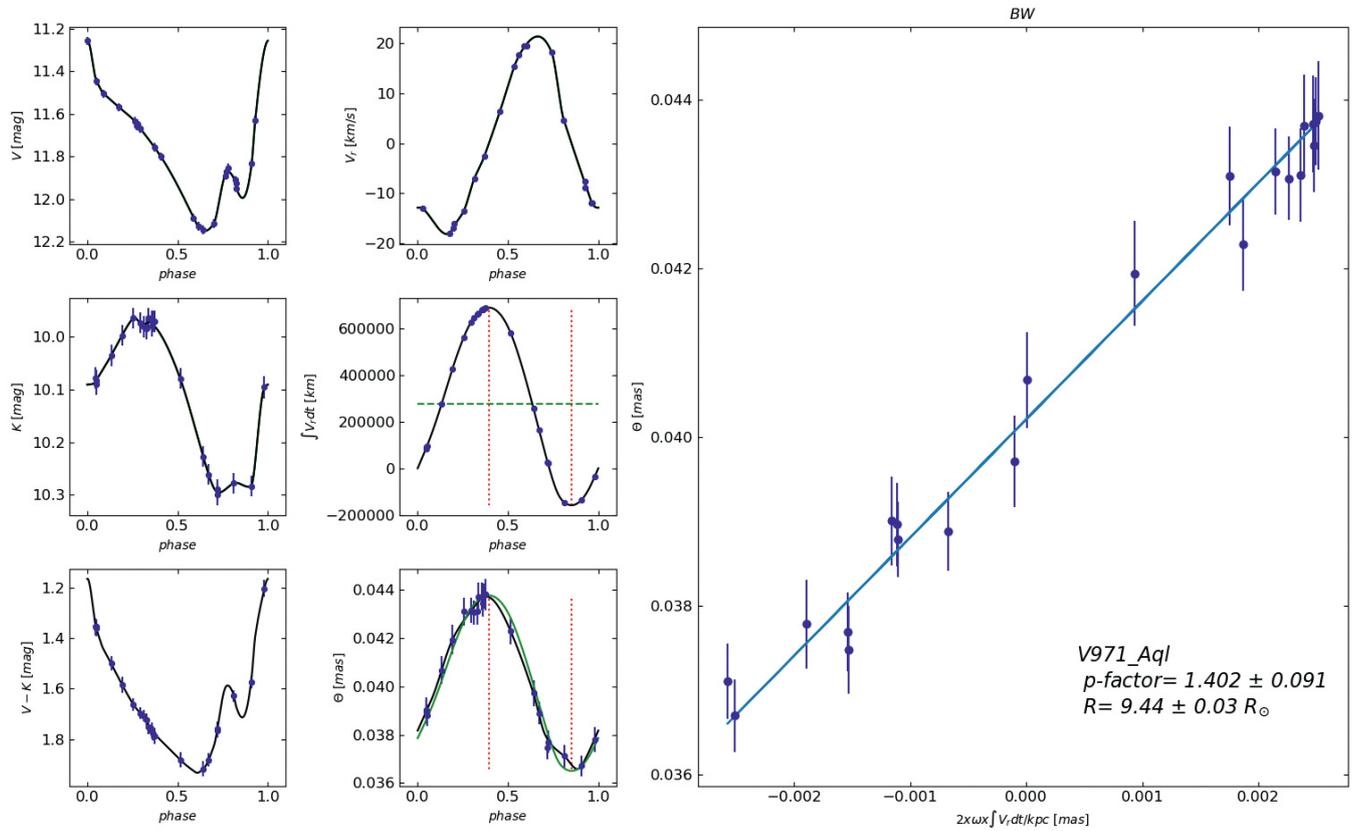

**Fig. 2.** The Baade-Wesselink analysis of V971 Aql.

**Louise Breuval[1,2]**

# The Influence of Metallicity on the Cepheid Leavitt Law

Cepheids are pulsating variable stars which play a key role as primary distance indicators thanks to the empirical relation between their pulsation period and intrinsic luminosity, the period-luminosity (PL) relation (Leavitt & Pickering 1912). This law is used to calibrate the brightness of type-Ia supernovae (SNe Ia) in nearby galaxies, which is in turn adopted to measure the distance to remote galaxies in the Hubble flow. This method, known as the extragalactic distance ladder, provides the current expansion rate of the Universe, the Hubble constant (Riess et al. 2021).

One of the most important remaining systematic uncertainties on the distance ladder is the influence of metallicity on the PL relation. It has been demonstrated that the chemical composition of Cepheids impacts their intrinsic brightness. Therefore, the difference in metallicity between Cepheids used to calibrate the PL relation and Cepheids in SNe Ia host galaxies must be taken into account by including a corrective term in the PL relation.

To measure this effect, we compared the Leavitt law in the Milky Way, where Cepheids are mostly metal-rich ([Fe/H] = +0.08 dex), with its Large Magellanic Cloud (LMC) and Small Magellanic Cloud (SMC) counterparts, where Cepheids are more metal poor ([Fe/H] = –0.41 dex and [Fe/H] = –0.75 dex respectively). In the Milky Way, we adopted Gaia EDR3 parallaxes corrected for the zero-point offset, after performing a quality selection based on the RUWE parameter provided in the Gaia catalog. For Magellanic Cloud Cepheids, we adopted the distances recently measured by Pietrzyński et al. (2019) and Graczyk et al. (2020) with a remarkable precision using late-type detached eclipsing binaries. We slightly corrected these distances to take into account the inclination of the LMC and SMC, and therefore used individual distances to each Cepheid depending on their position in the Clouds.

In order to avoid the presence of outliers, we only selected Cepheids within a radius of 3° and 0.6° around the LMC and SMC center, respectively. Apparent mean magnitudes were obtained from well covered light curves and were corrected for extinction. After applying a Monte Carlo simulation to calibrate the PL relation in the three galaxies, we concluded in Breuval et al. (2021) that metal-rich Cepheids are intrinsically brighter than metal-poor ones. This is equivalent to including a negative metallicity term ($\gamma$) in the Leavitt law.

[1] Department of Physics and Astronomy, Johns Hopkins University, Baltimore, MD 21218, USA
[2] LESIA, Observatoire de Paris, Université PSL, CNRS, Sorbonne Université, Université de Paris, 5 place Jules Janssen, 92195 Meudon, France

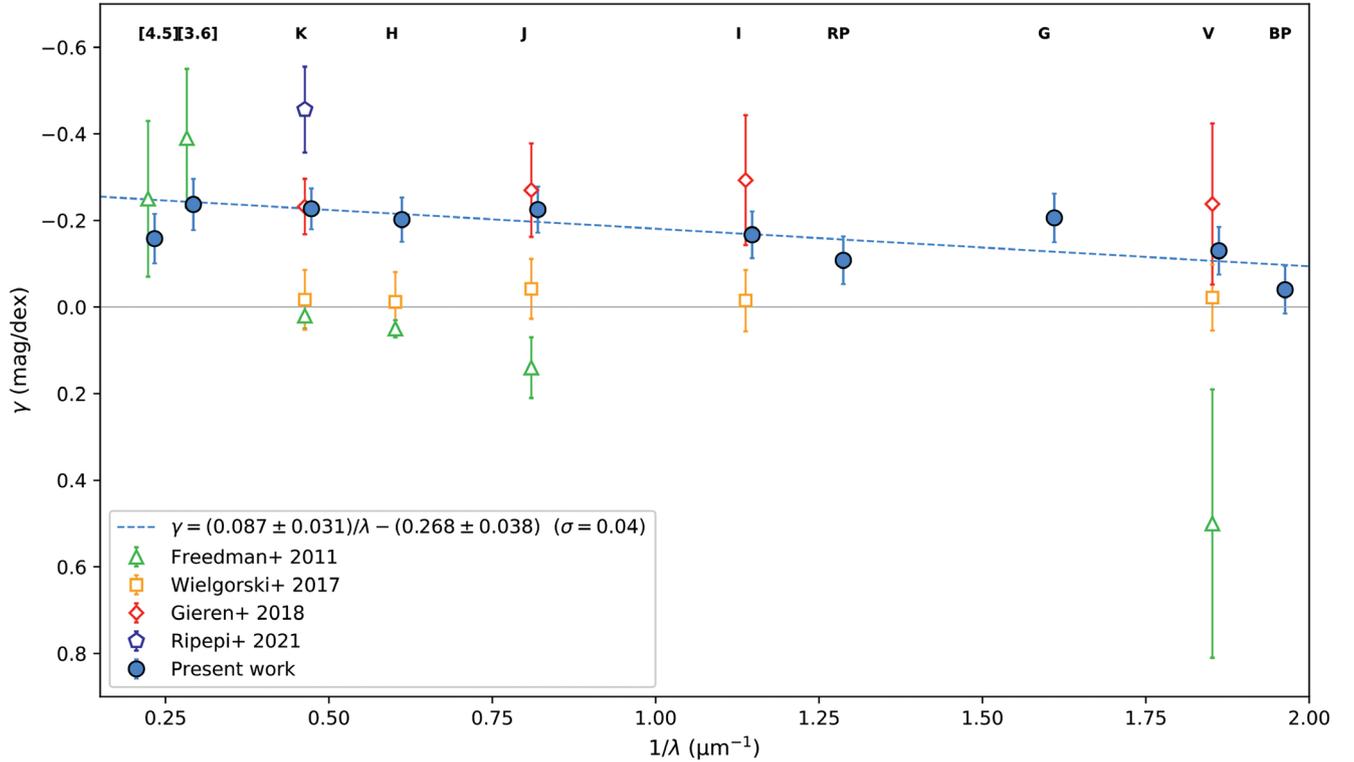

**Fig. 1.** The effect of metallicity on Cepheid brightness as a function of wavelength. Figure excerpted from Breuval 2021, PhD thesis, Université PSL.

The metallicity effect that we obtained in optical, NIR and mid-infrared bands is represented as a function of wavelength in Figure 1. A slight trend suggests that the metallicity effect is stronger (in an absolute sense) in the infrared than in the optical; in the mid-infrared and near-infrared, we obtained a significant metallicity effect of $\gamma \sim -0.20 \pm 0.05$ mag/dex, whereas in the optical it is also negative but weaker, with $\gamma \sim -0.10 \pm 0.06$ mag/dex.

These results agree well with the Baade-Wesselink analysis carried out by Gieren et al. (2018), although they derived a stronger metallicity effect in the optical with larger error bars. On the other hand, Wielgórski et al. (2017) obtained a metallicity effect consistent with zero and independent of the wavelength, based on a purely differential analysis between the LMC and SMC Leavitt law. A study by Freedman & Madore (2011) shows a similar trend although with a steeper slope and positive $\gamma$ values towards the optical. Finally, the most recent NIR estimate from Ripepi et al. (2021) suggests a very strong metallicity effect of -0.6 mag/dex. In conclusion, the size of this effect is still not fully established, but our results confirm its negative sign, meaning that metal-rich Cepheids are brighter than metal-poor ones from mid-infrared to optical wavelengths. Additionally, our study provides the most precise estimate to date of the metallicity effect on the Cepheid PL relation.

The PL calibration, although improved in recent years, still needs to be refined and several systematic effects must be better understood to overcome the remaining uncertainties in the extragalactic distance ladder. Publication of the next Gaia data releases should provide even more precise parallaxes for Cepheids and a better estimate of the zero-point. Regarding the metallicity effect, most studies still rely on mean metallicities in the three galaxies due to a lack of precision in the current individual abundance measurements. Therefore, precise metal abundance measurements for Galactic, as well as Magellanic Cloud, Cepheids are paramount. Cepheids in open clusters are very promising thanks to their precise average parallax: this sample will be observed in the Hubble Space Telescope photometric system for consistency with the measurements of Cepheids in SNe Ia hosts, which should considerably reduce the systematics related to the photometric zero point. Finally, the upcoming launch of the James Webb Space Telescope is expected na Finally, the James Webb Space Telescope is expected.

Ksenia Suchomska[1]

# Physical and orbital properties of Bulge late-type eclipsing binaries

One of the most important tasks in astronomy is understanding the basic physics and evolution of stars. Precise determination of physical parameters of stars, such as mass, radius, luminosity or metallicity, gives us an opportunity to better understand stellar structure and evolution, as well as to understand the evolution of the galaxies they are located in.

Although modern theories of stellar evolution are able to predict these parameters at every stage of evolution with high-accuracy, such predictions were calibrated mainly based on main-sequence stars, giving us no grounds to think that they are also correct for well-evolved stars. It is also worth mentioning that the comparison of more and more precise observations with theoretical models of stars shows some inaccuracies, which means that we are in need of deeper examination and refinement of some theoretical aspects, such as the theory of convection and the effect of overshooting.

It is worth noticing that stellar evolutionary models rely strongly on mass, radii or effective temperature, therefore the accuracy of determination of those parameters plays a major role. As was noted by Torres et al. (2010), only parameters determined with a precision higher than 3% can provide sufficiently strong constraints such that models with inadequate physics can be rejected. Well-detached eclipsing binary systems, where the components are evolved stars, give us an opportunity to derive the physical parameters of each component with such precision, meaning they are great candidates for such analyses. Calculations of their orbits allow us to directly determine the masses of their components with the required accuracy, which also gives us a chance to estimate other physical parameters.

We have to realize though, that detached double-lined eclipsing binaries (DEB SB2), where the components are two evolved stars, are not so easy to detect. This is mainly due to their long orbital period and relatively short eclipses. Nevertheless, the Araucaria Project, based on analysis of photometric data for over 50 millions stars in the center of the Milky Way, as well as in the Magellanic Clouds, collected by the OGLE project, selected over a dozen of these unique systems.

What is more, so far only a few such binaries in our Galaxy have been analyzed (Suchomska et al. 2015, 2019; Hełminiak et al. 2015, 2019).

In order to determine physical and orbital parameters of five selected systems in the Galactic bulge

[1] Nicolaus Copernicus Astronomical Center, Polish Academy of Sciences, Bartycka 18, 00-716 Warszawa, Poland

and disc, we collected both photometric and spectroscopic data using world class telescopes. The V-band and I-band optical photometry was obtained with the Warsaw 1.3 m telescope at Las Campanas Observatory during the third and fourth phase of the OGLE project (Udalski 2003; Soszyński et al. 2012, 2016), and the near-infrared J-band and K-band photometry was collected with the use of the ESO NTT telescope at La Silla Observatory equipped with the SOFI camera. As for the spectroscopic data, we used high resolution spectra obtained with the use of the Clay 6.5-m telescope at Las Campanas Observatory, equipped with the MIKE spectrograph and with the ESO 3.6-m telescope at La Silla Observatory, equipped with the HARPS spectrograph.

We derived absolute physical and orbital parameters of our systems using the Wilson-Devinney code (WD code, van Hamme & Wilson 2007; Wilson & Devinney 1971). The WD code allows us to simultaneously solve multi-band light curves and radial-velocity curves, which is essential when it comes to obtaining a consistent model of a binary. To measure radial velocities of the components we used the RaveSpan software (Pilecki et al. 2017) which uses the Broadening Function formalism. In the case of all of our investigated systems, both light curve and radial-velocity curve coverage was sufficient to perform the analysis (Figure 1).

We managed to determine the physical and orbital parameters of the binaries with an accuracy of ~2%

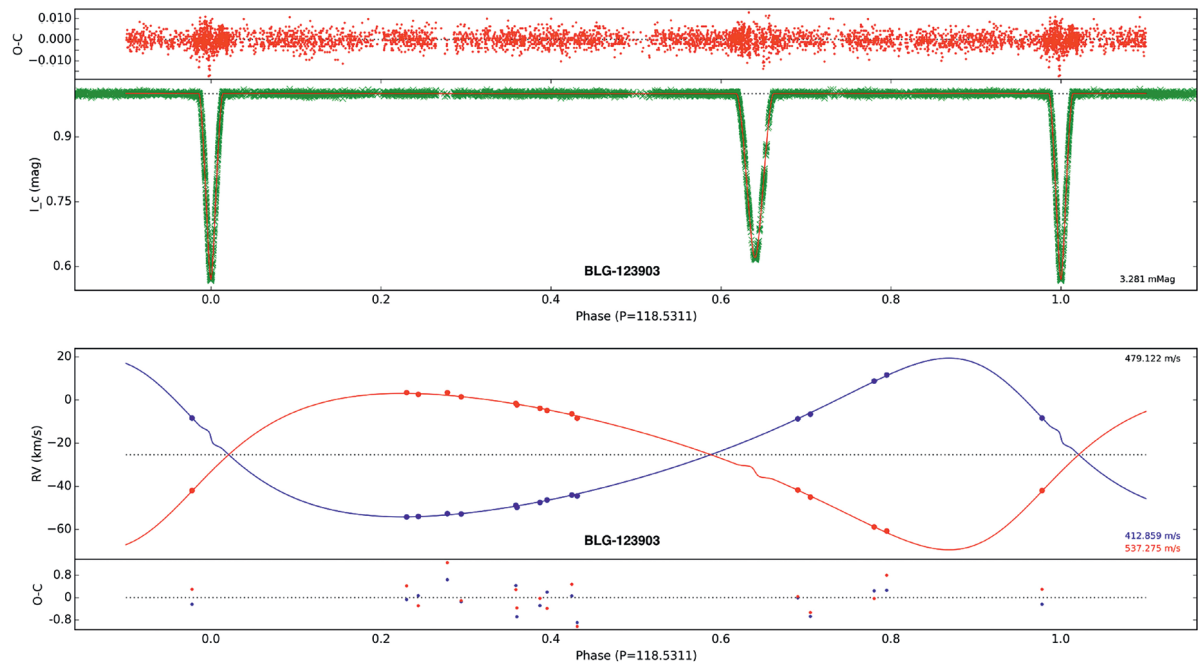

**Fig. 1.** The I-band light curve and radial-velocity solutions from the Wilson-Devinney code for OGLE-ECL-BLG-123903. The residuals of the fits are listed at the right ends of the panels.

(Table 1). We also measured the distances towards the systems using the surface brightness-colour relation, which is very well established for late-type stars by di Benedetto et al. (2005). The accuracy of our measurements was ~3%, which is comparable but slightly less accurate than the distances to the Large/Small Magellanic Cloud binaries obtained using the same method (Pietrzyński et al. 2009, 2013; Graczyk et al. 2012, 2014). This is caused mainly by the much higher interstellar extinction, as well as uncertainties in the infrared photometry for this sample of binaries. Therefore, there is significant room for improvement in the derived distances with more precise infrared photometry. Moreover, this method of distance determination also serves as an independent way of testing the distance and parallax determinations provided by the Gaia mission.

**Tab. 1.** Physical properties of the BLG-123903 system.

| Property | The Primary | The Secondary |
|---|---|---|
| Spectral type | K2 III | K2 III |
| $V^a$ (mag) | 17.607 | 17.709 |
| $I^a$ (mag) | 14.785 | 14.889 |
| $J^a$ (mag) | 12.697 | 12.804 |
| $K^a$ (mag) | 11.529 | 11.637 |
| $V-I$ (mag) | 2.822 | 2.820 |
| $V-K$ (mag) | 6.077 | 6.072 |
| $J-K$ (mag) | 1.168 | 1.167 |
| Radius ($R_\odot$) | $9.540\pm 0.049$ | $9.052\pm 0.060$ |
| Mass ($M_\odot$) | $2.045 \pm 0.027$ | $2.074 \pm 0.023$ |
| $\log g$ | $2.790 \pm 0.011$ | $2.841 \pm 0.012$ |
| $T_{\rm eff}$ (K) | $4780^b \pm 131$ | $4786^c \pm 180$ |
| $v \sin i$ (km s$^{-1}$) | $8.86 \pm 1.51$ | $8.86 \pm 1.44$ |
| Luminosity ($L_\odot$) | $43 \pm 7$ | $39 \pm 6$ |
| $M_{\rm bol}$ (mag) | 0.684 | 0.789 |
| $M_{\rm v}$ (mag) | 1.028 | 1.131 |
| $[{\rm Fe/H}]^b$ | $0.14 \pm 0.23$ | $0.39 \pm 0.28$ |
| $E(B-V)$ | $1.329 \pm 0.114$ | |
| Distance (pc) | $2953 \pm 59.1({\rm stat.})$ | $\pm 70.2$ (syst.) |

$a$−observed

$b$−atmospheric analysis

$c$−WD solution

The investigated sample of well-detached late-type eclipsing binary systems in our Galaxy nearly doubles the number of well characterized giant stars in the Milky Way. As already mentioned before, calculations of evolutionary status of evolved stars are sensitive to many parameters, and therefore the observation and analysis of additional systems, where the accuracy of the derived physical parameters is below 2%, is much needed.

Alexandre Gallenne[1,2]

# The role of interferometry in measuring high-precision distance and mass of binary stars

Long-baseline optical interferometry (LBI) is the only optical technique giving access to high-angular-resolution astronomy down to milli-arcsecond spatial scales. Interferometry combines the light coming from two or more telescopes to produce interference fringes, from which the astrophysical information is extracted (see e.g. Lawson 2000).

LBI only recently emerged, during the last decades, because of the complexity of the systems. Astronomers always want larger telescopes to gain collecting power and angular resolution, however, it is technologically difficult to construct a single large mirror telescope. The upcoming ~40m-class telescopes will bring huge improvement in light-gathering power compared to the current ~8-m class telescopes (~20 times better), but LBI will still provide an angular resolution > 3 times better at a given wavelength. This is because the resolution does not depend on the telescope diameter, but on the baseline between them. For instance, the longest baseline at the Center for High Angular Resolution Astronomy (CHARA, ten Brummelaar et al. 2005) is 330 m which provides a resolution of ~1 mas in the K band, while the current 8 m-class telescopes reach (in theory) ~30 mas and ~11 mas for the upcoming Extremely Large Telescopes at the same wavelength.

LBI is now delivering more and more outstanding results in several fields of astronomy, from exoplanets to black holes (see e.g. Gravity Collaboration et al. 2019, 2018). One of them is the project I am leading about precisely measuring the distance and masses of stars in binary systems using interferometry. The main advantages of using LBI are: (i) the capability to observe binary systems with small separations ($\lesssim$ 30 mas) and (ii) the capability to obtain high-precision astrometric measurements, as small as a few micro-arcseconds. The combination of astrometric measurements with radial velocities of both stars in the system offers a unique opportunity to precisely determine both the orbital parallax and dynamical masses in a geometrical and model-independent way. Precise distance determination is important for the cosmological distance scale, while precise masses are crucial to calibrate and test theoretical stellar evolution models.

[1] Universidad de Concepción, Departamento de Astronomia, Casilla 160-C, Concepción, Chile
[2] Unidad Mixta Internacional Franco-Chilena de Astronomía (CNRS UMI 3386), Departamento de Astronomía, Universidad de Chile, Camino El Observatorio 1515, Las Condes, Santiago, Chile

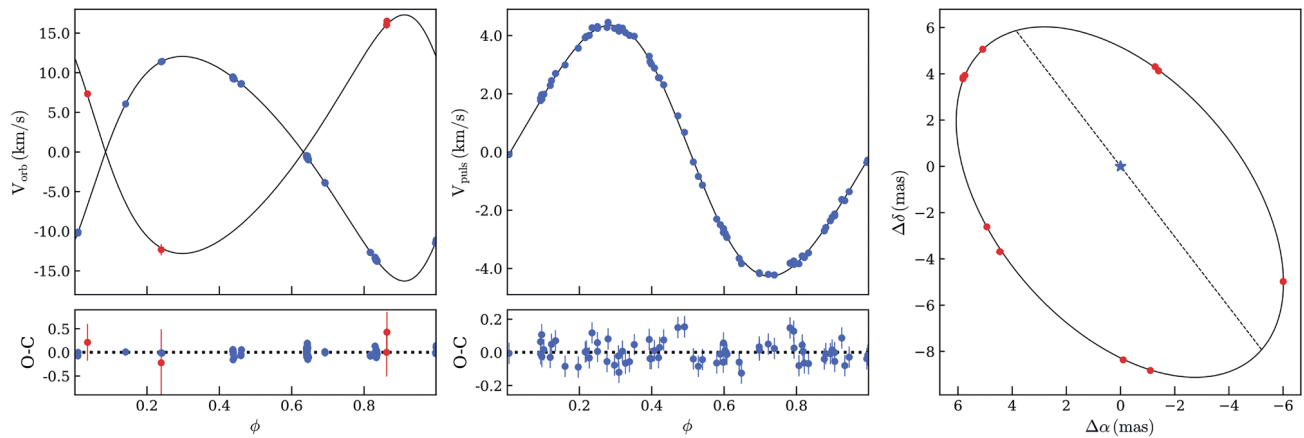

**Fig. 1.** Result of the combined fit for the Cepheid V1334 Cyg. *Left:* fitted (solid lines) and extracted primary (blue dots) and secondary (red dots) orbital velocity. *Middle:* fitted (solid line) and extracted (blue dots) pulsation velocity. *Right:* relative astrometric orbit of V1334 Cyg Ab.

I started this project in 2012 by observing the binary Cepheid V1334 Cygni with the MIRC instrument (Monnier et al. 2004) at the CHARA array. Long-term astrometric interferometry coupled with new ground- and space-based spectroscopic observations provided unprecedented results (Gallenne et al. 2018a). By simultaneously fitting the pulsation of the Cepheid, the orbital velocities of both stars and the relative astrometric motion of the companion, I was able to measure the most precise distance for a Cepheid (1%) and the most precise mass of a Galactic Cepheid (3%). This was possible thanks to high-precision data, ~35 μas precision in astrometry and ~40 m/s in radial velocities. With a semi-major axis of 8.5 mas, interferometry is necessary to spatially detect the companion around this Cepheid. The exquisite orbital fit is shown in Figure 1. The challenges for binary Cepheids are the contrast and separation between the components. The spectroscopic companions are in close orbit (< 50 mas) and are outshone by the Cepheids at wavelengths > 0.5 μm. The use of interferometry from the ground and ultraviolet spectroscopy from space with the Hubble Space Telescope is necessary for this project (see also Gallenne et al. 2013, 2014, 2015; Gallenne 2015; Gallenne et al. 2016a, 2019a).

Determining precise distance and masses for other types of binary stars is also essential in astronomy. In Gallenne et al. (2016b, 2018b), I measured the distance of the eclipsing system TZ Fornacis with a precision of ~0.4%, together with the mass of both components as precise as 0.05%. Again, the role of interferometry was critical because, as we can see in Figure 2, the separation between the two stars is less than 3 mas. I then reported similar precision in distance and mass for additional eclipsing systems in Gallenne et al. (2019b). These observations were performed with the VLTI/PIONIER recombiner which limited our precision of the distance to 0.35% due to the accuracy of the wave- length calibration (Gallenne et al. 2018b). The instrument VLTI/GRAVITY has a better accuracy of 0.02% in the high resolution mode (0.05% in medium resolution) thanks to a dedicated internal reference laser source. I have therefore started a new project with this instrument to observe 40 binary systems, together with new spectroscopic observations in order to measure distances and masses with a precision level of < 0.1%. Preliminary results already show that such a goal is possible.

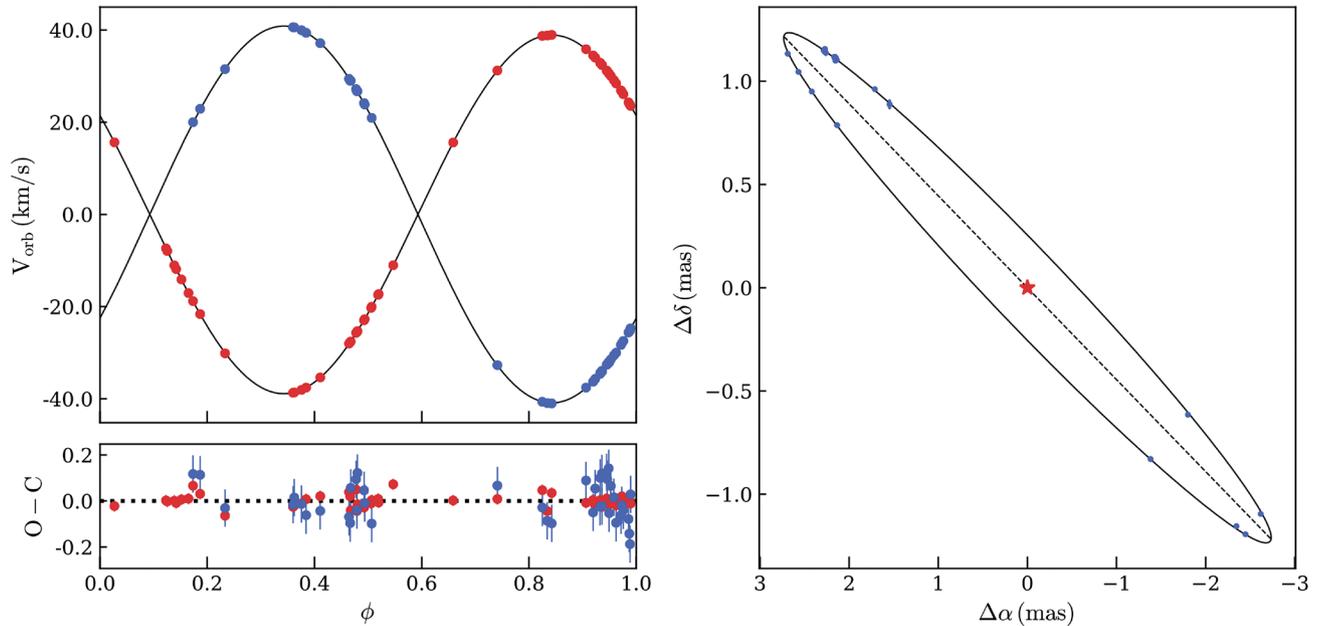

**Fig. 2.** *Left:* radial velocities of the primary (red) and the secondary (blue) star of eclipsing binary TZ For. *Right:* astrometric orbit of the secondary relative to the primary component.

The main limit in LBI to date is the sensitivity, preventing observations of faint objects (typically $K < 10$ mag). The advances recently made enable observations of targets as faint as $K \sim 18$ mag with the GRAVITY instrument, which is a big step for the interferometric community. In the next upgrades the sensitivity of GRAVITY+ will be enhanced to $K \sim 22$ mag, allowing new Galactic and extragalactic science cases and pushing the VLTI beyond the classical interferometric field. This will likely bring new insights in the extragalactic distance scale with, for instance, angular size measurements of the broad line region of active galactic nuclei, which will provide their geometric distance when combined with their linear size determined from reverberation mapping.

Grzegorz Pietrzyński[1,2], Wolfgang Gieren[2] , Dariusz Graczyk[1]

# Distance to the Magellanic Clouds accurate to 1% from eclipsing binaries

The Magellanic Clouds (MCs), our two neighboring galaxies, have been serving astronomers for more than 100 years as perfect laboratories in which to study many astrophysical processes and objects with high precision. In particular they provide a royal road to calibrate Cepheid variable stars and other stellar types as distance indicators, and as a result improve the determination of the extragalactic distance scale and the Hubble constant. Indeed, both Clouds possess large populations of classical Cepheids (Soszyński et al. 2017) and relatively small dust extinction (Górski et al. 2020) which helps to determine their brightnesses with high accuracy.

The condition for such absolute calibrations of stellar distance indicators in the Clouds is that we find a way to measure the distances to the Clouds very accurately. Accurate distances to these galaxies are extremely important not only for cosmology but also for many other fields of modern astrophysics. This is why more than 600 distance determinations to the MCs with different methods can be found in the literature (NED database, Mazzarella et al. 2007). However most of them suffer from relatively low precision and a lack of control of systematic errors affecting the measurements.

Detached eclipsing doubled-lined spectroscopic binaries offer a unique opportunity to measure stellar parameters like mass, luminosity, radius (Andersen 1991), and their distances very accurately in a quasi-geometri-cal way (Lacy 1977). Indeed, using high-quality radial velocity and photometric observations, standard fitting routines (e.g. Wilson and Devinney 1971) provide very accurate masses, sizes, and surface brightness ratios for the components of a double-lined eclipsing binary. The distance to the system follows from the dimensions determined this way, plus the absolute surface brightness, which can be inferred from the observed stellar colors using a precise empirical calibration of a suitable (near-infrared) surface brightness-color relation (SBCR, see Storm in this volume).

The first attempts made to use eclipsing binaries to determine the distance to the Large Magellanic Cloud (LMC) used relatively bright early-type systems (e.g. Guinan et al. 1998), for which an accurate empirical surface brightness-color relation is still unavailable. As a consequence, theoretical models need to be employed in such work, and this prevented the full realization of the potential offered by eclipsing binaries for a truly accurate measurement of their distances.

[1] Nicolaus Copernicus Astronomical Center, Polish Academy of Sciences, Bartycka 18, 00-716 Warszawa, Poland
[2] Universidad de Concepción, Departamento de Astronomia, Casilla 160-C, Concepción, Chile

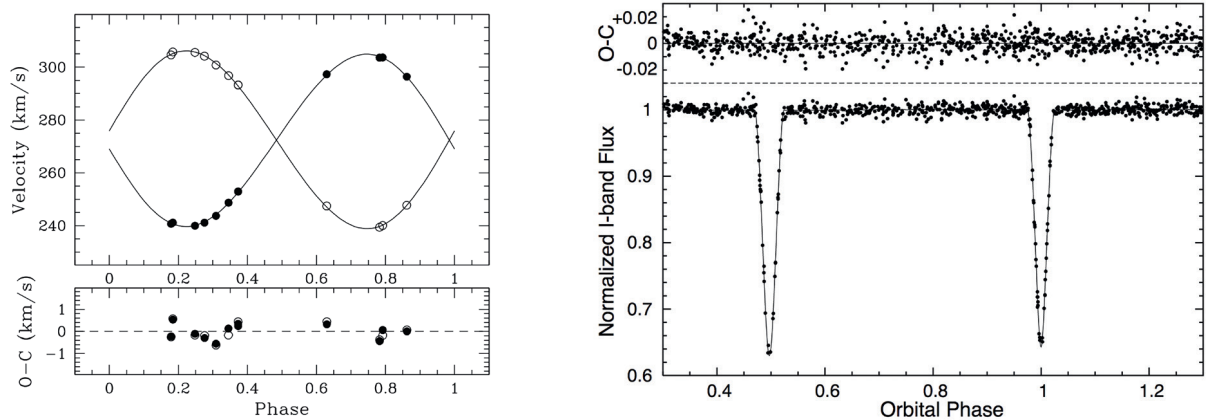

**Fig. 1.** Spectroscopic orbit (left panel) and the light curve (right panel) together with the best model obtained for our first eclipsing binary system in the LMC (Pietrzyński et al. 2009). Typically we obtained a 1-2% precision in stellar parameters and individual distances for all studied systems in the MCs.

In order to improve this situation we initiated a long-term Araucaria subproject focused on special eclipsing binaries in both MCs which consist of pairs of late-type red giant stars. In 2000 the first (out of some 20) observing proposals dedicated to these systems were submitted to ESO. In the following years we obtained an amazing amount of telescope time both at the European Southern and Las Campanas Observatories in Chile to carry out photometric and spectroscopic observations of highest quality of our selected late-type eclipsing binaries. Given the long orbital periods of our target systems (a few hundred days typically) the project finally needed almost two decades for its conclusion.

Our first distance determination to the LMC, precise to about 3%, was published by Pietrzyński et al.

(2009). It was based on one system only. Later Graczyk et al. (2012) presented the first distance determination to one of our late-type systems in the Small Magellanic Cloud (SMC). During the following years we finished the analyses of more systems in both Clouds which resulted in 2% and 3% distances to the LMC and SMC, respectively (Pietrzyński et al. 2013, Graczyk et al. 2014). These determinations were still based on small numbers of systems (8 and 5, respectively). The most important obstacle to improve on the accuracy of this method was the limited accuracy of the SBCR then available. Therefore we decided to improve the calibration of the SBCR using extremely high-precision near-infrared photometry of red clump stars in the Solar neighborhood (Laney et al. 2012, Gallenne et al. 2018) which resemble the component stars in our eclipsing target systems in the Magellanic Clouds. This way we were able to push the precision of our new SBCR calibration to 0.8% allowing us to improve both precision and accuracy of our distance determinations. Based on extended samples of 20 and 15 systems in LMC and SMC, respectively, we could improve the accuracy of the distances to our neighboring galaxies to 1% (LMC) and 2% (SMC) which now constitute the most accurate distances ever measured to these cornerstone galaxies in the nearby Universe. The details of this work can be found in the papers of Pietrzyński et al. 2019, and Graczyk et al. 2020.

Our quasi-geometrical distances to the Magellanic Clouds derived from our rare late-type eclipsing binary systems have yielded the best absolute calibrations for Cepheids, TRGB and other stellar distance indicators and provide a very solid zero point for the whole extragalactic distance scale. They are also crucial for establishing the metallicity effect on the major stellar distance indicators, most importantly for classical Cepheids (Wielgórski et al. 2017, Breuval et al. 2021). Currently all determinations of the Hubble constant based on Cepheids in tandem with the Tip of the Red Giant Branch (TRGB) and SN Ia methods rely on our Magellanic Cloud distances (e.g. Freedman 2021, Riess et al. 2021) showing that our eclipsing binary sub-project in the Araucaria Project has produced a breakthrough result for modern astrophysics.

Some current projects described in several articles in this book are related to further improvements on the SBCR and better extinction corrections in the Magellanic Clouds. They should result in a sub-percent LMC distance determination in the near future which will pave the way for finally pushing the accuracy of the Hubble constant to 1%, making a long-standing dream of astronomers come true.

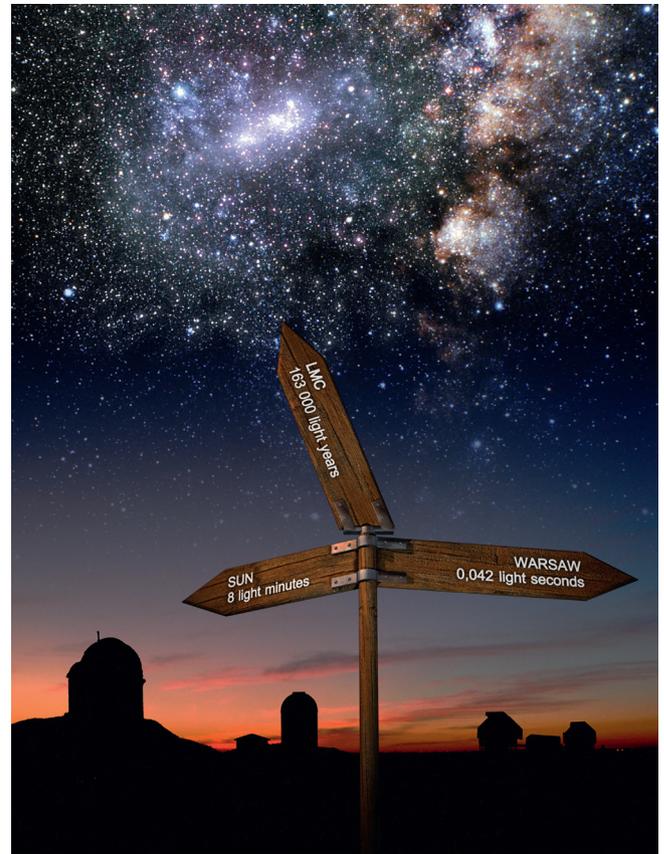

**Fig. 2.** An artistic view of the LMC distance determination.

Pierre Kervella[1]

# Multiplicity among classical Cepheids and RR Lyrae stars

As they are relatively massive stars, classical Cepheids (CCs) are commonly found in binary or multiple systems. However, due to the very high intrinsic brightness of these supergiant stars, the detection of their companions is difficult, and their multiplicity fraction is imperfectly known.

Binarity may also affect the evolution of CC progenitors e.g. through mass transfer. RR Lyrae stars (RRLs) are low-mass, old-population oscillating giants with short pulsation periods. Astoundingly, almost none of the 200,000+ stars classified as RRL were known with certainty to be in a binary system.

## Binarity from proper motion anomaly and resolved companions

The principle of our search for close-in orbiting companions is to look for a difference in proper motion (PM) vector between the long-term PM computed from the Hipparcos (epoch 1991.25) and Gaia Data Release 2 (GDR2, epoch 2015.5; *Gaia* Collaboration 2016, 2018) astrometric ($\alpha$, $\delta$) positions on the one hand (hereafter $\mu_{HG}$, for Hipparcos-Gaia) and the individual PM vector $\mu_{G2}$ from the GDR2 catalog on the other hand (Kervella et al. 2019b; see also Kervella et al. 2019a). For a single star, the long-term PM determined from the positions measured at the Hipparcos and Gaia epochs is identical to the short-term PM measured by each mission individually. For a binary star, the short-term PM includes in addition the tangential component of the orbital velocity of its photocenter. As the latter is changing with time over the orbital period of the system, a deviation appears between the short-term and long-term PMs of the star, due to the curvature of its sky trajectory. The proper motion anomaly (PMa), namely, the difference between the short-term and long-term PM, is therefore an efficient and sensitive indicator to detect non-single stars. We applied the PMa binary companion detection techniques to a sample of 254 CCs and 198 RRLs of the Milky Way (Kervella et al. 2019b).

We also searched for candidate common proper motion (CPM) companions of CCs and RRLs by identifying the angularly nearby stars in the Gaia catalog that share the parallax and proper motion of the targets. We then selected the gravitationally bound candidates by estimating their differential projected velocity with respect to the target, and comparing this velocity to the escape velocity. We conducted a CPM pair investigation on 456 CCs and 789 RRLs (Kervella et al. 2019c) based on the Gaia DR2 astrometry.

[1] LESIA, Observatoire de Paris, Université PSL, CNRS, Sorbonne Université, Université de Paris, 5 place Jules Janssen, 92195 Meudon, France

## Binary fraction of Classical Cepheids and RR Lyrae stars

We detected 57 new candidate CC binaries from their PMa signal (S/N > 3). We also found candidate resolved CPM companions for 27 CCs of our sample. While CCs have long been known to have a high binary fraction, our survey of nearby CCs indicates that more than 80% are probably in binary systems (Figure 1). Many CCs are also found in triple (e.g. SY Nor, AW Per, W Sgr, V0350 Sgr) or quadruple systems (e.g. δ Cep, SY Nor).

For our RRL sample, we detect 13 stars showing a significant PMa signal (S/N > 3), and 61 additional candidates with a mildly significant signal. We found in addition seven spatially resolved, probably bound systems with RRL primaries out of 789 investigated stars, and 22 additional candidate pairs with a lower probability. We report in particular new companions of three bright RRLs: OV And (companion star of F4V spectral type; Fig. 2), RR Leo (M0V), and SS Oct (K2V). The discovery of a significant number of RRLs in binary systems sheds new light on the long-standing

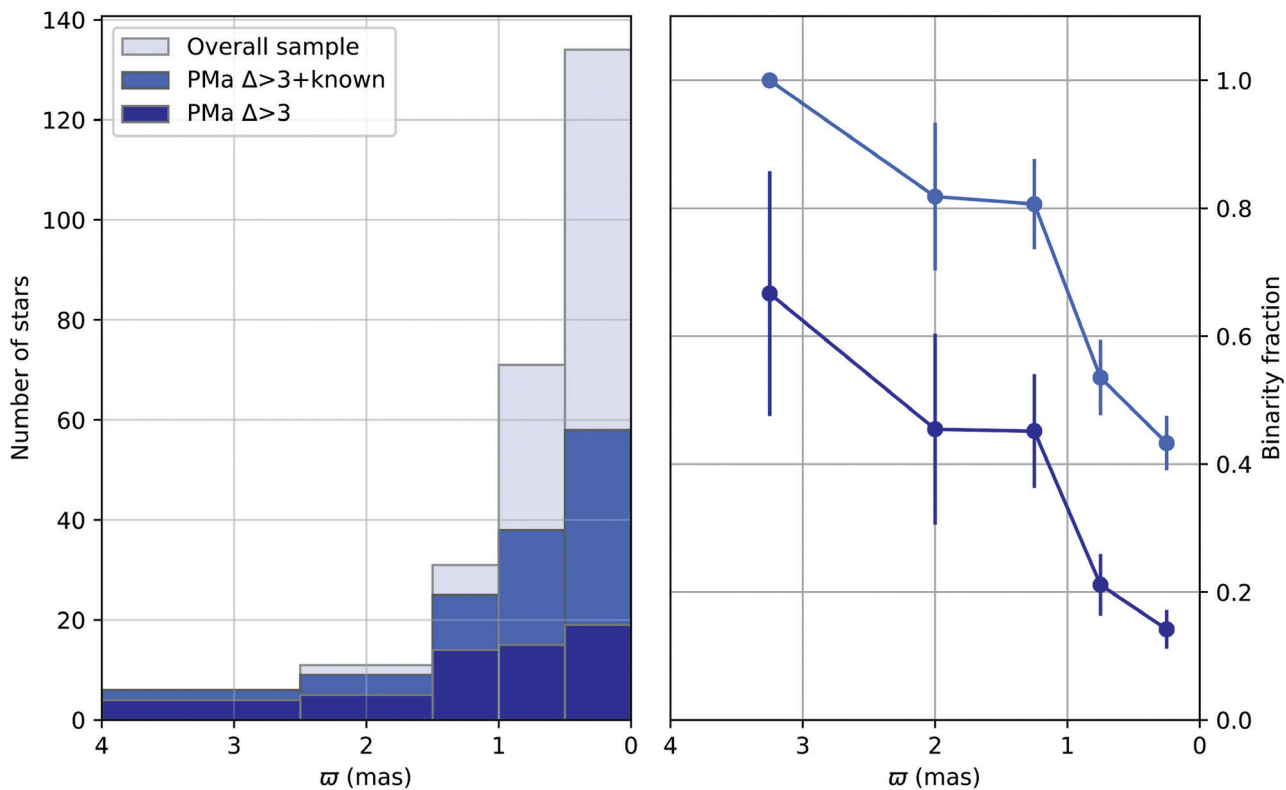

**Fig. 1.** Left panel: histogram of the Cepheids that show a PMa signal (S/N > 3, dark blue), with the additional stars classified as binaries in the Szabados (2003; medium blue) database and the overall sample (light blue) as a function of parallax. Right panel: binary fraction as a function of parallax (from Kervella et al. 2019b).

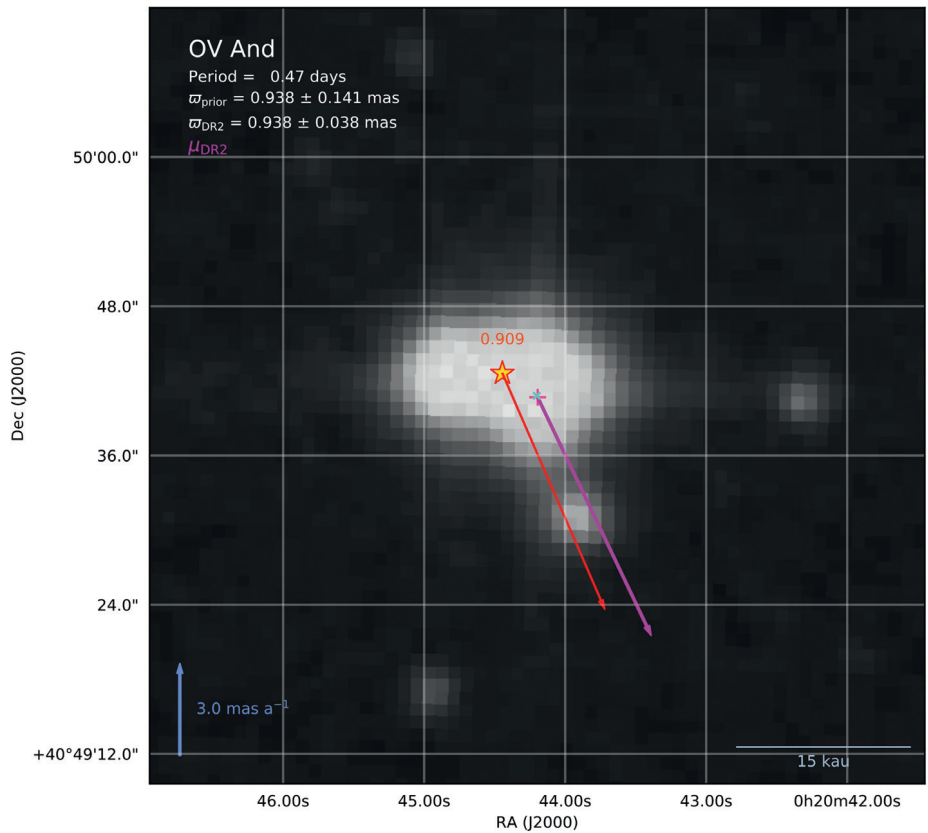

**Fig. 2.** The companion of the RR Lyrae star OV And. The bound candidate companion is marked with a yellow star and a red PM vector (from Kervella et al. 2019c).

mystery of their very low apparent binary fraction.

## Conclusion

Classical Cepheids have long been known to have a high binary fraction, and our survey of the PMa signal and CPM companions of nearby CCs indicates that their binary fraction is higher than 80%. A significant fraction (≈7%) of the nearby RRLs observed by Gaia also show indications of binarity in their proper motion, and 7 RRLs host resolved CPM companions that are likely gravitationally bound. The identification of companions of CCs presents the opportunity to evaluate their impact on the calibration of the period-luminosity relation. Moreover, as they are usually main sequence non-variable stars, the resolved companions provide us with valuable proxies to validate the uncertainties of the Gaia parallaxes (Breuval et al. 2020).

Gergely Hajdu[1]

# RR Lyrae variables in binary systems

RR Lyrae stars are prominent pulsating variables located on the Horizontal Branch of old (> 10 Gyr) stellar populations. Due to their use as distance indicators in the Local Universe, understanding their nature and possible limitations for distance determinations is paramount.

The mass of RR Lyrae variables is an indispensable constraint to test models of stellar evolution and pulsation. Binaries provide the most straightforward way to measure stellar masses, as the orbital parameters combined with Kepler's laws provide the masses of the components directly. Therefore, it is important to find and characterize RR Lyrae in binary systems.

Despite over 100 years of observations, and having discovered over 200,000 known members of this variable class, until recently only one RR Lyrae variable was known to reside in a binary system with high confidence (Liška et al. 2016). Unfortunately this system, TU Ursae Majoris, has a very long inferred orbit (~23.3 years), and no detected eclipses. Recently, candidates for RR Lyrae variables in wide binary systems were identified using data from the Gaia astrometric satellite using both proper motion anomalies (Kervell et al. 2019a) as well as directly resolved companions with common proper motions to the RR Lyrae variables (Kervella et al. 2019b). Nevertheless, the expected very long binary periods for such wide systems (decades to centuries) prevents the detailed characterization of the stellar obits. Hence, up to this point, no RR Lyrae mass measurement has been performed on any member of this class of variable stars.

Motivated by this situation, we have conducted an initial search for RR Lyrae variables in binary systems, using the Optical Gravitational Lensing Experiment (OGLE) light curves toward the Galactic bulge. In a binary system regular signals, such as the light curves of RR Lyrae, are modulated by the changing distance between the observer and the variable. When characterized well, this Light-Travel Time Effect (LTTE) gives the same information about the binary as radial velocity observations would in the case of single-lined (SB1) binary systems. Out of 1,952 analyzed RR Lyrae, 20 candidates for binary systems were identified (Hajdu et al. 2015), estimating the LTTE with Observed-Calculated diagrams (O-C; Sterken 2005). Together with later studies (Hajdu et al. 2018; Prudil et al. 2019), a total of 38 RR Lyrae binary candidates have been found toward the bulge at present.

We have undertaken a new, systematic search for RR Lyrae variables in binary systems using the OGLE

[1] Nicolaus Copernicus Astronomical Center, Polish Academy of Sciences, Bartycka 18, 00-716 Warszawa, Poland

survey data, taking advantage of the longer temporal coverage obtained since our original study. A larger number of sources (27,480) were analyzed compared to earlier studies, and each O-C diagram was inspected individually, selecting about 400 objects for further study. These were examined one by one, the uncertain cases discarded, and the final orbital parameters estimated for 87 high-quality RR Lyrae binary candidates using Markov-Chain Monte Carlo (MCMC) analysis. Out of these, 61 are new candidates, while 26 were announced previously. Twelve stars previously considered binary RR Lyrae candidates have been discarded during our analysis. A selection of O-C diagrams are shown for some of our best candidates in Figure 1.

This substantial sample of well-characterized binary RR Lyrae candidates allowed us to examine the distributions of their binary parameters for the first time ever. As shown in the top panel of Figure 2, no RR Lyrae binary candidate has a measured orbital parameter below 1000 days. This value provides a strong constraint on the lower limit of the binary period for future searches of RR Lyrae binaries. The middle panel of Figure 2 shows that the distribution of eccentricities is very peculiar: for most stars this value is between 0.05 and 0.6, but there is a large excess of stars in the eccentricity range between 0.25 and 0.3. The bottom panel of Figure 2 shows the mass-function distribution of our sample, which turns out to be apparently trimodal.

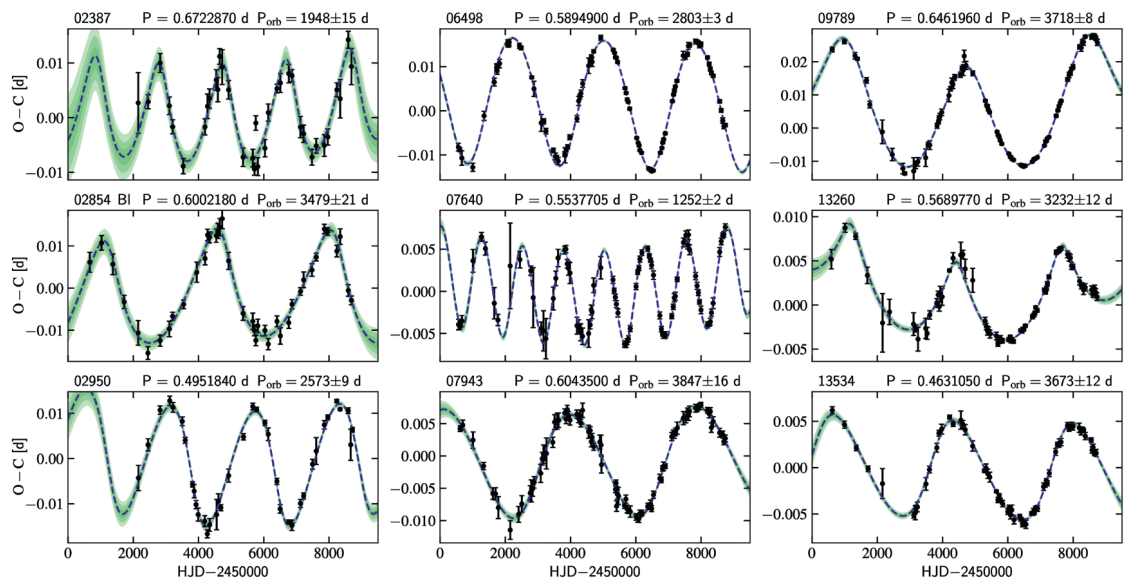

**Fig. 1.** Example O-C diagrams for some of our best RR Lyrae binary candidates. The measured O-C values and their uncertainties are shown with dots and error bars as a function of time, respectively. We interpret the cyclic variation as being caused by the LTTE, while the secular trends are caused by the slowly changing pulsation periods of the RR Lyrae themselves. Shaded green areas enclose 68.3, 95.4 and 99.7 per cent (equivalent to one, two and three standard deviations, respectively) of the MCMC solutions, in order of increasing transparency. Blue dashed lines show the final adopted mean binary solutions. Above each panel the following information is shown: the OGLE identifier of the variable (in the format OGLE-BLG-RRLYR-), "Bl" marks if a star is affected by the Blazhko effect, followed by the period used to construct the O-C diagram, and finally the estimated orbital parameter and its uncertainty.

While we don't know what the real masses of our RR Lyrae are, supposing an average mass of 0.65 $M_\odot$ and a uniform (isotropic) distribution of the inclinations for the binary systems, we can model the locations of these three modes with three populations of binary companions. The largest mass group have typical masses of 0.6 $M_\odot$, and probably contain a substantial fraction of both main-sequence and white-dwarf companions. The second, least significant group has a typical inferred mass of 0.2 $M_\odot$, and might signal an excess of M dwarf-type companions for RR Lyrae variables. The lowest-mass group has typical masses of 0.067 $M_\odot$ or about 70 $M_J$, meaning they cannot be be stellar objects, and must belong to brown dwarfs. If this group is confirmed with radial velocity measurements, it would mean the discovery of a hitherto unknown large population of brown dwarfs in wide binary systems belonging to the old stellar population of the galaxy.

The presented peculiarities do not appear to be caused by biases in the data or other effects, and therefore must originate from a combination of the initial binary parameters of the old stellar populations (Population II) and the binary (co-)evolution effects. We stress that further studies are needed both on the observational (binary statistics) and theoretical (binary evolution modeling) sides to fully understand the observed parameters of the RR Lyrae binary candidate population. These results were published in the Astrophysical Journal (Hajdu et al. 2021).

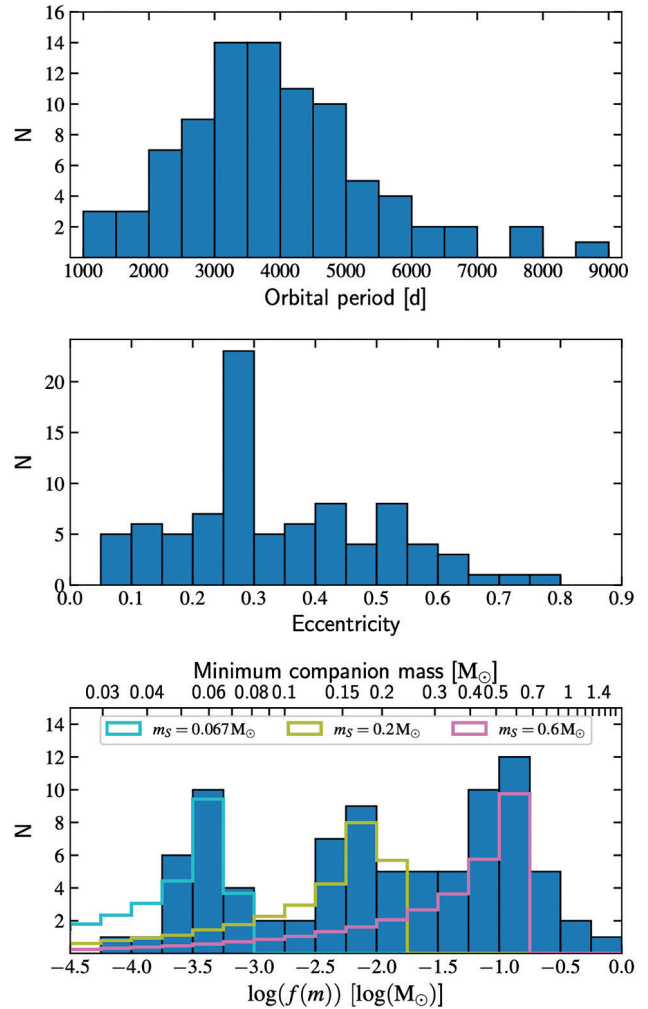

**Fig. 2.** Statistical parameters of the 87-candidate sample. Top: the orbital period distribution. Middle: the eccentricity distribution. Note the strong concentration of stars between 0.25 and 0.3. Bottom: The mass function distribution of the sample. The equivalent calculated minimum companion masses are denoted on the upper axis (calculated for $M_{RR} = 0.65\ M_\odot$ and i = 90 deg). Purple, yellow and cyan lines show the expected mass-function distributions for companion masses of 0.6, 0.2 and 0.067 $M_\odot$ (assuming $M_{RR}$ = 0.65 $M_\odot$, and an isotropic inclination distribution), respectively.

**Bogumił Pilecki[1]**

# High-precision astrophysics of Cepheids in binary systems

Classical Cepheids are one of the most important objects in astrophysics, crucial for various fields of astronomy like stellar oscillations and evolution of stars, and with enormous influence on modern cosmology. Since the discovery of the relationship between their pulsation period and luminosity (Leavitt & Pickering 1912), Cepheids have been extensively used to measure distances both inside and outside of our Galaxy.

In the past such measurements led to a breakthrough in our understanding of the structure and size of the Universe. Currently, the most reliable local measurement of the Hubble constant is based on supernovae distances calibrated with Cepheids (Riess et al. 2021).

Classical Cepheids are radially pulsating giant and supergiant stars that have already finished burning hydrogen in their cores and the majority exist on a so-called blue loop, where core helium burning takes place. To start pulsating as Cepheids on their evolutionary path, stars have to enter a narrow and relatively well-defined instability strip. For this reason, classical Cepheids are a sensitive probe of various properties important for evolutionary studies, such as mass, metallicity, overshooting, mass loss, and rotation. Cepheids also play an important role in the testing and development of pulsation theory.

Since the discovery of the binarity of Polaris in 1929 many Cepheids in binary systems have been identified

(Szabados 2003), but none were found in eclipsing systems. Given the lack of known eclipsing systems, all attempts to determine the precise physical parameters of these important stars had failed. Rough estimates (15-30% precision) for a few Cepheids were important, but their low quality prevented calibration of theoretical models with observational data.

The breakthrough came with the first confirmed Cepheid in an eclipsing binary system in the Large Magellanic Cloud (Pietrzyński et al. 2010). With subsequent advances in modeling, we determined its mass and radius with sub-1% precision (Pilecki et al. 2013, see also Figure 1). This was an important step in the resolution of the Cepheid mass discrepancy problem, forcing important adjustments in the evolution and pulsation theories (e.g. Marconi et al. 2013).Since this first discovery, we have analyzed several Cepheids in eclipsing binaries, providing complete sets of ~1%-precise physical parameters for 6 classical Cepheids (Pilecki et al. 2013, 2015, 2018). Based on these data we derived the first empirical period-mass-radius relation, which can serve

[1] Nicolaus Copernicus Astronomical Center, Polish Academy of Sciences, Bartycka 18, 00-716 Warszawa, Poland

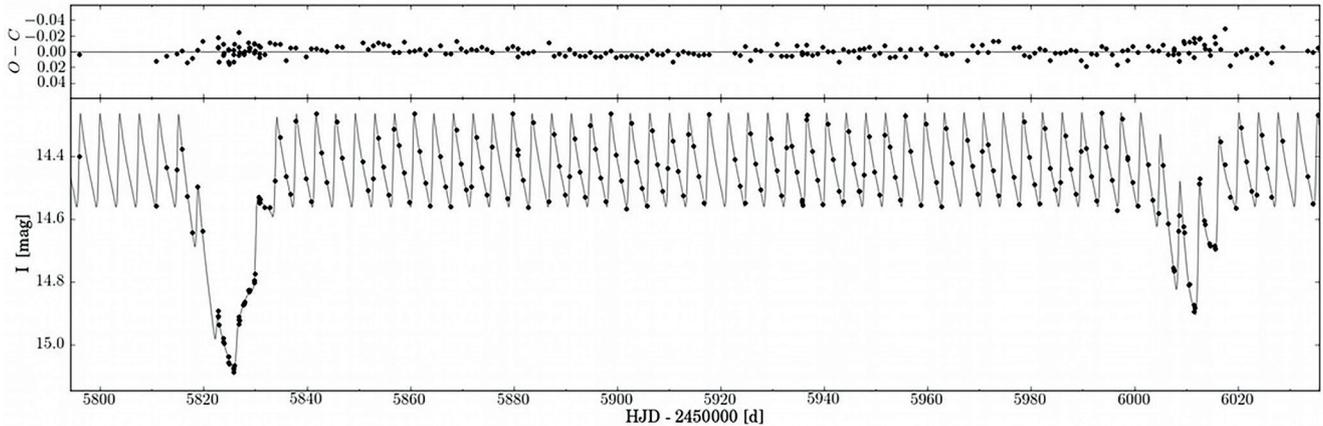

**Fig. 1.** Accurate light curve model of a Cepheid in an eclipsing binary system.

to determine masses of single Cepheids among other applications (Pilecki et al. 2018).

As all Cepheids in the analyzed systems were accompanied by giants (i.e. stars in a similar stage of evolution), the measured mass ratios were mostly close to unity but surprisingly, for one system, we found components with significantly different masses (Pietrzyński et al. 2011). This led to the conclusion that the system was a triple before, and that the Cepheid is a product of a merger (Neilson et al. 2015). We confirmed another of the studied systems to be a very rare binary composed of two Cepheids (Gieren et al. 2014). Such a system provides important constraints on the ages and chemical composition of the components. We also found one of the companions to the Cepheids to lie in the middle of the instability strip, yet it was not pulsating (Gieren et al. 2015). This is the best-defined object of this type so far.

Recently we proposed a hypothesis that most of the Cepheids that lie significantly over the period-luminosity relation are double-lined systems composed of a Cepheid and a giant companion (Pilecki et al. 2021, see Figure 2). In the preliminary study, we spectroscopically confirmed sixteen such systems in the Large Magel-lanic Cloud. Objects of this type are crucial for accurate measurements of Cepheid masses. In total, we expect to confirm at least 40 Cepheids with giant companions in this galaxy and many more in the Small Magellanic Cloud and in the Milky Way. Most importantly these Cepheids span a wide range of pulsation periods and thus, also of mass and other physical parameters.

Apart from classical Cepheids, we also observed and studied Type II Cepheids in eclipsing binary systems. Type II Cepheids are the less massive counter-parts of classical Cepheids. They exhibit a tight and well-defined period-luminosity relation, and therefore may also be used to measure distances. In spite of lower brightness as compared to classical Cepheids, they are important distance indicators because of their presence in old stellar systems, where classical Cepheids are not found.

Our analysis of one of these Cepheids showed that it actually belongs to another class of variables, probably to long-period Anomalous Cepheids (Pilecki et al. 2017). In a study of a second source, which belongs to the subclass of peculiar W Virginis stars, we found that in the past it transferred most of its mass to its com-panion. Contrary to what is normally assumed, this Cepheid is a young star despite its current very low mass of about 0.64 $M_\odot$ and an advanced evolutionary stage (Pilecki et al. 2018). Our modeling also revealed a structured (spiral or composed of rings) disk around the companion (see Figure 3). This system may be a progenitor of a Be-type star.

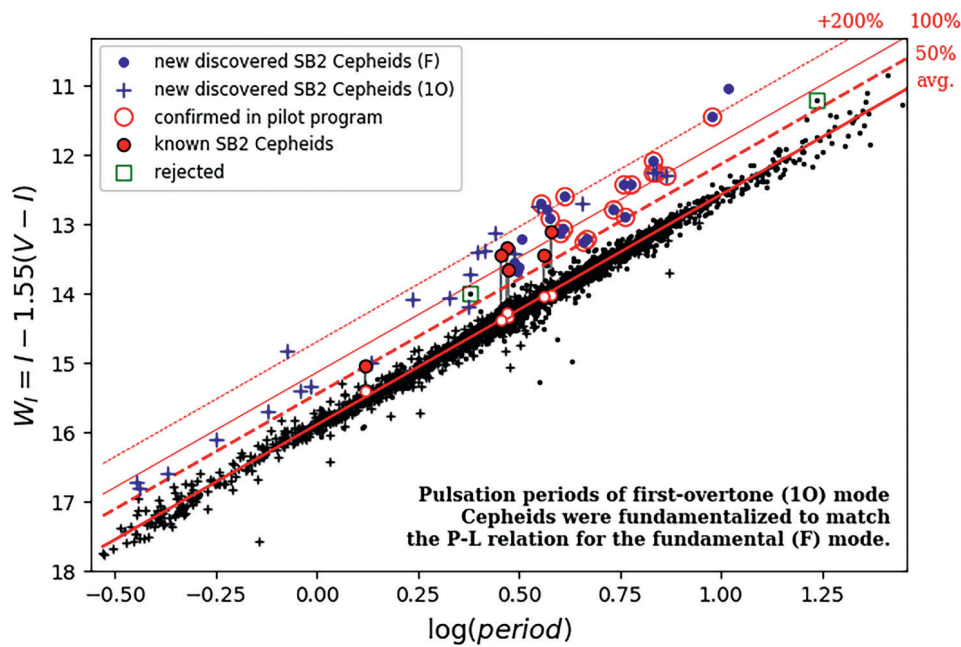

**Fig. 2.** Period-luminosity relation for Cepheids in the LMC. Cepheids that are at least 50% brighter than typical were shown to have giant companions.

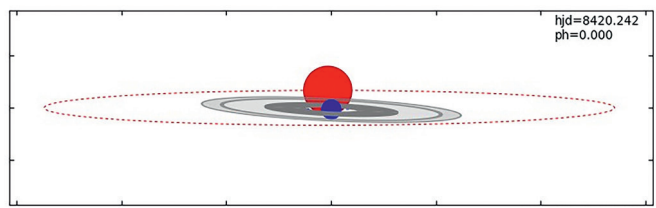

**Fig. 3.** The model of a system composed of a Type II Cepheid and a companion with a three-ring disk.

Jesper Storm[1]

# The Baade-Wesselink technique and the effect of metallicity on the Leavitt law

To get a good handle on the problem of the effect of metallicity on the Leavitt relation it is necessary to measure precise luminosities and metallicities for Cepheids not only spanning a large range of pulsation periods but also spanning a large range of metallicities. Nearby Cepheids for which parallaxes could be observed before the advent of Gaia were all of solar metallicity. Cepheids that are significantly more metal-poor are only found at larger distances in the outer parts of the Milky Way disk and in the Magellanic Clouds.

The Baade-Wesselink technique can be applied to radially pulsating stars to determine their individual distance and radius, and thus their luminosity, taking advantage of the additional information which can be gained from the pulsation. As the Baade-Wesselink method only requires light curves in a couple of suitable photometric bands as well as a radial velocity curve for a star, it can be applied to stars not only in the solar neighborhood but also in the Magellanic Clouds and beyond. This means that we can determine distances, radii and luminosities precisely for individual Cepheids covering a large range of metallicities.

We have employed the surface-brightness variant of the method as developed by Barnes & Evans (1976) but instead of using optical colors we use the near-infrared (V-K) color as pioneered by Welch (1994), creating the near-infrared surface brightness method (IRSB). The surface brightness, $F_V$, is related to the optical V-band magnitude and the stellar angular diameter, $\Theta$, through the relation $F_V = 4.2207 - 0.1V_0 - 0.5\log(\Theta)$. As shown by Welch (1994) Fv follows a tight linear relation with the (V-K) color, so with this relation the angular diameter can be determined at any given phase when light curves in V and K are available. At the same time, the variation of the stellar radius with time, $\Delta R$, can be determined from the radial velocity curve by integrating the pulsation velocity. As the angular diameter, $\Theta$, is geometrically related to the stellar radius, R, and the distance, d, as $\Theta = 2R/d = 2(R_0+\Delta R)/d$ and we have light curves to compute $\Theta$ and a radial velocity curve to compute $\Delta R$ we can solve for the radius $R_0$ and the distance d. Two critical ingredients required for the method to succeed are the calibration of the surface-brightness color relation, and the projection fac-tor, which is necessary to convert the observed radial velocities into pulsation velocities. The surface-brightness relation is an important relation in the Araucaria project, and much effort has gone into improving these relations, in particular for binary stars (e.g. Gallenne et al. 2018, Graczyk et al. 2021, and this volume). For the first relation we have adopted the relation determined by Kervella et al. (2004) which is based on interferometrically determined angular diameters of giants, including Cepheids. For the second relation we calibrated the method by requiring that the distances to Milky Way Cepheids with HST fine guidance sensor parallaxes from Benedict et al. (2007) and the distances to the Cepheids in the LMC do not depend on the pulsation period of the star (Storm et al. 2011). A significant effort has also been invested in understanding the physical reason for the dependence of the p-factor with period and the observed scatter (e.g. Nardetto et al. 2009, 2011, Gallenne et al. 2017, and this volume).

[1] Leibniz-Institut für Astrophysik Potsdam (AIP), An der Sternwarte 16, D-14482 Potsdam, Germany

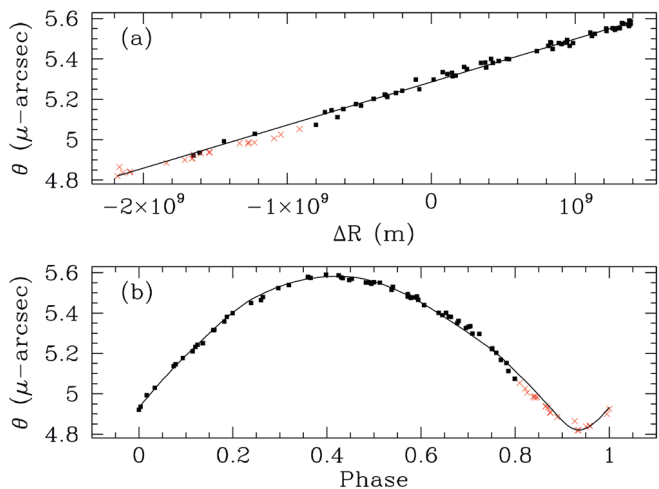

**Fig. 1.** In panel (a) the angular radius is plotted against the radius variation for the star OGLE-SMC-CEP1729. In panel (b) the angular diameter is plotted versus the pulsation phase. In both panels the points are based on the photometric data, and the curve is based on the radial velocity data.

Optical light curves have been meticulously compiled for individual stars over many decades but with the advent of large surveys like OGLE good data has also become available for thousands of fainter Cepheids in both the Milky Way and in the Magellanic Clouds. Similarly near-IR data has become much more readily available from e.g. Monson & Pierce (2011) and the VMC survey (Cioni et al. 2011). We have obtained additional near-IR data for quite a few Magellanic Cloud Cepheids using SOFI at ESO, but our main observational effort has been to obtain radial velocity curves for Cepheids in the LMC and SMC mainly with two ESO Large Programmes using the HARPS spectrograph on the ESO 3.6-m telescope at La Silla. This resulted in a total sample of 31 Cepheids in the SMC and 37 Cepheids in the LMC for which we could apply the IRSB method.

Using the distances from the IRSB analysis obtained by adopting mean metallicities from the literature for the SMC and LMC Cepheids respectively and the mean of the individual metallicities for the Milky Way Cepheids, we have determined Leavitt relations in the Wesenheit indices as well as in the V and K bands for each of the three metallicity samples. Fixing the slope for each relation we have determined the zero point shift between the three samples in each band. In case of the K-band we find that metal poor Cepheids are brighter and the zero point is shifted by $-0.23 \pm 0.06$ mag/dex while in the Wesenheit $W_{VI}$ index the effect is $-0.34 \pm 0.06$ mag/dex (Gieren et al. 2018). It should be stressed that the mean of the distances to each of the LMC Cepheids is $18.46 \pm 0.04$ mag (statistical error only) which is in excellent agreement with the eclipsing binary distance of $18.493 \pm 0.047$ mag from Pietrzyński et al. (2013) even if the IRSB method zero point is based on the HST parallax of nearby Cepheids.

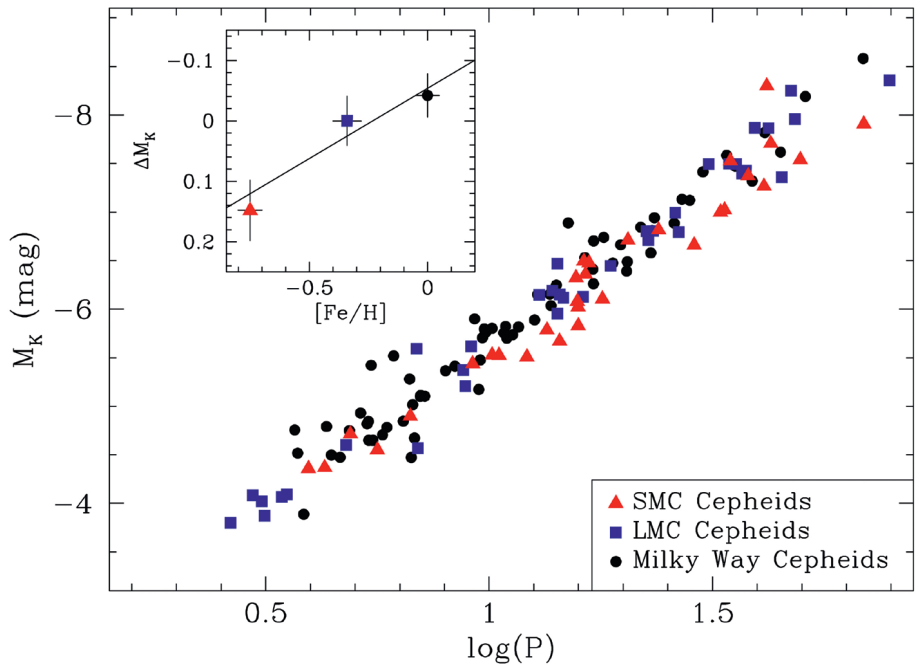

**Fig. 2.** Absolute K-band magnitude plotted versus the pulsation period for all the stars. The insert shows the resulting shift of the K-band zero point of the Leavitt relation with respect to metallicity.

**Nicolas Nardetto[1]**

# The Nice contribution: The projection factor of Cepheids

The projection factor of Cepheids is a key quantity used in the Baade-Wesselink method of Cepheid distance determination. *It is of high importance as there is a direct relationship between the distance of the Cepheid, the projection factor, the slope and the zero point of the PL relationship, and $H_0$.*

The principle of the Baade-Wesselink (BW) method is simple: it consists of comparing the linear and angular dimensions of the Cepheid in order to determine its distance by a simple division. Photometric measurements (associated with a surface brightness-color relation) and/or interferometric measurements provide the variation of the photospheric angular diameter of the star over the whole pulsation cycle, while the variation of the linear diameter is determined by a time integration of the pulsating velocity ($V_{puls}$) of the star. The determination of the latter, from the Doppler shift of the spectral line ($V_{rad}$), is extremely delicate and involves what is called the projection factor, $p$, defined by $V_{puls} = p \cdot V_{rad}$. This number alone summarizes all of the physics of the Cepheid atmosphere: limb darkening, velocity gradient and atmospheric dynamics. *To understand the projection factor is to understand the physics of Cepheids.*

For a Cepheid described simply by a uniform disk pulsating, the value of $p$ is 1.5 (whatever the pulsation phase). But actually, as already mentioned, the radial velocity of each surface element of the star is projected along the line of sight and weighted by the inten-sity distribution of the Cepheid. Thus, the limb-darkening of δ Cep reduces the p-factor significantly, and the so-called geometric projection factor ($p_0$, **step 1** in the Figure) varies from 1.36 to 1.39, depending on the wavelength considered in the visible range. The time variation of the p-factor, due mainly to limb-darkening variation, is neglected as it has no impact on the distance (Nardetto et al. 2006). However, a Cepheid is not simply a limb-darkened pulsating photosphere, it also has an extended atmosphere with various spectral lines (in absorption) forming at different levels from which we derive the radial velocity curve used in the BW method. Moreover, there is a velocity gradient in the atmosphere of the Cepheid, which can be measured from spectroscopic observations (**step 2**). Then, depending on the line considered, the amplitude of the radial velocity curve will not be the same and the resulting projection factor will be different. In the figure ($f_{grad}$, **step 3**), we show the impact of the atmospheric velocity gradient on the p-factor for a line forming rather close to the photosphere (line depth of about 0.1). The higher the line-forming region in the atmosphere, the lower the projection factor (lower up to 3% compared to $p_0$ in the case of δ Cep). The last correction

[1] Université Côte d'Azur, Observatoire de la Côte d'Azur, CNRS, Lagrange, CS 34229, Nice, France

on the projection factor ($f_{o-g}$, **step 4**) is more subtle. In spectroscopy, the radial velocity is actually a velocity associated with the moving *gas* in the line-forming region, while in photometry or interferometry, we probe an *optical* layer corresponding to the black body continuum (i.e. the layer from which the photons escape). A correction on the projection factor of several percent (independent of the wavelength or the line considered) has to be considered. A relation between the period of Cepheids and the p-factor has been established using this approach for a specific line (Nardetto et al. 2007). In 2009, during my 1-year post-doc in Concepcion, just before my hiring at CNRS in France, I devoted my time to adapt the approach described above to the cross-correlation method of radial velocity determination (Nardetto et al. 2009).

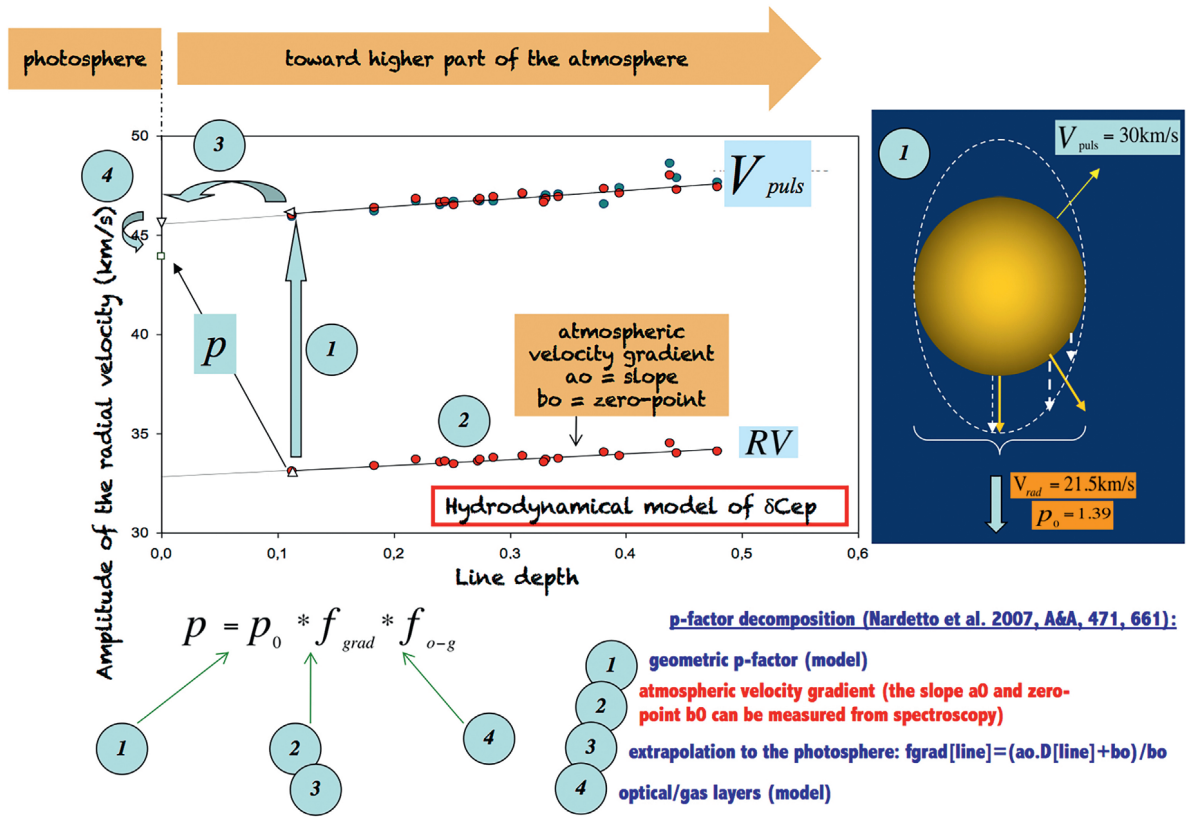

**Fig. 1.** Decomposition of the projection factor of classical Cepheids

In the Araucaria group, we used the Pp relation to apply the BW method and thus determine the distance of Galactic Cepheids (Fouqué et al. 2007; Storm et al. 2011a, 2011b). However, despite all these efforts, the technique was not up to the task and was not used by the SHOES project. *In 2009, Riess et al. wrote, rightly: "We have not made use of additional distance measures to Galactic Cepheids based on the BW method (...) due to uncertainties in their projection factors".*

Despite considerable efforts, there is no consensus on the projection factor today. The methods based on hydrodynamical models (Nardetto et al. 2004, Vasilyev et al. 2017) provide consistent results. The novel purely observational approach based on Cepheids in eclipsing binaries is also consistent with these theoretical values (Pilecki et al. 2013). However, all inverse Baade-Wesselink approaches (based on surface brightness color relations or interferometric observations) provide significantly different results for a Cepheid of the same period. *Even more confusing, they seem to show that there is no Pp relation (Trahin et al. 2021).* Because of the projection factor limit, the current precision on the BW distance is about 5-10%.

*But the p-factor requires a p-attitude, positive! Hope of gain decreases the pain.* Indeed, with the p-factor and the BW method it is possible to determine directly the distance of extragalactic Cepheids (and the galaxies containing them), individually, without going through the statistical Cepheid PL relation. Achieving this objective would allow us to make an accurate mapping of the galaxies in the local group, and thus to significantly improve (probably by an order of magnitude) the first stage of the distance scale in the universe. Hope is allowed. *Several avenues of research have not been explored yet at the level of the photosphere, the atmosphere and the environment of Cepheids.*

**Rolf-Peter Kudritzki[1,2]**

# Flux Weighted Gravity – Luminosity Relationship: A New Distance Indicator

A key goal of the Araucaria project is the determination of accurate distances to galaxies using a wide range of different stellar distance indicators. During the course of the collaboration we have developed a new very precise distance determination method, the flux weighted gravity – luminosity relationship (FGLR) of blue supergiant stars (BSG).

BSG are massive stars, in the range of 15 to 60 solar masses, which finished central hydrogen burning on the main sequence and crossed the HRD on their way to the Hayashi-limit of red supergiants. In an effective temperature range 7900K < $T_{eff}$ < 25000K, they emit most of their enormous luminosity as visual light. With absolute visual magnitudes up to -10 mag BSG are true beacons in the universe and many magnitudes brighter than the standard extragalactic distance indicators such as Cepheids or TRGB stars. This allows for detailed spectroscopic studies of individual BSG in galaxies out to 10 Mpc with present day telescopes to determine effective temperatures, gravities, chemical composition, reddening and reddening law (details of the spectroscopic analysis method are provided in the references given below). With the ground-based 30 to 40-m telescopes of the future and further optimized multi-object spectrographs and detectors 50 Mpc will be possible.

The crucial quantity for BSG distance determinations is the flux-weighted gravity, $g_F = g/T^4$, a combination of stellar gravity g and effective temperature T, where T is measured in units of $10^4$K. Kudritzki et al. (2003, 2008) discovered that BSG show a tight relationship between flux weighted gravity and absolute bolometric magnitude $M_{bol}$, the FGLR. BSG cross through the HRD at constant mass and luminosity, which means that at each stellar mass $g_F$ is constant during the evolution. The FGLR is then a consequence of the power-law relationship between stellar luminosity and stellar mass. The most recent calibration of the FGLR has been provided by Urbaneja et al. (2017) through a comprehensive spectroscopic study of 90 BSG in the LMC and is shown in Figure 1 in a slightly updated form as obtained by Sextl et al. (2021) accounting for the new precision distance to the LMC obtained by Pietrzynski et al. (2019).

[1] Institute for Astronomy, University of Hawaii, 2680 Woodlawn Drive, Honolulu, Hawaii 96822, USA
[2] Universitätssternwarte, Ludwig-Maximilian-Universität München, Scheinerstr. 1, 81679 München, Germany

Some 50 nights of ESO VLT/FORS and Keck/LRIS observing time have been used to investigate BSG in ten galaxies so far to determine distances and stellar metallicities (for the latter see also the contribution by Fabio Bresolin in this volume): NGC 300 (Kudritzki et al. 2008), WLM (Urbaneja et al. 2008), M33 (U et al. 2009), M81 (Kudritzki et al. 2012), NGC 3109 (Hosek et al. 2014), NGC 3621 (Kudritzki et al. 2014), M83 (Bresolin et al. 2016), NGC 55 (Kudritzki et al. 2016), IC 1613 (Berger et al. 2018), NGC 2403 (Bresolin et al. 2022, in prep.). FGLR distance moduli based on this work and the FGLR of Figure 1 are given in Table 1 (see also Sextl et al. 2021).

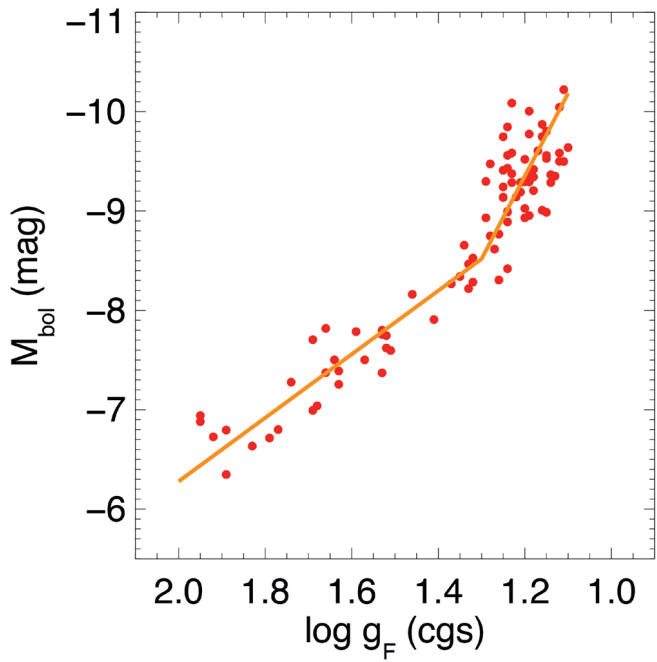

**Fig. 1.** The observed FGLR of 90 BSG (red) in the LMC. The two-component regression used for extragalactic distance determinations is shown in orange. Data are from Urbaneja et al. (2017) adopting the LMC distance by Pietrzyński et al. (2019). From Sextl et al. (2021).

**Tab. 1.** FGLR distances.

| Name | m-M (mag) |
|---|---|
| NGC 3621 | $28.95 \pm 0.11$ |
| M83 | $28.41 \pm 0.11$ |
| M81 | $27.62 \pm 0.08$ |
| NGC 2403 | $27.40 \pm 0.08$ |
| NGC 55 | $26.79 \pm 0.08$ |
| NGC 300 | $26.34 \pm 0.06$ |
| NGC 3109 | $25.57 \pm 0.07$ |
| WLM | $25.05 \pm 0.06$ |
| M33 | $24.97 \pm 0.07$ |
| IC 1613 | $24.37 \pm 0.11$ |

As can be seen from Table 1, FGLR distances are fairly accurate and comparable in accuracy with Cepheid and TRGB distances. However, compared with these two and other well established purely photometric distance determination methods the FGLR method has three fundamental advantages. First, and most importantly, interstellar reddening and extinction are precisely determined for each individual BSG through a comparison between observed and modeled SED data, the latter calculated after T, g, and metallicity were determined as the result of the spectral line fitting. In cases where NIR data are also available, this includes a constraint of the reddening law for each individual object. Second, stellar crowding and blending are significantly less important as a source of systematic uncertainty as, for instance, for Cepheids, where blending bias corrections of a significant amount have to be applied for galaxies at larger distances. Third, effects of metallicity can be accounted for if needed, because the stellar spectroscopy provides accurate BSG metallicities. However, we note that no significant metallicity dependence of the FGLR has been found so far in agreement with stellar evolution theory (Meynet et al. 2015). Figure 2 shows an example of a FGLR distance determination fit for one of the galaxies of Table 1.

Recently, Kudritzki et al. (2020) have discovered that the FGLR distance determination method is not restricted to BSG only. As it turns out, all stars in a wide range of stellar masses, from 0.3 to 20 solar masses and in a wide range of evolutionary phases from zero-age main sequence to the beginning of the red-giant stage follow an extremely tight FGLR (see Figure 3). This can be used to determine distances with ten percent precision and for investigations of the detailed structure of the Milky Way (Kudritzki et al. 2020). In this way, the old dream of stellar astronomers, to tell accurate stellar distances from the inspection of the spectrum, which started with the seminal work by Annie Jump Cannon (Cannon 1918, Cannon & Pickering, 1918), seems to have come true.

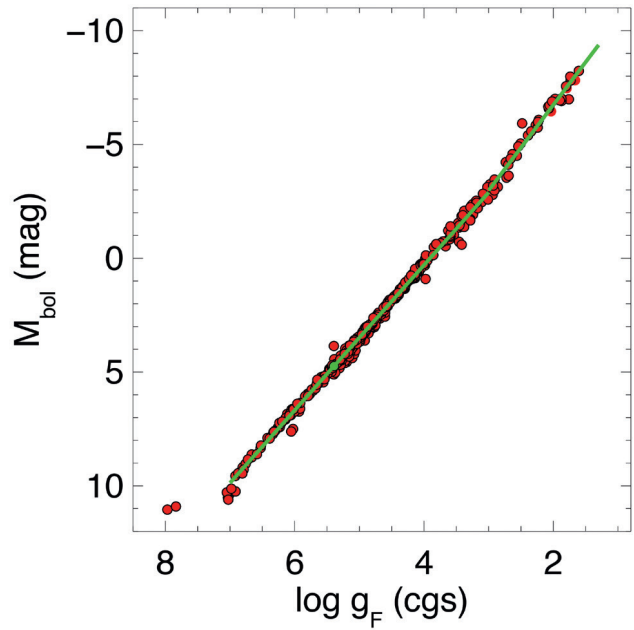

**Fig. 3.** The FGLR of 445 stars in the mass range from 0.3 to 20 solar masses in evolutionary phases from the zero-age main sequence to the beginning of the red-giant stage. The stars have precision stellar parameters and are from the Gaia benchmark sample (Heiter et al. 2015) or detached eclipsing binaries (Graczyk et al. 2018 and DEBcat catalog by Southworth 2015). The two-component regression allows for distance determinations with 10 percent precision. From Kudritzki et al. (2020).

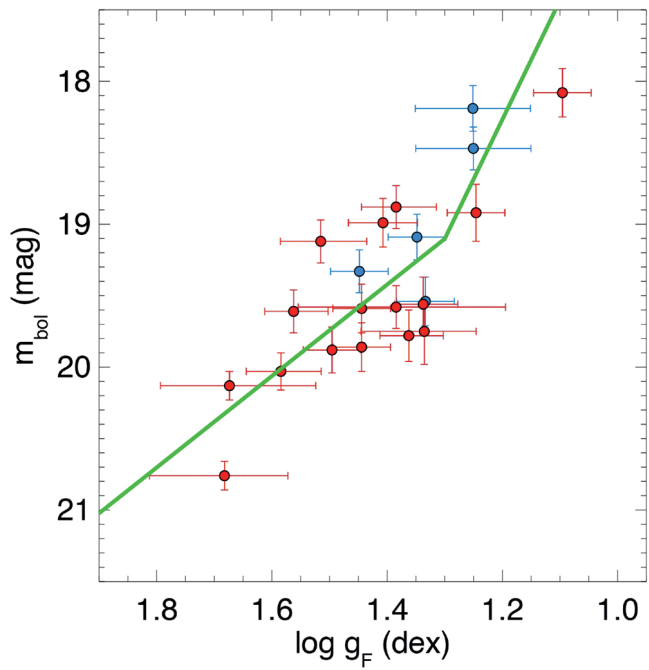

**Fig. 2.** FGLR distance determination of the galaxy M81. The observed FGLR is based on quantitative Keck LRIS spectroscopy. Early spectral type BSG are shown in blue and late type BSG in red. The LMC regression of Figure 1 (green) is shifted in distance modulus until a minimum of the scatter around the relationship is found. This leads to a value of $(m-M) = 27.62 \pm 0.08$ mag for M81.

**Nicolas Nardetto[1]**

# The Nice contribution: CHARA/ VEGA and the SBCR of early-type stars

During my post-doc in Concepcion in 2009, Grzesiek Pietrzyński showed me Figure 1 of di Benedetto and told me something like: *"Look at this surface brightness color relation (SBCR), we are missing data for early-type objects with V-K color lower than 1, which is exactly what we would need to derive the distance of M31 and M33"*. Twelve years later, the progress is significant as we have a SBCR accurate at the 2.5% level for B stars. But let's go back to the beginning.

The eclipsing binary method of distance determination is simple: The linear diameter (r) can be determined for a detached eclipsing binary from the analysis of its light curve and spectroscopic orbit with a precision of the order of 1% (Andersen et al. 1991), while the angular diameter ($\theta$) can be computed from a surface brightness-color relation, i. e. $m_0 = S - 5 \log (\theta)$ where S is the surface brightness and m0 is the intrinsic magnitude in a photometric band. S depends only on the color of the star. The main source of error in this method of distance determination of eclipsing binaries comes almost entirely from the surface brightness-color relation. For the distances of the LMC and SMC, it is necessary to measure the angular diameter of KIII type stars, while for the distance of M31 and M33, it is necessary to calibrate the SBCR for O or B type stars. Thus, calibrating the SBCR for early-type stars paves the way to derive the distance to M31 and M33, which provides crucial constraints on the Hubble constant determination. It is interesting to mention that the current distance considered by the SHOES project (Riess et al. 2016, 2021) which is accurate to 4.4% for M31 (Vilardell et al. 2010) is not based on an SBCR relation but depends on models. One method to calibrate the SBCR is to use interferometric observations.

In order to derive the angular diameter of early-type objects, which are bright but far from us, we need high angular resolution (typically 0.3-0.8 mas). In 2009, at the Nice observatory, the CHARA/VEGA instrument was developed (Mourard et al. 2009). The CHARA array with its 330 m baseline and the VEGA instrument operating in the visible domain are the perfect combination to reach 0.3 mas of angular resolution.

Around 2012, several observing campaigns have been conducted on the CHARA/VEGA instrument in order to derive the angular diameter of B spectral type stars (V-K < 0 mag). Challouf et al. (2014) first obtained a precision of 7-8% on the SBCR by compiling already

[1] Université Côte d'Azur, Observatoire de la Côte d'Azur, CNRS, Lagrange, CS 34229, Nice, France

existing data and by observing 8 stars with CHARA/VEGA. A careful methodology was carried out in order to handle the extinction problem. Also, such calibration requires the activity of the star to be taken into account as it potentially has an impact on the interferometric and/or photometric measurements. This means that we have to discard as many binaries or multiple systems, fast-rotation stars, and stars with winds or complex environments as possible. For this reason, Challouf et al. (2015, 2017) also quantified the effect of fast rotation on the SBCR using dedicated models.

More recently, thanks to a rigorous methodological approach in the selection process of stars, Salsi et al. 2021 identified 18 non-active stars from V-K of −1 to 0.6 mag. These stars were observed intensively in 2019 and 2020 with CHARA/VEGA. The last observations of VEGA before decommissioning in December 2020 were devoted to this program. Using these data and a careful analysis, Anthony Salsi has improved the relation to reach a precision level of the order of 2.5% (see the figure). In this work we could also compare with the inverse approach (see Taormina et al. 2019, 2020). The results are consistent within the uncertainties, but one has to take care about conversion issues between K and $K_S$ photometry in the SBCR calibration. This SBCR could already be used to derive the distance of extragalactic early-type eclipsing binaries. However, our goal is to further improve, in terms of precision/accuracy, the SBCR using the next generation visible interferometer CHARA/SPICA.

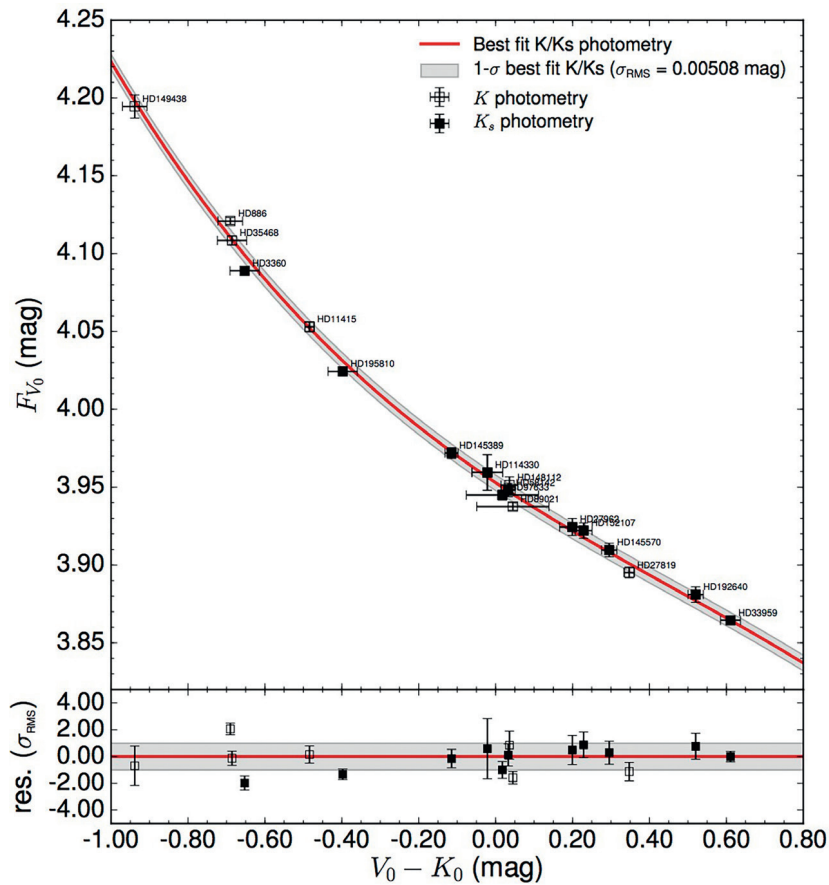

**Fig. 1.** The surface brightness color relation for early-type stars calibrated using CHARA/VEGA interferometric observations in the visible domain (Salsi et al. 2021).

SPICA (Stellar Parameters and Images with a Cophased Array) is an instrument operating in the visible domain, fibered, with 6 telescopes and equipped with a new generation detector, which takes advantage of the new adaptive optics of the CHARA interferometer (SPIE paper, Mourard et al. 2018). It will be possible to measure the angular diameter of a star with a precision of 1 % in less than 30 minutes of observation. With this instrument we already identified 20 dwarf and sub-giants stars with V-K < 0 mag. They are B stars with angular diameters from 0.2 to 0.5 mas. The key point is that with its 6 telescopes CHARA/SPICA will detect stellar activity, which will allow us to verify that the selected stars are indeed not active. Among the 20 stars, 13 have infrared photometry more precise than 0.025 mag (or 2%). Dedicated infrared homogeneous observations are required to improve the calibration (for instance from Armazones). With all these improvements in hands, we expect to reach a 1-2% precision on the SBCR or early-type stars, which will open the way to derive the distance to M31 and M33 empirically with a few percent precision.

Mónica Taormina[1]

# Toward Early-type Eclipsing Binaries as Extragalactic Milestones

After decades of great effort, a precise 1% distance to the Large Magellanic Cloud (LMC) was obtained by Pietrzyński et al. (2019) using detached eclipsing binaries composed of late-type giants. Currently, the challenge is to measure distances to other galaxies in the Local Group in a similar way, which is necessary to improve the accuracy of the determination of the Hubble constant.

In that regard, double-lined detached eclipsing binaries (DEBs) composed of early-type stars provide a unique way to achieve such a goal. They are bright enough and can be easily used out to distances of 1Mpc, e.g. in the M31 and M33 galaxies.

The first to use extragalactic early-type DEBs as distance indicators were Guinan et al. (1998), who analyzed such a system in the LMC galaxy. After that, distances to similar objects in the LMC (Ribas et al. 2000, Fitzpatrick et al. 2003), M31 (Ribas et al. 2005), and M33 (Bonanos et al. 2006) were determined. However, these distances relied on surface brightnesses predicted by model atmosphere theory, once the effective temperatures of the components were derived from quantitative spectroscopy or spectrophotometry. The problem is that the atmospheres of hot massive stars are affected by strong deviations from local thermodynamic equilibrium, severe metal-line blanketing, and the hydrodynamics of stellar winds. All these effects need to be taken into account and, even with the enormous progress in modeling the atmospheres of hot massive stars that has been made, the jury is still out on how accurate the surface brightness predictions are. In addition, massive early-type stars, just born in regions of heavy star formation, are usually severely affected by interstellar reddening with reddening laws that deviate from those usually applied (Maíz Apellániz et al. 2014, 2017; Urbaneja et al. 2017). The best example of how difficult it is to determine a distance to an early-type DEB, is the spread of distance moduli from 18.20 to 18.55 mag determined to the same eclipsing binary HV 2274 in the LMC (e.g. Guinan et al. 1998; Udalski et al. 1998; Nelson et al. 2000).

To overcome these problems one has to accurately determine the surface brightness of early-type stars, find the dependence on e.g. V-K color, and construct a precise surface brightness-color relation (SBCR). In our project, we investigated the light curves of all blue (V-K < 0.0 mag) binaries in the LMC and selected nine which, thanks to their brightness and light curve shape, are best suited for distance determination. Although all systems from our sample were expected

[1] Nicolaus Copernicus Astronomical Center, Polish Academy of Sciences, Bartycka 18, 00-716 Warszawa, Poland

to be of B-type, our preliminary analysis showed that some of them are O-type binaries, which are even more valuable for the calibration of the SBCR. The goal is to model the selected systems to obtain the precise radii of the components and use the known distance to the LMC to obtain their surface brightness. Then, with the use of observed V- and K-magnitudes, we will be able to improve the calibration of the SBCR focusing on an extension towards hot stars.

We have already acquired spectroscopic and infrared observations for all nine systems and modeled their light and radial velocity curves. After a detailed analysis, we obtained the properties of five of these systems and the physical parameters, including masses and radii, of their components (Taormina et al. 2022, in prep., see also Figure 1). These measurements are the most precise that have been ever obtained for extragalactic early-type binary systems. All analyzed binaries have been confirmed to be well-detached, although some components were found to be slightly deformed by tidal forces. Components of two systems were found to still rotate super-synchronously, which means that the tidal forces have not yet slowed down their rotation significantly.

The blue part of the surface brightness-color relation is very sensitive to the reddening towards the systems. To obtain reliable results we determined the reddening using three different methods. First, we measured the equivalent widths of interstellar sodium lines, which can be translated into values of reddening (Munari & Zwitter 1997), second, we used the recalibrated reddening map of Haschke et al. (2011), and third, we fitted the spectral energy distribution for each system. We found a good agreement between these three methods. The reddening values for systems in the sample ranged from 0.07 to 0.19 mag.

We performed an evolutionary analysis for two of the systems (Taormina et al. 2019, 2020). In both cases, it was impossible to fit evolutionary tracks with parameters typically assumed for massive early-type stars for the measured masses. The predicted luminosities were too high. Our results confirmed that the so-called "mass discrepancy" problem, already noted in the literature for Galactic objects, is also present in the LMC galaxy. We proposed a possible solution to this problem, which led to models consistent with our observations. From spectroscopic analysis, we determined independent distances to these two systems, which resulted to be within error bars from the current best distance determination to the LMC (49.59 kpc).

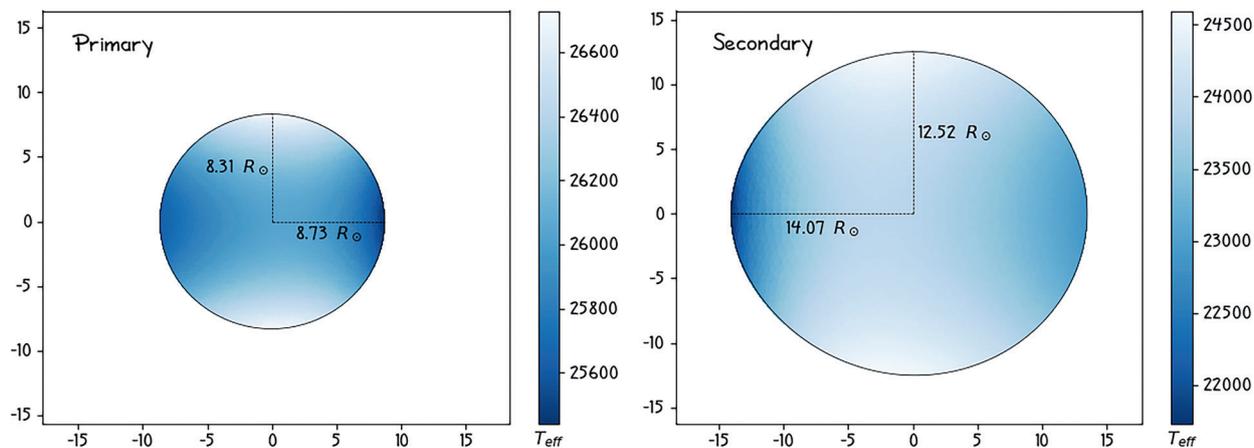

**Fig. 1.** Example of the surface temperature distribution and the shape of the primary (left) and secondary (right) component of one of the B-type systems (OGLE LMC-ECL-22270) as seen from the side. The same scale is used in both panels. Polar and maximum (toward the mass center) radii are shown.

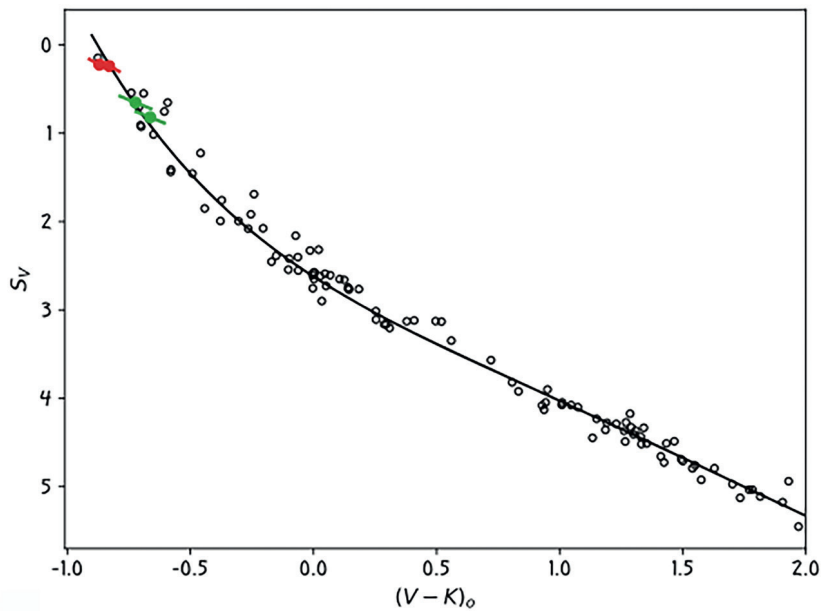

**Fig. 2.** Stellar surface brightness–color relationship. Interferometric measurements (small open circles) and the corresponding fit (solid line) from Challouf et al. (2014) are shown together with our results for OGLE LMC-ECL-22270 (green) and the O-type binary OGLE LMC-ECL-06782 (red).

Finally, we tested the best available (at that time) calibration of the surface brightness-color relation for early-type stars obtained from interferometric observations by Challouf et al. (2014) that is uncertain to 8%. Our results were found to be in good agreement with that relation (Taormina et al. 2020, see also Figure 2). We expect to obtain an accuracy of 2% once the analysis of all the systems is done in a uniform manner.

**Fabio Bresolin[1]**

# Metallicities of Nearby Galaxies

During the course of the Araucaria Project we have explored in detail the chemical composition of nearby galaxies. Our motivation is the potential dependence of the properties of the distance indicators we use on their metallicity. In addition, the scrutiny of the chemical abundances of galaxies provides essential clues about their evolutionary status.

While spectra of long-established stellar distance indicators, such as Cepheids and eclipsing binaries, can directly provide their metal content, this can reasonably be accomplished only for objects residing in the Milky Way and the Magellanic Clouds. At considerably larger distances the primary extragalactic metal tracers of the young population are giant H II regions, which are bright and numerous in spiral and irregular galaxies. For decades the spectral analysis of these ionized gas clouds has afforded an indispensable tool to probe the chemical abundances of galaxies.

The Araucaria approach to the study of the metal content of galaxies focuses on the additional, complementary investigation of the young and bright stellar constituent of galaxies: the blue supergiant stars. This strategy allows us to address the study of extragalactic metallicities by adopting distinct methods and line diagnostics. Since the B- and A-type supergiants are the visually brightest stars in hydrostatic equilibrium, we can resort to relatively low-resolution (around 5 Å FWHM) spectroscopy at 8-to-10-m class telescopes to quantitatively analyze stars located well beyond the confines of the Local Group, out to distances of 6-8 Mpc (Bresolin et al. 2001; Kudritzki et al. 2013, 2014).

The derivation of the chemical abundances of supergiant stars and ionized gas relies on different techniques and spectral features. For example, modeling the atmospheres of massive stars, accounting for non-LTE effects, is essential for the blue supergiant analysis. The oxygen abundance of the H II regions yields a proxy for their metallicity [Z], utilizing the solar O/H value and assuming that the other elements scale following the solar pattern. The metallicity of the blue supergiants is instead calculated from the analysis of spectral lines originating from multiple elements, including C, N, O, Mg, Si (B types), Mg, Si, Ti, Cr and Fe (A types).

The comparison between nebular and stellar (supergiant) abundances helps us to address some serious difficulties in deriving metallicities from H II region emission lines. First is the inability to implement the classical, so-called *direct* method, which involves the flux of weak auroral lines, at high metallicities (approaching the solar value). This makes it necessary to use *strong-line* methods of abundance determination, which however are affected by large systematic uncertainties. Second is the systematic abundance difference one obtains by using metal recombination lines (seldom observed for extragalactic targets, due to their weakness) instead of collisionally excited lines (the standard diagnostics).

[1] Institute for Astronomy, University of Hawaii, 2680 Woodlawn Drive, 96822 Honolulu, USA

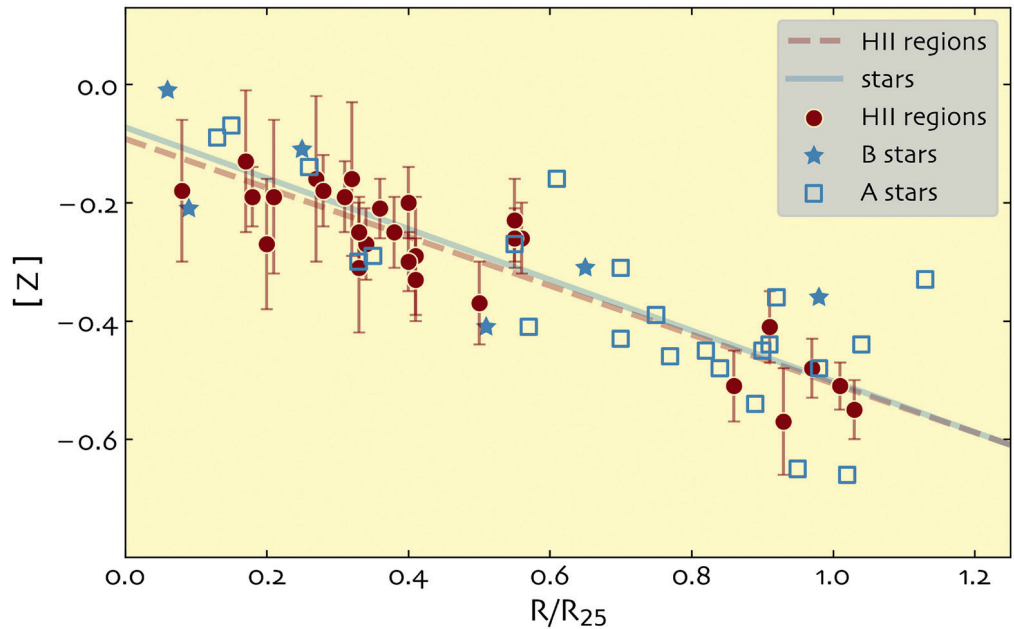

**Fig. 1.** The radial abundance gradient in NGC 300 determined from both H II regions and blue supergiants (adapted from Bresolin et al. 2009).

For several galaxies included in the Araucaria sample we have been able to compare stellar and nebular metallicities, the latter obtained via the direct method. For irregulars, thanks to the low value and homogeneity of the metallicity, a few stellar and nebular targets suffice to carry out a meaningful comparison. As an example, in IC 1613 we found excellent agreement in the oxygen abundances of two ionized nebulae and nine early-B supergiants (Bresolin et al. 2007).

In the case of spiral galaxies the comparison is performed by looking at the radial distribution of the chemical abundances derived independently from stars and ionized gas clouds. Examples include NGC 300 (Bresolin et al. 2009), M81 (Kudritzki et al. 2012), M83 (Bresolin et al. 2016) and NGC 2403 (Bresolin et al. 2022, in prep.). In Figure 1 the stellar and H II region radial metallicity gradients in NGC 300, determined from 29 and 28 individual objects, respectively, display an impressive degree of agreement.

The state of the art is illustrated in Figure 2, adapted from Bresolin et al. (2022, in prep.). The stellar metallicities for 10 of the 15 targets included in this plot were measured by us using the VLT and Keck telescopes. We find that the stellar metallicities agree well with the nebular metallicities derived using the direct method, across a wide [Z] range. The measurements based on the metal recombination line analysis in H II regions appear instead to diverge as the metallicity decreases. Overall, these results give us confidence in the validity of the techniques we employ to derive the chemical abundances of star-forming galaxies.

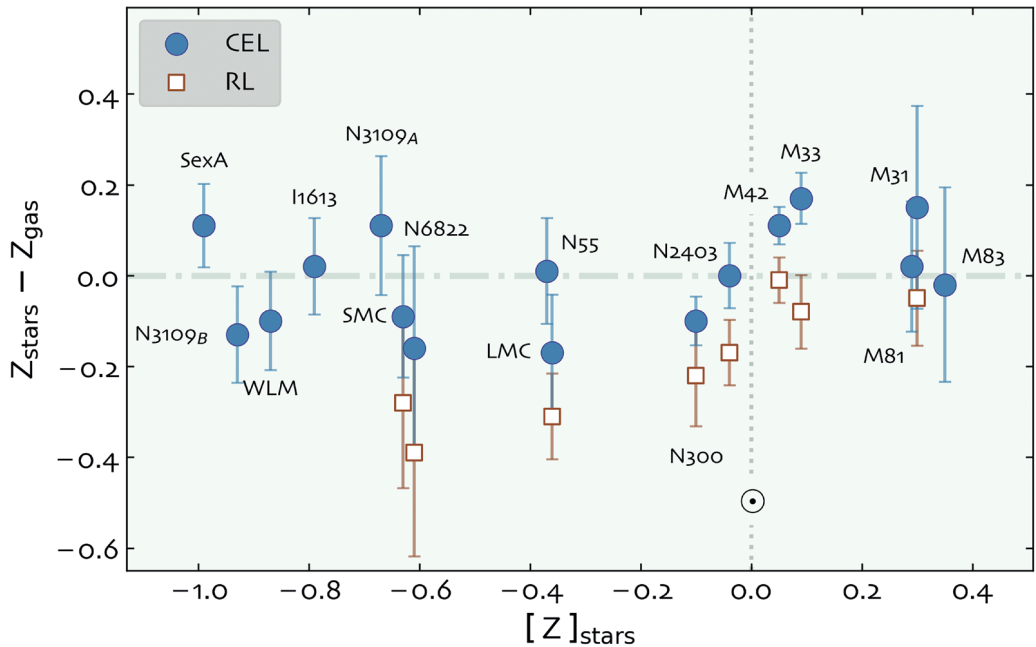

**Fig. 2.** The metallicity difference between supergiants and H II regions as a function of stellar metallicity, as derived for 14 galaxies and the Orion nebula. The gas metallicity has been measured for all objects with the direct method applied to the collisionally excited lines (CELs). The metallicity based on the detection of oxygen recombination lines (RLs) has been measured for a subset of eight targets (adapted from Bresolin et al. 2022, in prep.). A correction of +0.1 dex has been applied to the gas abundances, in order to account for dust depletion.

**Radosław Smolec[1]**

# Modeling of classical pulsators

As excellent distance indicators, classical pulsators are among the most important stars for stellar astrophysics. Classical Cepheids are the inevitable rung of the cosmic distance ladder. RR Lyrae stars serve as excellent tracers of old stellar populations, providing insight into the evolutionary history of the Milky Way and its local neighborhood.

Applications of these stars go far beyond distance determination. For example, RR Lyrae stars are used to determine metallicity and chemical evolution history of the hosting system through empirical photometric metallicity formulae. Classical Cepheids, through period-age relation, provide a clock for age tomography.

While based on firm theoretical grounds, the use of classical pulsators as distance indicators is nearly exclusively based on empirically calibrated period-luminosity relations. Possible metallicity dependence of these relations is studied through observations-calibration and comparison of different metallicity environments (see e.g. Breuval's contribution in this book). Theoretical modeling seems to play a secondary role.

As evolved, core helium-burning stars, classical pulsators are a magnifying glass for stellar evolution theory, which reveal several disturbing problems. Focusing on classical Cepheids, the most challenging problem seems to be the mass discrepancy problem: masses of Cepheids as determined from stellar-evolution theory are overestimated as compared to masses resulting from stellar-pulsation theory. Thanks to Ar-aucaria's analysis of a growing number of Cepheids in eclipsing binary systems (see Pilecki's contribution in this book) we know that pulsation theory is right! Part of the discrepancy was due to the poor opacity tables, but even after their revision, the evolutionary and pulsation masses are discrepant at 20% level (Keller 2008). Proposed solutions to the mass discrepancy problem through inclusion of three different physical processes just highlight the intrinsic shortcomings of stellar evolution and pulsation theories. Two scenarios rely on additional mixing beyond the convective hydrogen-burning core. Bringing fresh hydrogen into the core, mixing prolongs the main sequence lifetime and increases luminosity at a given mass. Proper determination of the convective core boundary represents a real challenge in 1D stellar evolution codes in which convective energy transfer is described through a local mixing length theory (see e.g. Paxton et al. 2019). Convective core overshooting, an ad hoc additional mixing beyond the convective boundary, is thus commonly invoked in stellar evolution calculations. Not only may it solve the mass discrepancy problem (e.g. Prada Moroni et al. 2012) but it seems essential to reproduce many other observations, including the observed width of the main sequence.

[1] Nicolaus Copernicus Astronomical Center, Polish Academy of Sciences, Bartycka 18, 00-716 Warszawa, Poland

The models that invoke overshooting usually neglect rotation. But stars do rotate! While giant Cepheids are slow rotators, their main sequence progenitors were not. Rotation not only affects the hydrostatic structure of the model, but leads to additional mixing through a few rotationally induced instabilities that bring several free parameters to the theory (see e.g. Heger et al. 2000). As such, rotation acts similar to overshooting. It was invoked as another solution to the mass discrepancy problem (see Anderson et al. 2014). However, including rotation in 1D evolutionary codes needs some simplifications and there is no unique method of its implementation. The additional mixing relies on the treatment of composition gradients, that for a typical stellar stratification stabilize against radial mixing (which brings another free parameter to the theory).

Another approach addressing the Cepheid mass discrepancy problem aims to reduce mass at a given luminosity through pulsation driven mass loss (e.g. Neilson et al. 2011). Modeling of underlying shocks in extended, pulsating atmospheres, is strongly simplified though.

Mass discrepancy is not the only disturbing problem for Cepheids. Sometimes getting a blue loop in an evolutionary track turns out to be a challenge, not only at the low and high mass ends (a persistent problem), but even for moderate masses loops may behave in an erratic way. Clearly, modeling of this evolutionary phase is very challenging and in fact relies on proper numerical and physical treatment of earlier evolutionary phases, in particular on modeling the chemical stratification left by the retreating convective core at the end of the main sequence.

Despite being the most simple pulsating stars, pulsating radially and typically in a single pulsation mode, the nonlinear nature of their large-amplitude pulsation poses a challenge for stellar pulsation theory. Among the phenomena that are still lacking satisfactory explanation are Blazhko modulation in RR Lyrae stars (see however Kollath 2021) and the mode selection problem (what are the conditions and what drives double-mode pulsation; see e.g. Smolec 2013). Through the past few years, thanks to all-sky surveys and precise photometry of space telescopes, classical pulsators revealed a more complex nature and exposed new challenges. Low-amplitude periodic modulations appear to be common, even in fundamental mode classical Cepheids (see Smolec 2017). Excitation of low-amplitude non-radial modes may be common in first overtone pulsators (e.g. Netzel & Smolec 2019 and references therein). These new phenomena still await explanation or elaboration of the existing theory (e.g. Dziembowski 2016). While with the inclusion of time-dependent convection into nonlinear pulsation codes clear progress was made in reproducing light and radial velocity curves, the most challenging problems (cause of modulations, mode selection problem) were not solved. A somewhat hidden cost is the introduction of several free parameters to the theory. Radiative energy transfer is still oversimplified in most of the nonlinear codes (diffusion approximation is commonly used).

Further progress of evolution and pulsation theories requires going beyond 1D and clearly needs a better description of convective energy transfer and related phenomena, crucial for both evolution and pulsation. Thus it requires enormous computing power, development of new numerical techniques and entirely new tools. The progress is slow: some 2D and 3D pulsation models are far from a state at which they can be routinely confronted with observations (see however Geroux & Deupree 2015). For evolution theory, the efforts go into inclusion of results of 3D hydrodynamic simulations into traditional 1D structure equations (see e.g. Arnett et al. 2015). Needless to say these 3D simulations are often simplified and still far from stellar parameter regimes (see e.g. Kupka & Muthsam 2017). We will thus stay with 1D tools for years to come. Im-provements may come from calibrating several free parameters of the existing 1D models. For this purpose we need new top-quality observations and precise observational determinations of physical parameters of stars. Here Araucaria provides and will continue to provide invaluable input. Homogeneous sets of light curves across different passbands and spectroscopy for classical pulsators will improve existing empirical relations (like period-luminosity) and provide constraints for models, both evolutionary and pulsation models. Precise physical parameters of double-lined eclipsing binaries (see e.g. Graczyk, Suchomska, and Taormina's contributions in this book) will provide constraints for chemical mixing efficiency and extent at various evolutionary stages. Homogeneous light curves across different passbands and from different environments from Cerro Armazones will help to calibrate nonlinear pulsation theory.

Paulina Karczmarek[1], Radosław Smolec[2]

# Binary Evolution Pulsators
# - a new class of pulsating stars

In 2011 a promising candidate for an RR Lyrae star in an eclipsing binary system was found (Soszyński et al. 2011). The object, named OGLE-BLG-RRLYR-02792, exhibits RR Lyr-like pulsations with a period of 0.627 d and eclipses with a period of 15.24 d (Figure 1ab). The discovery was astonishing, because not even one case of an RR Lyrae (RRL) star in an eclipsing binary system had been known before, and consequential, because physical properties of both components (especially their masses) could be determined from simultaneous analysis of the light and radial velocity curves. A model-independent mass measurement of an RRL star would be vital to constrain the stellar pulsation and evolution theories.

However, the dynamical mass of the pulsating component of OGLE-BLG-RRLYR-02792 turned out to be only 0.26 $M_\odot$ (Pietrzyński et al. 2012), less than a half of the mass required for helium ignition, and therefore completely at odds with the predictions of theoretical models of RR Lyrae stars (Bono et al. 2003). The presence of a cooler (by 2300 K), fainter (by about 2 mag in the V-band), and more massive companion ($M_{comp}$ = 1.67 $M_\odot$) is a clue that mass transfer had to occur in the past. Indeed, the system evolved from a close binary with an initial orbital period of 2.9 d, underwent the mass-transfer episode from the initially more massive primary to the secondary, and, as a result, the primary became a hot low-mass helium-core star with a thin hydrogen envelope. Its temperature and luminosity placed it in the same area of the instability strip (IS) in the Hertzsprung-Russell diagram that is occupied by RRLs (Figure 1c). This discovery set up a new class of pulsating stars. They are referred to as Binary Evolution Pulsators (BEPs) to underline the interaction between components, in the form of the mass transfer, which is mandatory to peel off the mass of one component and evolve it through the IS.

Linear pulsation calculations show that BEPs are similar to classical pulsators: this new class of pulsating stars should exhibit radial pulsations driven through the opacity mechanism acting in the hydrogen-helium partial ionization zone (Smolec et al. 2013). OGLE-BLG-RRLYR-02792 is a radial fundamental mode pulsator. With precise determination of physical parameters and well sampled light and radial velocity curves it is a unique target for nonlinear pulsation modeling. The best matching nonlinear convective pulsation model (Figure 2) matches the overall shape of the ob-served light and radial velocity curves well. However, the phase lag between the radial velocity curve and the light curve is significantly smaller in the model. Non-linear modeling provides an insight to the origin of the bump observed in the middle of the ascending branch of the radial velocity curve (Figure 2, bottom panel). It is caused by the 2:1 resonance between the fundamental mode and the linearly stable second overtone – the same mechanism that causes the Hertzsprung bump progression in classical Cepheids. A nonlinear pulsation model survey, conducted for a range of physical parameters expected for BEPs, reveals a clear bump progression and shows that the 2:1 resonance may be crucial in shaping the light and radial velocity curves of BEPs. The resulting subtle structures in the light and radial velocity curves can be used to distinguish BEPs from RR Lyrae-type stars.

[1] Universidad de Concepción, Departamento de Astronomia, Casilla 160-C, Concepción, Chile
[2] Nicolaus Copernicus Astronomical Center, Polish Academy of Sciences, Bartycka 18, 00-716 Warszawa, Poland

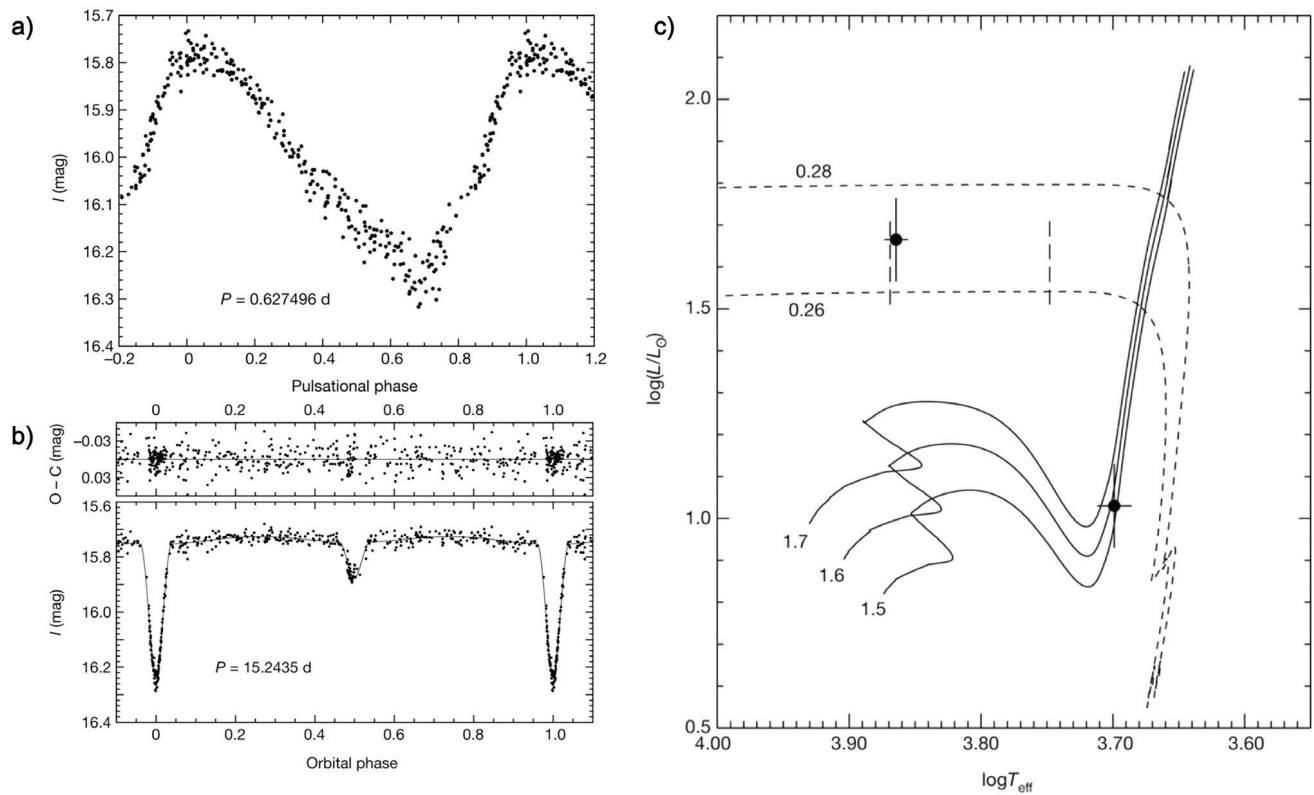

**Fig. 1.** Light curve of OGLE-BLG-RRLYR-02792, folded with a) the pulsational period and b) the orbital period. c) Location of the BEP and its companions (black solid circles) on the Hertzsprung-Russell diagram. Figures excerpted from Pietrzyński et al. (2012).

The serendipitous discovery of the first BEP was possible due to its eclipses, which means that other BEPs in non-eclipsing binaries could have been mis-classified and, as a result, contaminate the sample of genuine RRLs. The consequences for distance and age determinations using significantly contaminated RRL samples would be as follows: (i) unresolved BEPs are brighter than RRLs, which would make the distances determined from the near-infrared RRL period-luminosity relation shorter than in the reality; (ii) globular clusters and galaxies hosting unresolved BEPs would be considered younger than if they hosted genuine RRLs. In order to estimate the percentage of BEPs among RRLs (i.e. *contamination value, c*) we created a synthetic population of BEPs using the binary population synthesis code StarTrack (Belczynski et al. 2002, 2008). We allowed BEPs to be created in the entire luminosity range of the IS, i.e. $\log(L/L_\odot) = 1-4$, and discovered that BEPs can contaminate not only RRLs ($c_{RRL}$ = 0.8%) but also more luminous classical Cepheids ($c_{Cep}$ = 5%), however the contamination values are low and should not have a significant impact on distance and age determinations. Population synthesis confirmed that BEPs have low masses (0.2-0.8 $M_\odot$), and predicted that they should be found in binaries with orbital periods in range 10-2500 d (Karczmarek et al. 2017). These findings are purely theoretical and need to be verified by observations of many more BEPs; unfortunately only one confirmed BEP is known up to date, and one RRL-like BEP candidate (OGLE-LMC-RRLYR-03541) awaits a spectroscopic confirmation of its nature. Perhaps brighter, Cepheid-like BEPs can prove easier to detect and/or reclassify in the future.

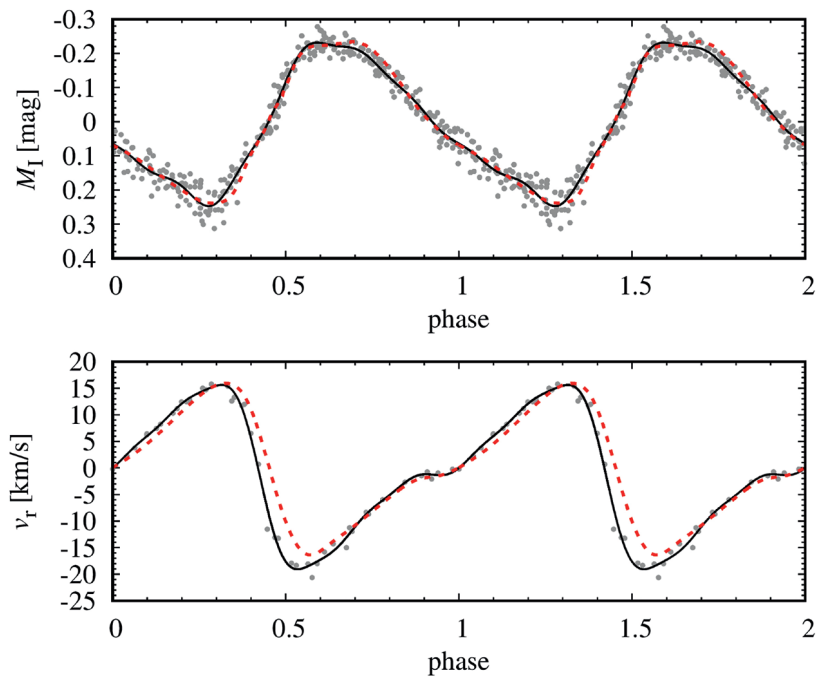

**Fig. 2.** I-band (top) and radial velocity (bottom) curves for OGLE-BLG-RRLYR-02792. In both panels gray dots are observations and black solid lines are the corresponding Fourier series fits. Best matching nonlinear pulsation model is plotted with a red dashed curve.

Weronika Narloch[1]

# Non-pulsating stars in the instability strip of classical Cepheids

Pulsating stars occupy specific regions on the Herzsprung–Russell (HR) diagram. Several related classes of classical pulsators (stars periodically expanding and shrinking in the radial direction) populate the so-called instability strip (IS, e.g. Cox 1974; Gautschy & Saio 1996). It is believed that, during its evolution, when a star enters the IS it develops this instability and starts to pulsate. This general conviction, however, does not seem to explain the behaviour of all the stars in the IS.

As early photometric studies have shown, a significant fraction of stars located in the IS region occupied by Cepheid variables are photometrically stable at the level of tens of mmag (e.g. Fernie & Hube 1971; Percy et al. 1979). Butler (1998) noticed that among 15 photometrically stable stars from the IS, four showed no variability on a specified level, while the rest of them manifested a variety of behaviours. He concluded that observed peaks in periodograms could be due to, e.g. a companion planet (or star) or non-radial pulsations. Pilecki et al. (2018) analysed an eclipsing binary system OGLE-LMC-CEP-4506 composed of a classical Cepheid and a secondary component, which turned out to be a non-pulsating red giant residing in the center of the classical IS. These stars turned out to be similar in terms of masses, radii, and colors, with the Cepheid being more evolutionarily advanced. Several recent studies reported the existence of non-pulsating stars also in the IS of other classes of pulsating variables. Guzik et al. (2013, 2015) found stable objects lying in the IS of δ Sct and γ Dor stars in Kepler data. Rozyczka et al. (2018) showed that the variable star V4, being a member of the globular cluster M10, is a constant star and not a RR Lyr-type star as was suspected. These few examples prove the existence of non-pulsating stars in the IS and raise the questions: How many are there? What is the nature of these objects?

To seek an answer to the first question, we decided to select the candidates for the non-pulsating stars residing in the Cepheid IS in the Large Magellanic Cloud (LMC) using a combination of OGLE (Optical Gravitational Lensing Experiment) BVI photometry (Udalski et al. 2000) and Strömgren photometry obtained by our group. The Strömgren photometric system is composed of four medium-band filters (*uvby*) defined specifically for stellar astrophysics (Crawford 1987). Par-ticularly useful in our study was *(b-y)* colour which is a good estimator of the temperatures of stars, as well as *c1=(u−v)−(v−b)* index which is sensitive to surface gravity. The data used in the study were collected within the Araucaria Project (Gieren et al. 2005) during a single night with a 4.1-m SOAR (Southern Astrophysical Research) Telescope placed in Cerro Pachón in Chile and equipped with a SOI (SOAR Optical Imager) camera.

[1] Universidad de Concepción, Departamento de Astronomia, Casilla 160-C, Concepción, Chile

As a first step, we defined empirical boundaries for the Cepheid IS using the LMC Cepheids from the OGLE Collection of Variable Stars. We employed an edge detection technique on the previously derredened catalogue and defined the blue and red edges of the IS. We selected the stars lying within these boundaries on the OGLE VI colour-magnitude diagram (CMD) with magnitudes no fainter than the faintest Cepheid. Next, we excluded from the sample stars identified as Cepheids (see Figure 1). We then cross-matched the OGLE photometry with our Strömgren photometry and plotted the selected sample of potential candidates for non-pul-sating stars together with identified Cepheids on the *c1-(b-y)* relation (see Figure 2). We used the theoretical *c1-(b-y)* relations from static atmosphere models of Castelli & Kurucz (2003) for specific surface gravity (log(g), adjusted accordingly to match our Cepheids) to further select stars lying in a relatively wide range of possible log(g) values of Cepheids, i.e., between log(g) = 1 and log(g) = 3. In the final step, we applied data from the *Gaia* Data Release 2 (DR2) catalogue (Gaia Collaboration et al. 2016, 2018) to reject stars with proper motions within a box of ±1 mas/yr around the average proper motion of the LMC. The rejected stars are most probably giants from our Galaxy. In the end, we were left with 19 reliable candidates for non-pulsating stars. These candidates account for approximately 30 percent of the whole sample of objects (Cepheids and non-pulsating stars) located in the classical IS in the LMC. Comparison of this result with the CMD based on Strömgren photometry put seven of the candidates out of the IS edges, reducing previous percentage value to 21 percent.

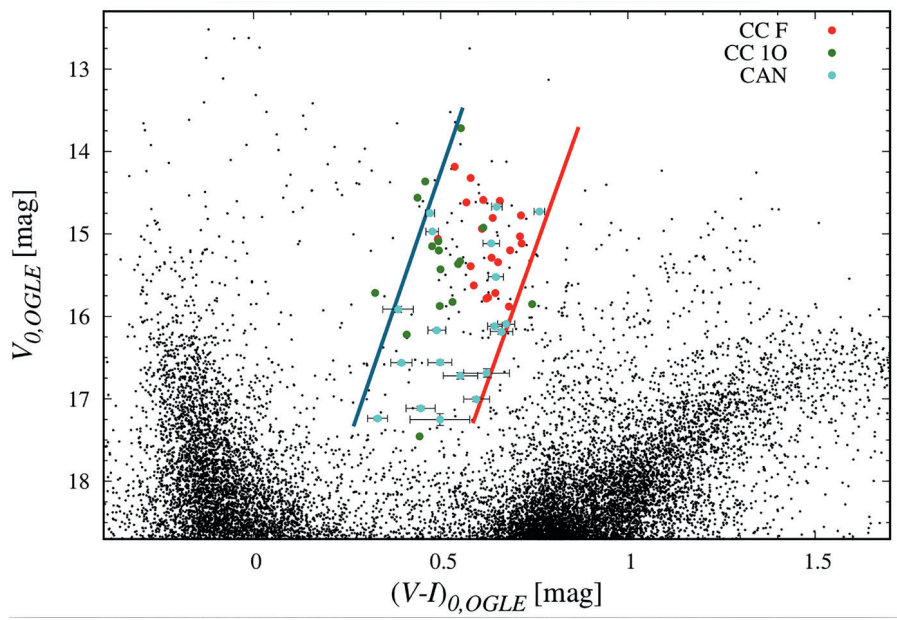

**Fig. 1.** Reddening-corrected OGLE CMD for analysed fields (black points) with marked Cepheid variables in the fundamental (red dots) and first-overtone (green dots) modes and candidates for non-pulsating stars (cyan dots). Source: Narloch et al. (2019).

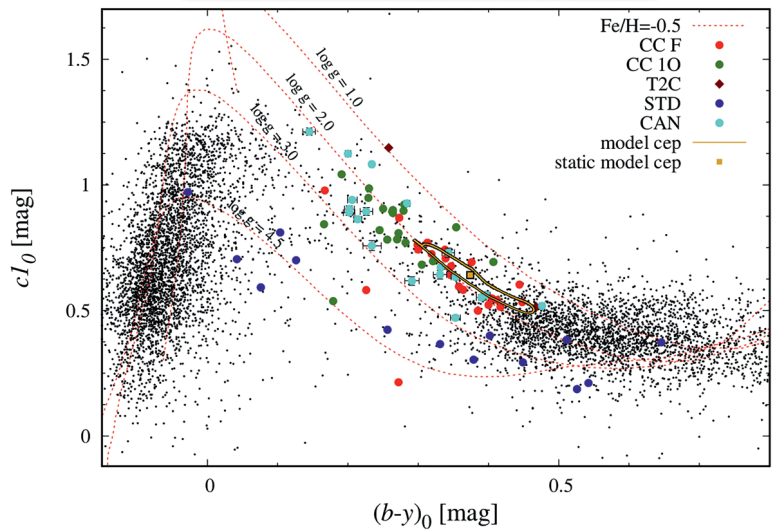

**Fig. 2.** Reddening-corrected two-colour diagram brighter than 19 mag from our four fields in the LMC with marked classical Cepheids (red and green dots), Type II Cepheids (dark red diamond), candidates for non-pulsating stars (cyan dots), standard main-sequence stars used for standardization (blue dots), and theoretical $c1$-$(b$-$y)$ relations for metallicity [Fe/H] = –0.5 dex. The yellow loop marks a model of a Cepheid during its pulsation cycle, while a yellow square shows a static model for this Cepheid. Source: Narloch et al. (2019).

The analysis of candidates' OGLE light curves with an arbitrarily set signal-to-noise ratio of 4 as a threshold for reliable detection of significant periodicity in the frequency spectrum, confirmed that no significant variability was detected at the level of a few mmag. We also investigated the possibility of blending due to binarity and crowding. Blending can affect the light of the star causing it to fall into or out of the IS. Some authors (e.g. Mochejska et al. 2000) show that this effect can be significant in the case of Cepheids. In the extreme case, all our candidates could be blends and we are not able to resolve this issue without high quality spectroscopy. The effect of crowding was easier to estimate by doing an artificial star test, which showed that about one third of our candidates could be blended, but the remaining ones still account for about 21 percent.

Follow-up work would focus on answering the next intriguing question about the nature of selected objects. What could be the cause of their lack of pulsations? Are they binary systems lying, in fact, outside of the IS? Or maybe they pulsate in non-radial modes? The observational premises for such explanations would be very valuable for theoretical knowledge. We hope that spectroscopy of our candidates will help to reveal the secret of these puzzling objects or at least reject some of the possible explanations.

Cezary Gałan[1]

# Atmospheric parameters and abundances of eclipsing-binary-system components from high-resolution spectra

Simultaneous modelling of the light and radial velocity curves of eclipsing-binary systems can provide good values for the temperature ratio (in addition to a nearly complete set of the orbital parameters), but cannot provide the temperature of the individual particular components very well, and information about the metallicity is often uncertain. This is because limb-darkening coefficients are not adjusted during light curve fitting and additional external information is needed to set the effective temperature scale.

Thus, we launched a program to derive, in an independent way, the atmospheric parameters of eclipsing binary system components, and additionally to derive the chemical composition of their atmospheres. With the next-phase goal of deriving additional constraints to improve calibrators for the surface brightness–color relation, we aim to test the evolutionary status and the mass loss/mass transfer in these objects, and possibly also the convective core overshooting in F-type stars, as well as the issues concerning synchronization of rotations.

## Methods

We used high-resolution HARPS spectra, disentangled with the RaveSpan code (Pilecki et al. 2017), and the "Grid Search in Stellar Parameters" GSSP software package (Tkachenko 2015). The code uses the spectrum synthesis method by employing the SYNTHV LTE-based radiative transfer code (Tsymbal 1996). We used the LLMODELS grid of atmosphere models (Shulyak et al. 2004) provided with the GSSP code, suitable for the parameters of our targets.

We generally searched for four free parameters: metallicity [M/H], effective temperature $T_{eff}$, microturbulent velocity $\xi$, and projected rotational velocity $V_{rot} \sin i$, however in some cases (Graczyk et al. 2021) we also searched for one more free parameter: macroturbulent velocity $\zeta$. In selected cases abundances were also calculated for ~30 individual chemical elements. We relied mainly on the "binary" version of the code (as the most suitable for our sample). GSSP calculated synthetic spectra for each set of grid parameters and compared them with the normalised observed spec-trum. The $\chi^2$ function was calculated to judge the goodness of the fit. The code searched for the minimum $\chi^2$, and yielded the best-fit parameters together with the values of reduced $\chi^2$ and the $1\sigma$ uncertainty level in terms of $\chi^2$. In our HARPS spectra the continuum level was not reconstructed well enough in the region of hydrogen Balmer lines; therefore, the appropriate regions of H$\alpha$, H$\beta$, and H$\gamma$ lines were excluded from the analysis. We also skipped the regions contaminated by the lines from water and oxygen molecules in Earth's atmosphere (see Figure 1).

[1] Nicolaus Copernicus Astronomical Center, Polish Academy of Sciences, Bartycka 18, 00-716 Warszawa, Poland

## Atmospheric parameters

The input values of the parameters of the system components ($T_{eff}$, log g, $V_{rot}$ sin i) were taken from the results of prior modelling with the use of the Wilson-Devinney (W-D) code, which simultaneously fit the light and velocity curves of both components. The surface gravities were fixed on the values from the W-D solution, while the five free parameters were: [M/H], $T_{eff}$, $\xi$, $\zeta$, and $V_{rot}$ sin i. The procedure used to find the solution was described by Graczyk et al. (2021).

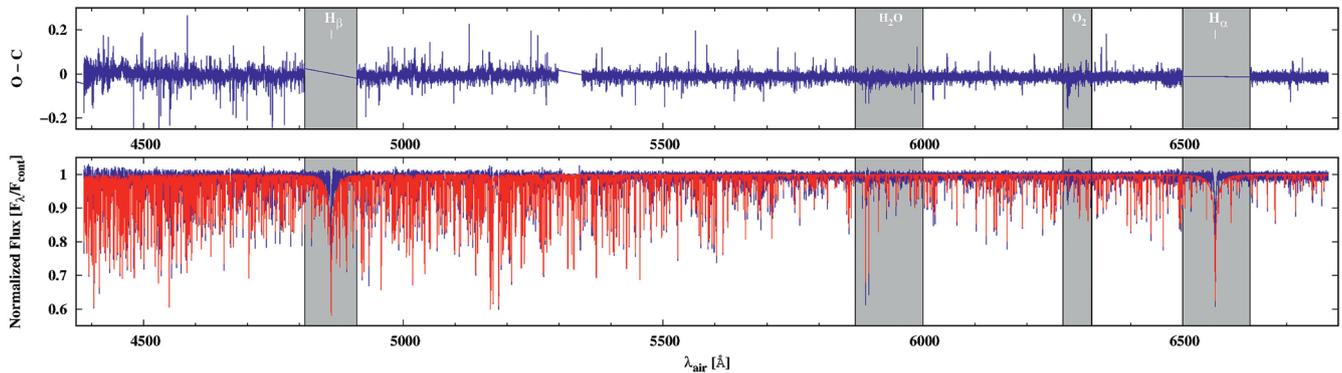

**Fig. 1.** The HARPS spectrum of FM Leo (blue) compared to the synthetic spectrum (red). The gray shaded areas contaminated by the transmission of water or $O_2$ molecules from Earth's atmosphere, as well as those with Belmer's series hydrogen lines endowed with wide wings (preventing robust reconstruction of the continuum level), were not used in calculations.

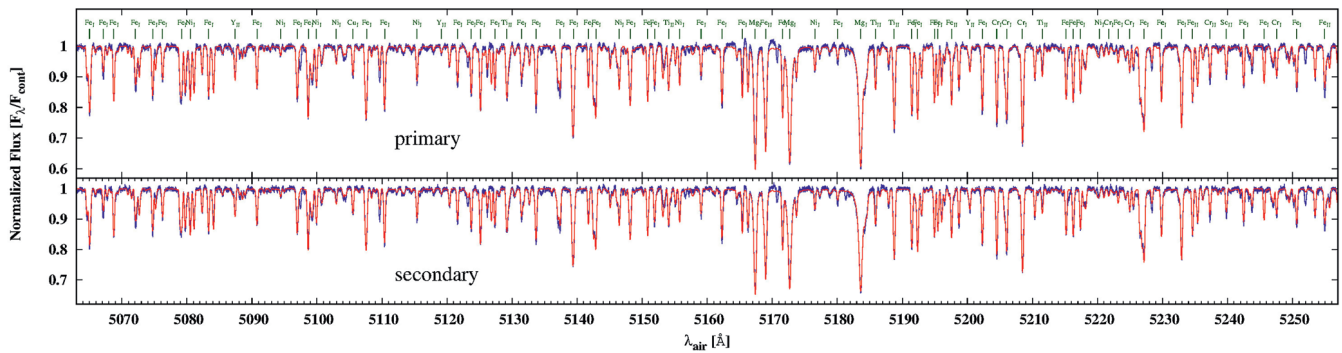

**Fig. 2.** The 5063-5257 Å region of uncorrected, disentangled spectra of the primary and the secondary components of FM Leo with selected spectral lines identified. The red lines denote the atmosphere models fit to the observed spectra (blue).

## Exemplary results

So far we have derived the atmospheric parameters for the components of 16 eclipsing binary systems (Graczyk et al. 2021; Graczyk et al. in prep.; Suchomska et al. in prep.), and abundances in several cases (Gałan et al. in prep.) Roughly one hundred more systems await a similar analysis. The resulting final parameters for FM Leo are shown in Table 1 and the comparison between the observed spectrum and synthetic fit is shown in Figure 2. The resulting final temperatures for components of our studied binary systems were consistent with those derived from photometric colors with an accuracy generally better than $1\sigma$ (Graczyk 2021; Graczyk in prep.) In most cases, we found that stars rotate synchronously within errors ( $< 3\sigma$), with those velocities which can be derived from known orbital periods and radii of the components.

**Tab. 1.** The best-fit parameters for FM Leo, and abundances of selected elements derived on the scale of $\log \epsilon (X) = \log(N(X)/N(H)) + 12.0$, together with 1 σ uncertainties for the primary (Prim.) and secondary (Sec.) components.

| Parameter | Unit | Prim | Sec. | Element | Prim. log $\epsilon$(X) [dex] | Sec. log $\epsilon$(X) [dex] |
|-----------|------|------|------|---------|------------------------------|------------------------------|
| [M/H] | dex | $-0.10 \pm 0.06$ | $-0.10 \pm 0.06$ | Fe | $7.29 \pm -0.05$ | $7.29 \pm 0.06$ |
| $T_{eff}$ | K | $6425 \pm 88$ | $6418 \pm 96$ | Ti | $4.74 \pm -0.11$ | $4.75 \pm 0.12$ |
| ξ | km/s | $1.61 \pm 0.15$ | $1.49 \pm 0.17$ | Cr | $5.54 \pm -0.12$ | $5.53 \pm 0.14$ |
| ζ | km/s | $5.8 \pm 1.2$ | $4.3 \pm 1.5$ | Ca | $6.25 \pm -0.17$ | $6.22 \pm 0.20$ |
| $V_{rot} \sin i$ | km/s | $11.0 \pm 0.6$ | $11.0 \pm 0.6$ | Si | $7.24 \pm -0.19$ | $7.24 \pm 0.21$ |
| log g [a] | dex | 4.1 | 4.2 | Ni | $6.07 \pm -0.12$ | $6.08 \pm 0.13$ |

[a] Fixed on the values from the W-D solution.

**Weronika Narloch[1]**

# Age-metallicity relation of star clusters in the Magellanic Clouds

Galaxies are ancient objects in the Universe composed of stars, gas, and dust. The great diversity of their parameters, e.g, shapes or chemical composition, is proof of their fascinating history of creation and interactions. Stars are born in stellar clusters. The metallicities and ages of the latter follow the age-metallicity relation (AMR) which traces the chemical enrichment in time. This fact makes star clusters perfect tracers of the chemical evolution history of their host galaxies.

The Large and Small Magellanic Clouds (LMC and SMC) are a pair of nearby neighbors of our Galaxy – the Milky Way (MW). The SMC is a dwarf irregular galaxy located at a distance of $(m-M)_{SMC}$ = 18.977 mag (Graczyk et al. 2020). The LMC is classified as a Magellanic spiral galaxy (SB(s)m type) at a distance of $(m-M)_{LMC}$ = 18.477 mag, as measured recently by the Araucaria project with a precision of 1% (Pietrzyński et al. 2019). Their proximity makes them important targets for extensive astrophysical studies, among which chemical evolution is one of the most crucial. The chemical enrichment history can be traced through the AMR analysis of star clusters, a large and diverse type of system which is harbored by both galaxies.

The Strömgren-filter data used in the study were collected during 6 nights in 2008 and 2009 with the SOI (SOAR Optical Imager) camera mounted on the 4.1-m SOAR (Southern Astrophysical Research) Telescope located in Cerro Pachón in Chile. The Strömgren pho-tometric system consists of four medium-band filters *uvby* (sometimes plus Hβ) defined in a way such that they are very useful in stellar astrophysics (Crawford 1987). The *y* filter transforms very nicely to the *V* filter from the Johnson-Cousin photometric system, the colour *(b-y)* is a good estimator for stellar temperatures, and the index *m1* is sensitive to stellar metallicities. There exist several calibrations of the *m1* index with metallicity in the literature. In this study we used one presented by Hilker (2000), calibrated for red giant stars and valid in a wide metallicity range, from –2.0 dex up to 0.0 dex.

In the first step of the selection of star clusters, Galactic foreground stars with high proper motion values taken from Gaia catalogue (Gaia Collaboration 2016b; 2021; Lindegren et al. 2021) were rejected. The equatorial coordinates of the centers of clusters and their radii were adopted from the updated catalogue of Bica et al. (2020) in the case of the SMC and Bica et al. (1999) in the case of the LMC. Stars lying within the adopted ra-dius were classified as cluster members and those outside this area as field stars. The reddening for a given cluster or field was calculated as an average of values taken from two reddening maps: Górski et al. (2020) and Skowron et al. (2021). In the next step, metallicities for individual stars fulfilling the adopted calibration were calculated, and based on this, the mean metallicity of a given cluster or field was determined. Finally, the ages of star clusters were estimated using two sets of isochrones: first one from the Dartmouth Stellar Evolutionary Database (Dotter et al. 2008), and second one from the Padova database of stellar evolutionary tracks and isochrones available through the CMD 3.3 interface (Marigo et al. 2017) based on PARSEC (Bressan et al. 2012) and COLIBRI (Pastorelli et al. 2019) evolutionary tracks with a fixed mean distance to the gal-axies. This way we studied 35 star clusters located in 29 fields in the SMC and 147 star clusters distributed in 80 fields in the LMC. For many of them we obtained both mean metallicities and ages. For the remaining ones no stars were available for metallicity calculation, as those clusters were too young, but we estimated the age from the isochrone fitting.

[1] Universidad de Concepción, Departamento de Astronomia, Casilla 160-C, Concepción, Chile

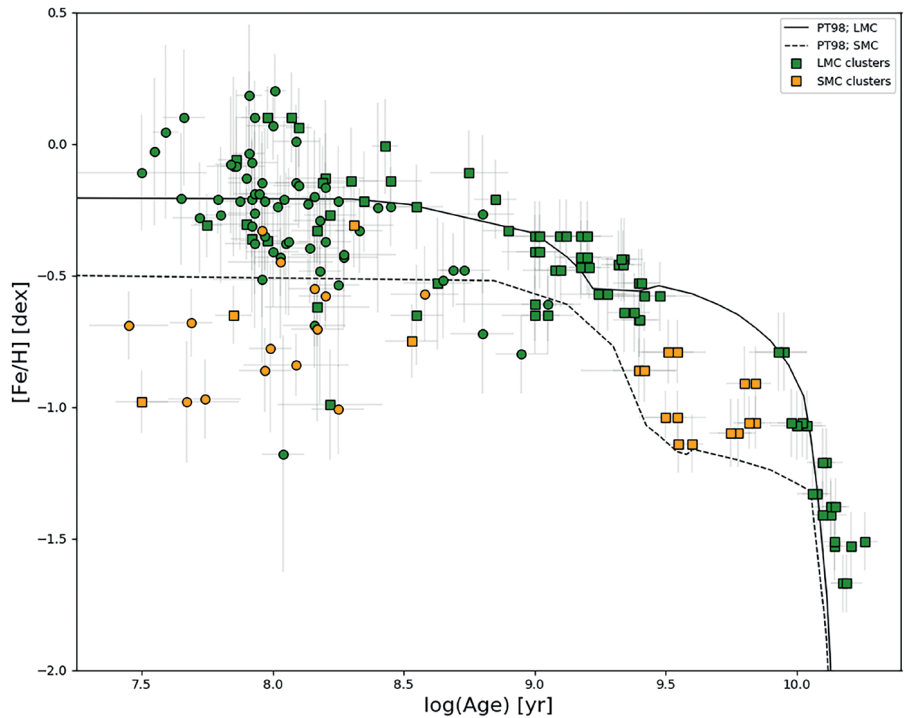

**Fig. 1.** Comparison of the AMR of the LMC (green) and the SMC (orange). Overplotted are Pagel & Tautvaišvienė (1998) bursting models (solid and dashed lines for the LMC and the SMC, respectively). Source: Narloch et al. (2022, in prep.)

The spatial distribution of the metallicity and ages of star clusters in the SMC is typical for irregular galaxies. Older and more metal-poor star clusters tend to occupy the outer regions of the galaxy, while younger and more metal-rich clusters cumulate in the center. Also, there is a lack of metal-poor clusters older than about 10 Gyr in the SMC. The analogous distribution in the LMC reveals a much more complicated structure compared to its neighbour. Ancient, metal-poor globular clusters are distributed more or less uniformly in the area of the galaxy. The LMC bar region is dominated by young metal-rich clusters which, however, can be also found in the non-bar regions–mostly in the Constellation III and 30 Doradus regions. Intermediate-age clusters are located in the non-bar regions.

The general picture that emerges from the resulting AMRs of the SMC and LMC (see Figure 1) suggests that these two galaxies could have interacted in the past. The oldest star clusters in the LMC were formed during the first 3 Gyr, rapidly increasing the metallicity from about –1.7 dex to about –1 dex (e.g. Pagel & Tautvaišvienė 1998). After that initial dynamic period of time, the cluster formation in the LMC stopped, and was followed by a long period of quiescence, which continued for about the next 6 Gyr (e.g. Harris & Zaritsky 2009). During that time no star clusters are observed in the AMR and the chemical enrichment is minor. Interestingly, the end of the initial star formation epoch in the LMC was simultaneously the beginning of star formation activity in the SMC. There are not many star clusters older than 10 Gyr observed in this galaxy, as the literature indicates. A burst of star cluster formation in the SMC appears about 7.5 Gyr ago and starts at a similar metallicity level as in the LMC at that time. Then the metallicity in the SMC drops slightly, perhaps as fresh metal-poor gas starts to take part in the star formation. The next burst of chemical enrichment in the SMC appeared about 3.5 Gyr ago (e.g. Parisi et al. 2014) and was followed by a similar burst in the LMC about 3 Gyr ago (e.g. Pagel & Tautvaišvienė 1998). This event raised the current metallicity value in the SMC to about –0.70 dex, while the average metallicity of numerous young star clusters in the LMC, located mostly in the bar region, is about –0.25 dex. The stars created after the burst in the LMC are now centrally concentrated, suggesting evolution in the outside-in direction, whereas in the SMC they seem to be distributed in a ring-like structure (Harris & Zaritsky 2009).

Studies of the AMR of galaxies can reveal their fascinating chemical histories and complicated interactions, and allow us to trace the possible scenarios of their formation.

**Megan Lewis[1]**

# Maser-derived parallax measurements of Miras

Period-luminosity relationships for a variety of pulsating variable stars are effective tools for determining and anchoring distance determinations in local galaxies, and here we address the goal of adding Mira variables to the ensemble of variable stars with established high-precision period-luminosity relations.

---

Like all pulsating stars, Miras come with their own advantages and disadvantages as distance-indicators. They are very bright (with absolute $K_S$ magnitudes between roughly -6.5 and -8.5, Whitelock et al. 2008) and extremely populous as they form from stars with low to intermediate progenitor masses in the range of 0.8 to 8 solar masses. However, their periods are long, on the order of a year (e.g. Glass et al. 1995), requiring long time baselines to determine their variability type and periods, and their circumstellar envelopes can complicate the interpretation of optical and infrared emission from these sources (e.g. Qin et al. 2018).

Perhaps the most interesting aspect of the Mira period-luminosity relation with regards to the Araucaria Project is that distances to Mira sources can be derived from maser parallax measurements, and thus the period-luminosity relationship can be derived independently of both Gaia-derived distance measurements and distance estimates to the LMC. Although these maser parallaxes also pose a number of observational hurdles, they are important to pursue because these objects are often too large, too red, and too bright for optical parallaxes to be accurate, and because these maser parallaxes could be the basis of a completely independent measurement of distances to local galaxies (i.e., no dependence on the Gaia parallax zero-point offset, distance measurements to the LMC, projection factors, etc.).

The key to these independent parallax measurements is radio-frequency maser emission from the circumstellar envelopes around Mira stars. The envelopes of oxygen-rich Miras can harbor SiO, $H_2O$, and OH masers at stellar radii on the order of 10, 100, and 1000 AU respectively. Maser parallaxes are a completely geometrical measure of absolute distance that take advantage of this bright and compact radio emission from the envelopes around Mira variables. The radio wavelengths of these masers allow observations of objects with high extinction, including Miras in the Galactic bulge and Miras embedded in thick circumstellar envelopes. Additionally, Very Long Baseline Interferometry (VLBI) instruments can probe these frequencies and provide better than milliarcsecond angular-resolution

[1] Nicolaus Copernicus Astronomical Center, Polish Academy of Sciences, Bartycka 18, 00-716 Warszawa, Poland

observations of maser spots. Thus there is great potential to derive distances to Galactic Mira sources via this method, providing information on Mira luminosities as well as Galactic dynamics and maser properties.

This technique has been successfully applied to about twenty evolved stars, roughly a dozen of which are Mira variables, with parallax uncertainties between 2 and 20% (Xu et al. 2019 and references therein). These parallax measurements barely scratch the surface of maser-bearing Miras in the Galaxy, and the stellar hosts of SiO masers in particular, which have relatively thin circumstellar shells, are extremely abundant with over ten thousand SiO hosts identified by the Bulge Asymmetries and Dynamical Evolution (BAaDE) survey in the Milky Way (Stroh et al. 2019, Lewis et al. 2020). Thus the BAaDE survey database is the natural starting point for target selection for new maser parallax measurements.

From this it is clear that the difficulty in obtaining maser parallax measurements is not in identifying suitable targets; the issue is instead identifying a suitable reference frame from which to derive the precise absolute position of these targets. The traditional way of doing this is via phase referencing using dedicated phase calibrators on the sky (usually background quasars) to remove the effects of the atmosphere on the phase of the targets. However, these dedicated phase calibrators are rare at high frequencies (for example at 43 GHz where SiO maser transitions occur) and the real astrometric (and parallax) measurement challenge for almost all Galactic SiO maser sources is to identify suitable calibrators.

Although this is a challenging endeavor, there are a number of techniques being pursued to make future parallax measurements of SiO masers, while, at the same time, exploring the nature of SiO maser emission and new observational techniques. For example, observations attempting "shared astrometry" for SiO masers, where the positions and proper motions of many masers in a single field are used to determine precise positions in multiple reference frames with limited use of phase calibrators, are underway. High-resolution studies of these masers will work towards both revealing the nature of SiO maser emission and establishing an independent anchor for the Mira period-luminosity relation.

**Nicolas Nardetto[1]**

# The Nice contribution: The CHARA/SPICA interferometer

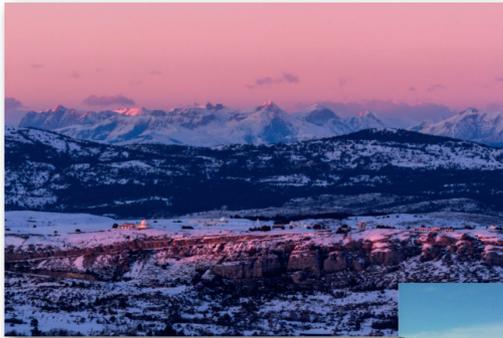

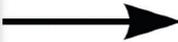

remote observations

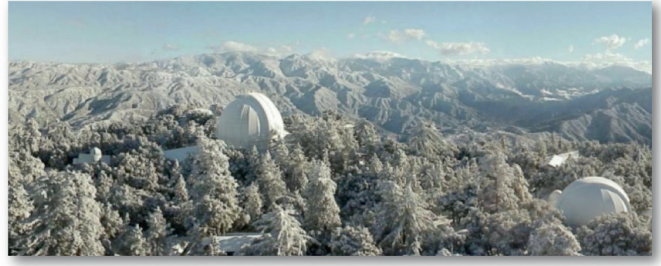

CHARA Interferometer, Mt Wilson, California, USA
Credit: CHARA group

Calern Observatory near Nice, France
Credit: Guillaume Doyen

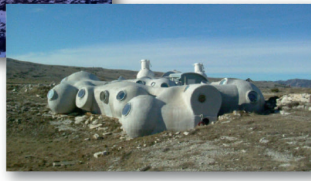

GI2T: Grand Interféromètre à 2 Télescopes
(remote control room of CHARA/SPICA)
Credit: Nicolas Nardetto

One of the famous places where interferometry was born is the Calern Observatory near Nice (left part of the image). Two optical interferometric instruments in particular were built in this beautiful place: the *Interféromètre à 2 Telescopes* (I2T) (Labeyrie 1975a, Labeyrie 1975b) and the *Grand Interférométre à 2 télescopes* (GI2T) (Mourard, 1988; Mourard et al. 1994). Later, the GI2T instrument was moved at the focus of the CHARA interferometer at Mount Wilson in California (ten Brummelaar et al. 2005) and the VEGA@CHARA instrument had its first light in 2009 (Mourard et al. 2009). The VEGA@CHARA instrument was operating in the visible domain with the possibility to use 4 telescopes among the 6 of the CHARA array with baselines up to 330 meters. This instrument led to about 50 refereed publications in stellar physics (https://lagrange.oca.eu/fr/publications-vega). Then, CHARA@VEGA was decommissioned in December 2020. Recently, the 6 telescopes of the CHARA array have been equipped with adaptive optics. In order to take advantage of these adaptive optics, early in 2015, it was decided to start the development of a new instrument, namely CHARA/SPICA, that should have its first light by the end of 2022.

[1] Université Côte d'Azur, Observatoire de la Côte d'Azur, CNRS, Lagrange, CS 34229, Nice, France

CHARA/SPICA (Stellar Parameters and Images with a Cophased Array) is an instrument operating in the visible domain (as CHARA/VEGA), fibered, with 6 telescopes and equipped with a new generation of detector. It also has a spectrograph which allows it to perform spectro-interferometry up to a spectral resolution of about 13000. This instrument with a baseline of 330 meters, in the visible domain, has a typical angular resolution $(R_a)$ of 0.5 milli-arcsecond (mas): $R_a[mas] = 250 \times \Lambda[\mu m]/B[m]$. The project is presented in two SPIE papers: Pannetier et al. 2020, Mourard et al. 2018. A kick-off meeting was organized in January 2019 in Nice, which allowed us to constitute a scientific group currently composed of about 80 researchers, including many researchers from the Araucaria group. The ERC ISSP starting in 2021 (PI: D. Mourard) is at the heart of the CHARA/SPICA program and consists of a first phase to observe 1000 stars (800 measurements of angular diameters with 1% of precision and 200 images) distributed on the whole HR diagram over a period of 4 years (2022-2025) at a rate of 50 nights of observations per year. These observations aim at determining the fundamental parameters (radius, mass, effective temperature, and age) of 1000 stars taking into account stellar activity, i.e. binarity, rotation, wind and environment. The interferometer is nevertheless limited to objects brighter than 8 to 10 magnitudes in the visible domain. In order to overcome this limitation, the goal is to ensure that the calibration of the CHARA/SPICA Surface Brightness Color Relations (SBCR) on the whole HR diagram will be achieved, and properly feed the Araucaria objectives.

The CHARA/SPICA instrument will be useful in the context of the Araucaria project, i.e. for the calibration of the distance scales in the universe. We can list the following objectives.

## Surface Brightness color relations

Firstly, one of the goals of the ISSP project is to calibrate the SBCRs all over the HR diagram. With basically 2 stars to be observed per sub-spectral channel in [B0, M3] and per class [V, IV, III], the idea is to cover a large domain in temperature and to calibrate the SBCR as a function of the surface gravity (log g) following the work done by Salsi et al. (2020, 2021). The final objective is to reach a precision and accuracy on the SBCR of late-type giant stars lower than 1%, in order to improve the Araucaria distance to the LMC. For this, high precision homogeneous photometry and spectroscopy are necessary. A complementary objective is then to observe metal poor stars, calibrate their SBCR in order to empirically check their dependency on metallicity in a very homogeneous way.

Secondly, the objective is to continue the work initiated by Challouf et al. (2014) and Salsi et al. (2021) in order to calibrate the SBCR of early type stars (O, B0, B1, B2) with a precision of 1%. This is a very difficult goal as these stars are highly active (binarity, rotation, wind, and environment). However, such stars exist, and they can be selected carefully. Also, with the image capabilities of SPICA (6 telescopes), it will be possible to check *a posteriori* if they are standard or not. Ultimately, the fast rotation can be corrected using models as done by Challouf et al. (2015). Such a SBCR will be used to determine the distance to M31 and M33 with 1-2% precision and accuracy.

## Cepheids

Firstly, the SPICA@CHARA instrument will have the capability *in principle* to measure the limb-darkening of a few Cepheids as a function of the pulsation phase. This would constrain their geometric projection factors for the first time. Also, Nardetto et al. (2006) has shown from a hydrodynamical model of delta Cep, that the limb-darkening variation has no impact on the distance. It will be possible to verify this assumption with CHARA/SPICA. Secondly, with SPICA@CHARA, we will have the opportunity to calibrate the SBCR of Cepheids in a very precise and homogeneous way, and compare it with the SBCR of standard stars outside the instability strip. This will allow us to improve the InfraRed Surface Brightness method developed by W. Gieren and collaborators in the Araucaria team. Thirdly, the precise measurements of visibilities of SPICA at short spatial frequencies will allow better characterization of the environment of Cepheids in the visible domain, which is probably a key to understanding their physical nature. Such an environment has been detected around Delta Cep in the visible domain already with VEGA@CHARA (Nardetto et al. 2016) but with poor precision. With SPICA it will be possible to improve the precision and enlarge such studies to many Cepheids. These observations will help to improve the circumstellar environment models of Cepheids (Hocdé et al. 2020a, 2020b, 2021).

In addition to this, SPICA@CHARA will be extremely useful to characterize Cepheids in *binaries*. But not only eclipsing binaries, also standard visual detached binaries will be observed in order to characterize the fundamental parameters of their components.

In conclusion, if SPICA@CHARA reaches its expected performance it should deeply contribute to the Araucaria objectives, that are:

- reach a sub-percent precision on the distance LMC taking into account metallicity effects on the SBC,
- reach a 1-2% on the distance of M31 and M33,
- unlock the difficult problem of the projection factor of Cepheids.

Rolf Chini[1,2,3]

# Observatorio Cerro Armazones
# – history, present, and future

## History

The history of OCA dates back to 1996 when I took over the chair of Astrophysics at the Ruhr-University in Bochum (RUB). There, I came into contact with a 1.5-m prototype of a new-technology hexapod telescope (HPT) with a thin active Zerodur mirror, a hexapod-supported CFK mount and a CFK mirror cell. Since this unfinished project was always ridiculed by the German community, I felt obliged to continue this never-ending story and bring it to an end – with whatever result. After three years of testing the HPT in the Botanical Garden in Bochum it was found worthy to be brought to a decent observing site.

This required raising the corresponding funds and finding a place with the existing infrastructure. The Ministry of Science and Research and the Academy of Science and Arts, both in Northrhine Westfalia, and RUB supported the project generously. The Gamsberg in Namibia as well as La Silla and Cerro Tololo in Chile were tested for their suitability and in particular the help from potential partners. Eventually – due to the cooperative attitude of the American colleagues – Cerro Tololo was selected. A contract was formulated and agreed but never signed by the Bochum administration because of a tiny legal formality.

During my ongoing search in Chile, I met a Chilean colleague who passed his PhD at the Astronomical Institute in Bochum and who was leading the Astronomy Department at the Universidad Católica del Norte (UCN) in Antofagasta. He invited me to see their small observatory at the foot of Cerro Armazones. The terrain was donated to UCN by the Chilean government and it was protected by law for exclusive scientific research – a fact which secured astronomical observations from being affected by possible future mining activities.

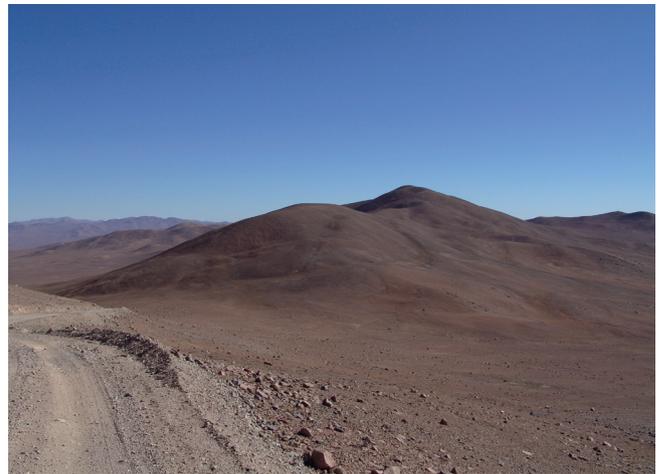

**Fig. 1.** A pristine hill of 2873 m as seen from Cerro Armazones (October 16, 2004).

[1] Nicolaus Copernicus Astronomical Center, Polish Academy of Sciences, Bartycka 18, 00-716 Warszawa, Poland

[2] Astronomisches Institut, Ruhr-Universität Bochum, Universitätsstrasse 150, D-44801 Bochum, Germany

[3] Instituto de Astronomía, Universidad Católica del Norte, Avenida Angamos 0610, Antofagasta, Chile

It was not my first visit to this remote place, but seeing suddenly some containers, a generator, a water tank and two small domes appearing in the middle of nowhere convinced me to bring the HPT to the Atacama Desert.

The UCN observatory was located by a gravel road leading through a trough between Armazones (3046 m) and an unknown side hill of 2873 m with the consequence that large parts of the sky were blocked. Therefore, the option for the new HPT home was either Armazones or its side hill. Although the top of Armazones was already accessible via a tough path, eventually I voted for the slightly lower side hill without a name and without a road to be our future stairway to heaven. Later on, the hill was called Cerro Murphy to acknowledge the tremendous support from Prof. Miguel Murphy at UCN.

The civil works started in 2005 with cutting the top of the mountain to provide a small flat plateau. A rough dusty road of about 12 km was built to shorten the existing way to the UCN observatory by 33 km. A large part of this road was later used by ESO to construct the paved road to the ELT site. Eventually a steep, winding access path of about 500 m connected the site with the rest of the world.

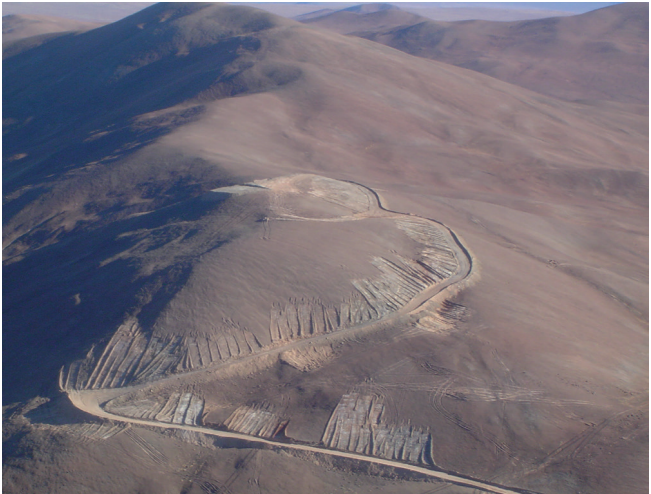

**Fig. 2.** Plateau and access road (July 8, 2005).

Originally, the observatory was designed for the HPT and two smaller auxiliary telescopes. A serious concern during the planning was to change the environment as little as possible and especially to adapt the energy supply to this "green" idea. Therefore, the constructions were limited to a main building and a compact auxiliary building for solar batteries and an emergency generator. The main building comprises two living/bedrooms with bathrooms, a salon and a kitchen. A control room with a separate computer room and an air-conditioned room for a spectrograph house the technical equipment. Attached to the east and west sides of this central building are two smaller telescope housings, each with a roll-up roof for the installation of two auxiliary telescopes.

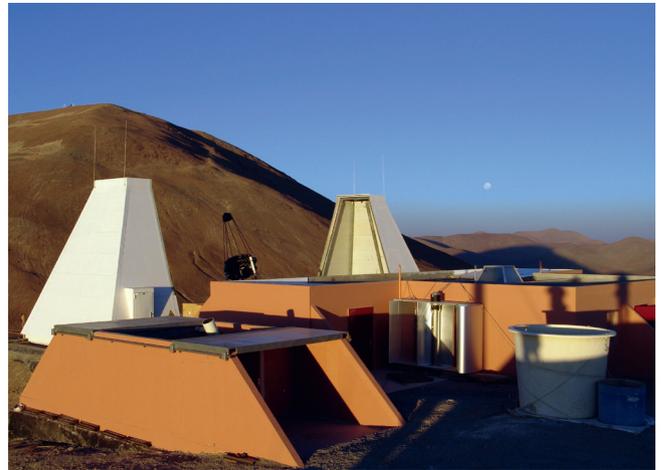

**Fig. 3.** Buildings at sunset (March 3, 2007).

The architectural design was severely influenced by the dome of the HPT which was a pyramid of 10 m height. Therefore, the geometry of inclined walls is found again in the main building and the battery house. At the same time, these slopes also served as the basis for 100 PV modules which generate the energy for the observatory. Likewise, hot water is produced by two solar heat modules. During operation – with an increasing number of telescopes – the need for additional energy support, particularly at night, led to the installation of three windmills.

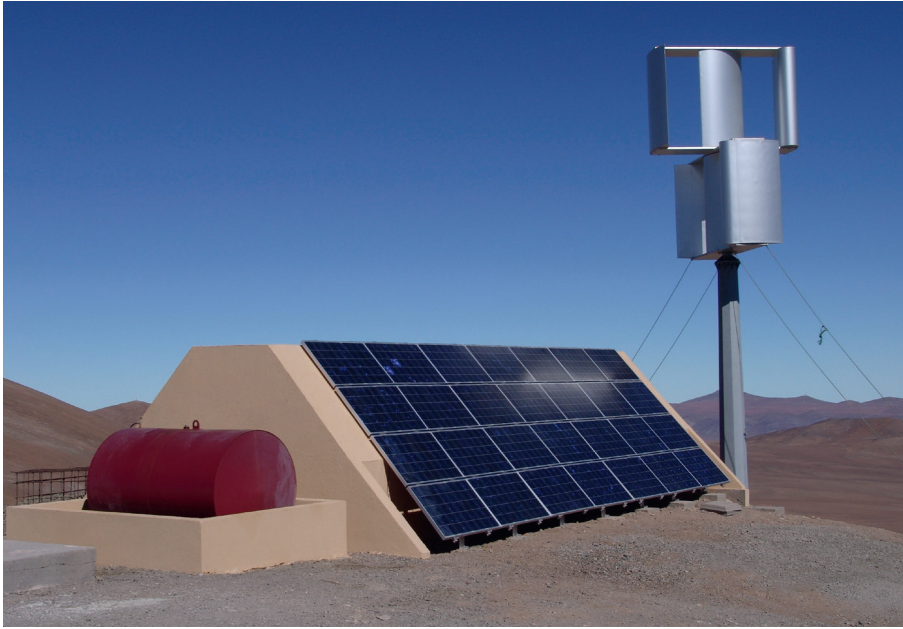

During the construction activities the German Aerospace Center (DLR) joined the project to host an additional telescope in support of the CoRoT mission. Therefore, another small building with a sliding roof was added west of the main building.

Although the HPT was not yet fully assembled, the official inauguration took place on September 28th 2006 in the presence of the German Ambassador in Chile, the president of the Chilean Astronomical Society, the rectors of UCN and RUB, a number of authorities from both universities and numerous colleagues from ESO and CTIO and DLR.

## Present
While the previous section describes the original planning and construction phase, the "present" section deals with the development of the observatory up to its current status.

The astronomical work started with the HPT, equipped with BESO, a clone of the ESO high-resolution spectrograph FEROS. The handling of the HPT proved to be extremely complex: six legs for pointing and tracking, six support legs for the M2 control, and 36 piezo-actuators for the active main mirror – all this kept two observers fully occupied during an observing night. Nevertheless, thousands of valuable multi-epoch spectra were secured – predominantly of high-mass stars, with the final result that most (if not all) early-type stars are multiples.

In August 2006, i.e. only a few days before the inauguration, a 40 cm Newton telescope equipped with a 3k

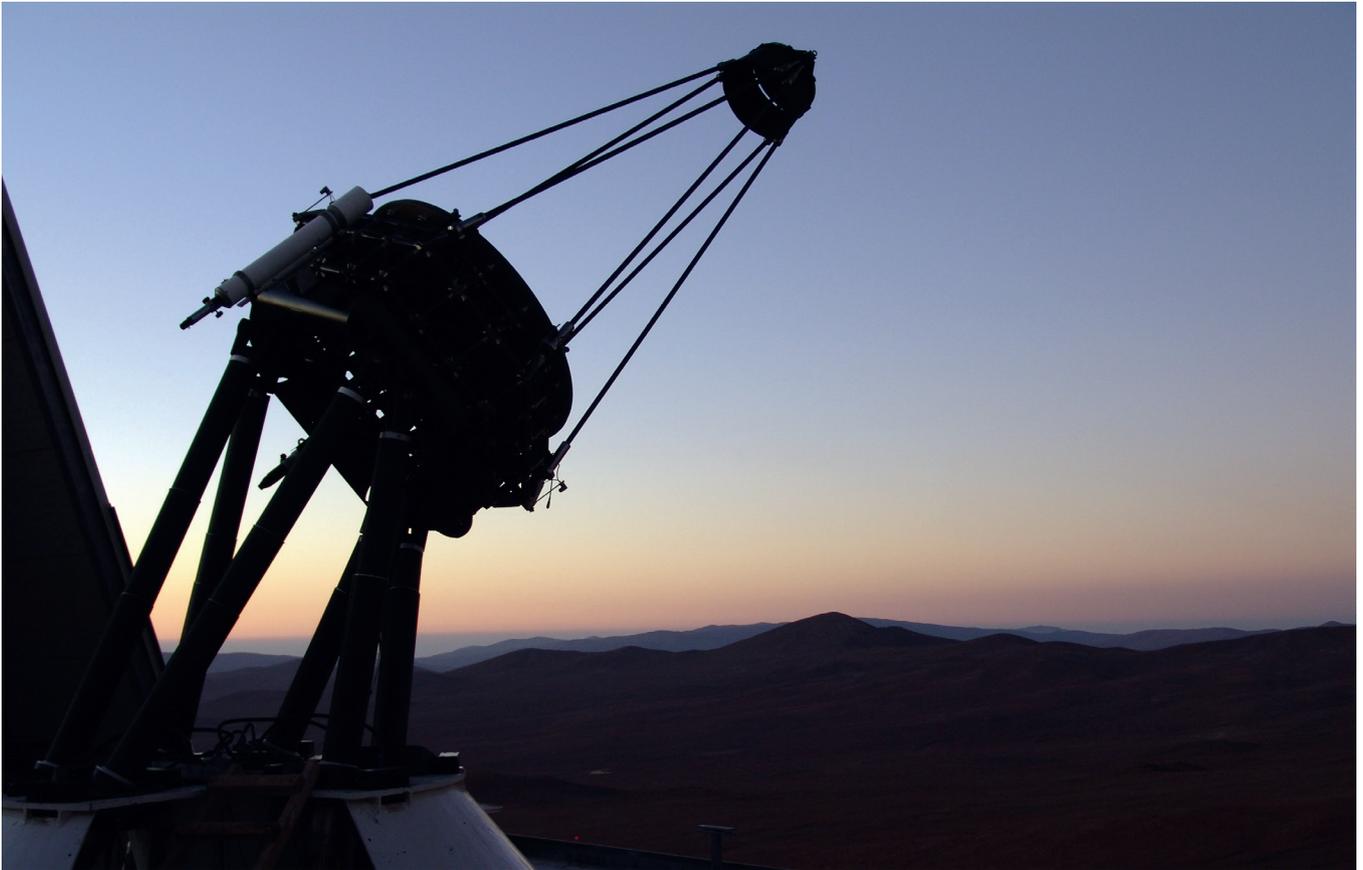

**Fig. 5.** The Hexapod Telescope (March 3, 2007).

x 2k CCD camera and a FoV of 41′ x 27′ was installed in the western telescope room of the main building. It was named VYSOS 16 according to its major purpose, namely a **V**ariable **Y**oung **S**tellar **O**bject **S**urvey in prominent Galactic star forming regions.

In November 2006 the DLR installed a 25 cm telescope (BEST II) which complemented their equipment to hunt for extra-solar transiting planets also in the southern hemisphere; BEST I had already been working, since 2001, in the northern hemisphere.

In May 2008 a Takahashi TOA-150 refractor was installed in the eastern telescope room of the main building. A second Takahashi TOA-150 refractor was bought and installed in August 2010 at the same mount as the previous one, transforming VYSOS 6 into a double system that allowed simultaneous observations in two filters. Both telescopes were equipped with 4k x 4k CCD cameras giving a FoV of about 2.5°. This twin refractor conducted – among many other projects – a multi-epoch *r* and *i* survey of the southern galactic disk (GDS) with the prime aim to find new, low-mass pre-main sequence stars and to monitor their light curves. Up to now, this survey contains light curves for 16 million stars including about 70.000 variables. Surprisingly, 62.000 of them were new detections, which means that almost 90% of variable objects in the Milky way were unknown before this survey. Part of the data were compiled into the largest astronomical image ever, consisting of 46 billion pixels with 196 GB (http://gds.astro.rub.de).

After the optical regime was covered satisfactorily by the HPT and the VYSOS telescopes, our main branch of research, namely star formation, asked for completion in the IR regime. Therefore, the IRIS project came into play where a new 80 cm Nasmyth IR telescope was funded by RUB. Its installation took place in February 2010. The 1k x 1k IR camera was donated by the Institute of Astronomy at the University of Hawaii and provides a FoV of 1' x 1'. So far IRIS has contributed to a number of projects, e.g. stars in the solar neighborhood, star forming regions, Cepheids and AGN.

In the framework of an ESO/European project (EVALSO) Cerro Murphy was connected by a 1 Gbit/s glass fiber in October 2010. This innovation allowed for remote control of the telescopes from Bochum and provided a fast data transfer from Chile to Germany (60 GB per night). Up to this time, the observatory was rather isolated from the rest of the world and observers had to take their data home on hard disks in their carry-on baggage.

Year 2011 became a fateful year for the observatory: An agreement was signed between ESO and the Chilean government, including the donation of 189 km² of land around Cerro Armazones for the installation of the ELT, as well as a concession for 50 years relating to the surrounding area. Suddenly OCA was on ESO terrain with all the political and juristical consequences – including a possible shutdown of the observatory and the dismantling of the telescopes. One afternoon, the ESO Director General, Prof. de Zeeuw and the Director of Paranal, Dr. Kaufer, came over for a coffee klatch with Chilean cookies. On this occasion, I showed them around the observatory and presented some spectra of O stars, which proved the multiplicity of these objects. Both impressed the visitors in such a way that Prof. de Zeeuw decided that this green observatory should continue its valuable work in the future.

In 2017 the Leibniz Institute for Astrophysics in Potsdam (AIP) joined to install a robotic 30 cm Zeiss refractor to support the satellite mission PLATO. Because at that time the Hexapod-Telescope had accumulated several technical problems which could not be repaired due to a lack of replacement parts, it was decided to remove both the telescope and the pyramid and to use its foundations for the AIP telescope. During 2019 a 5 m Ashdome was constructed at the location of the former pyramid and the wide-field camera (BMK) saw first light in September 2019. The camera has 10k x 10k pixels, providing a FoV of 13.6° x 13.6°.

Also in 2017 there was an agreement "for the use of observing time at the 0.8 m infrared telescope (IRIS) on Cerro Armazones" between RUB and CAMK which was the starting point for a fruitful collaboration between the two institutions and which led finally to the transition of the observatory to CAMK. The final contract between ESO, CAMK and RUB was signed on January 17, 2020. It just so happens that the birthday of the new observatory coincides with my own.

## Future

During the coming three years the observatory will undergo a phase of renewal and expansion. With respect to the telescopes, VYSOS 6 (RoBoTT), VYSOS 16 (BMT) and BEST II will be decommissioned. IRIS will move into a new classical 4.1 m dome east of the BMK dome on the former HPT platform. The sliding roof of the VYSOS 16 building will be removed and equipped with a classical dome. After that a new 60 cm Ritchey Chretien telescope (AZ 600) from ASA with a Cassegrain focus will move into this refurbished housing. An-other new Ritchey Chretien telescope (AZ 800) with an 80 cm mirror and two Nasmyth foci will be installed in a second 4.1 m dome west of the BMK.

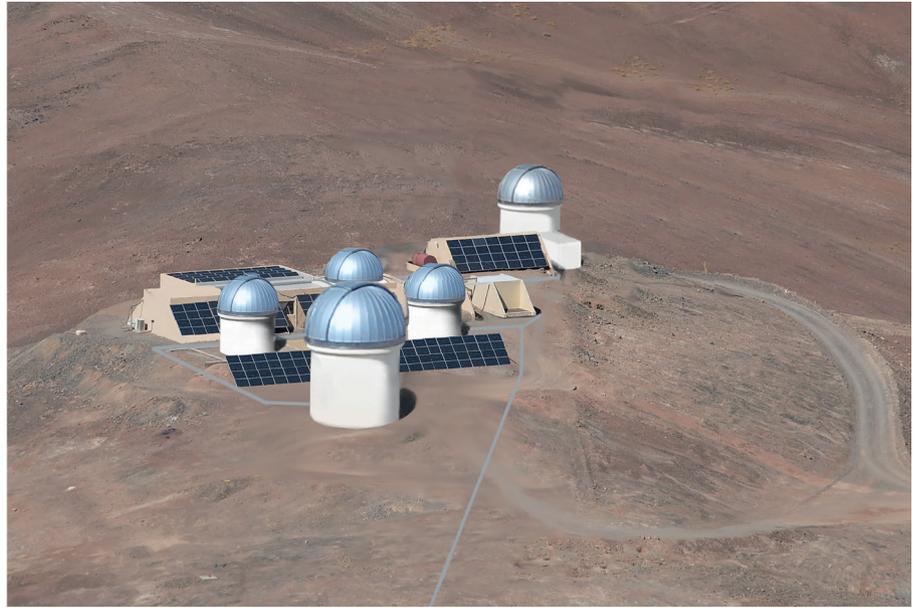

**Fig. 6.** Future view of OCA.

Southwest of the battery house a 7.5 m classical dome with a 1.5 m Ritchey Chretien telescope (AZ 1500) will be assembled; it has four foci including two Nasmyth foci. North of the dome there is an adjacent air-conditioned extension room which will host BESO; the spectrograph will be moved from its current room in the main building to act as the prime instrument at the new 1.5 m telescope.

The so far described installations are planned to be finished by the end of 2022. Parallel to this, the development of a new 2.5 m telescope is being pushed forward, with the goal of putting it into operation in 2024. Its location will be northeast of the current installations. Altogether, there will be a battery of new instruments, comprising wide-field optical cameras, IR cameras, and spectrographs of various resolving powers.

With respect to the infrastructure, the old solar batteries have to be replaced by new Li batteries. The large number of new telescope locations requires the construction of new ducts for electricity cables and data lines, replacing the old tube connections. The windmills will be removed while the power of the solar panels will be increased. Due to the removal of BESO some internal walls of the building will be displaced. The net-result will be a larger control room, a larger kitchen at the expense of the computer room, and a new guest room with a bath. Last but not least there is hope that OCA will be connected to the Chilean electricity grid within the coming years.

In summary, the variety of new telescopes and instruments will tremendously strengthen the observing capabilities within the Araucaria project. The accuracy of the cosmic distance ladder can be improved significantly by various monitoring projects, leading to an unprecedented knowledge of the Hubble constant.

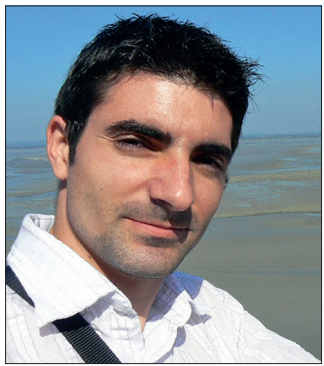

Alexandre Gallenne

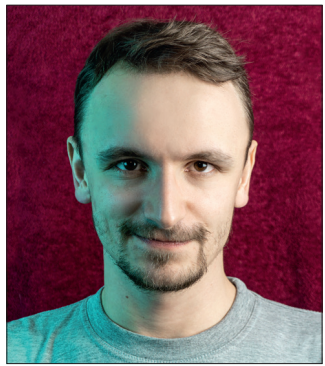

Bartek Zgirski

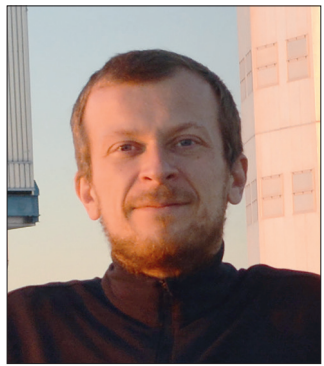

Bogumił Pilecki

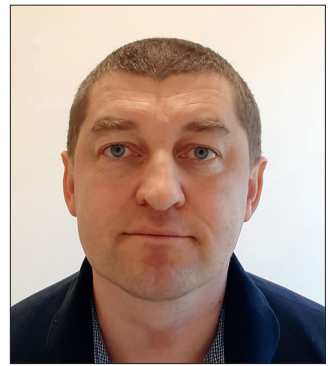

Cezary Gałan

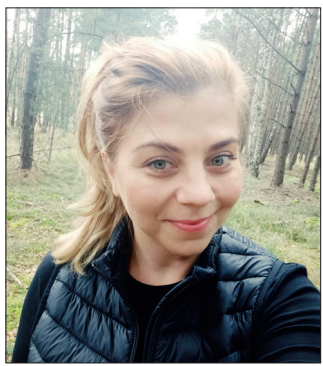

Ksenia Suchomska

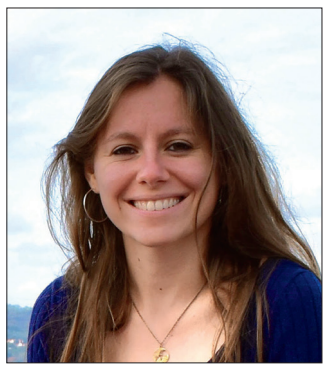

Louise Breuval

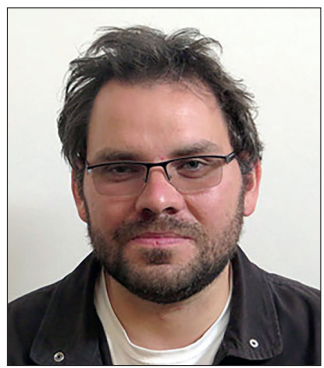

Marek Górski

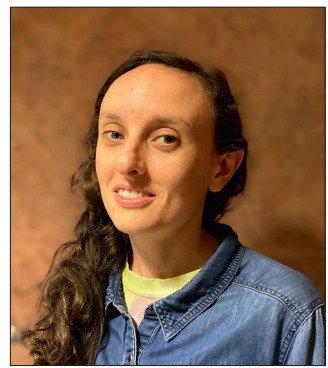

Megan Lewis

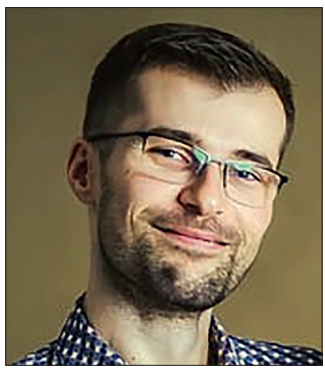

Piotr Wielgórski

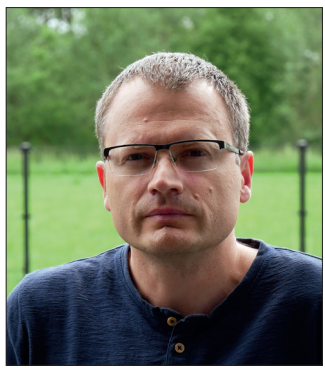

Radosław Smolec

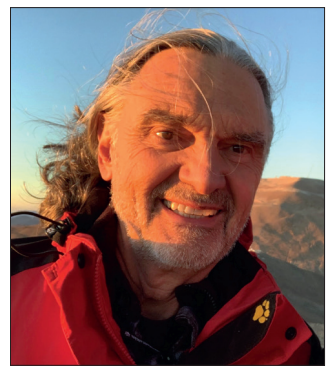

Rolf Chini

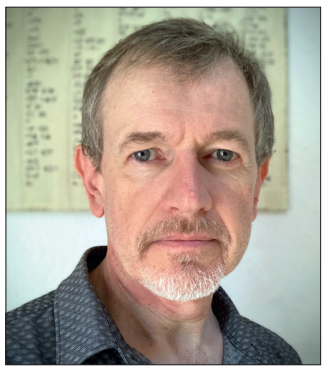

Fabio Bresolin

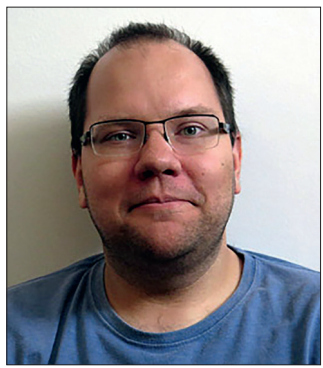

Gergely Hajdu

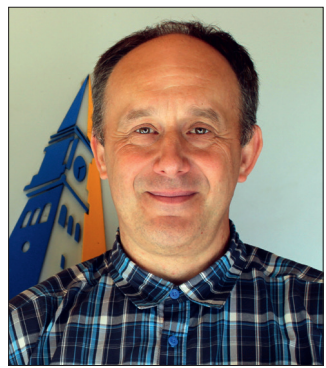

Grzegorz Pietrzyński

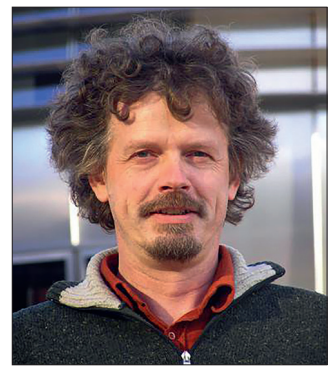

Jesper Storm

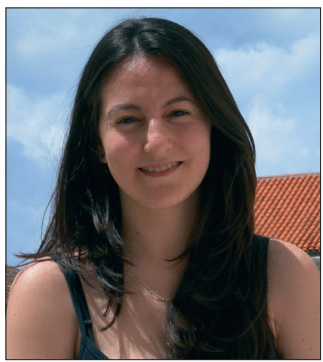

Mónica Taormina

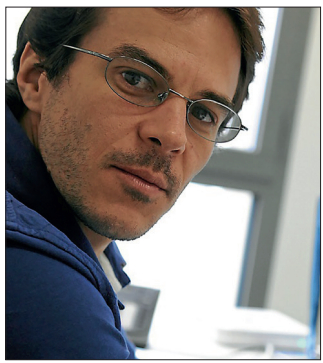

Nicolas Nardetto

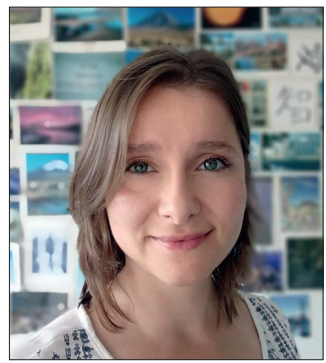

Paulina Karczmarek

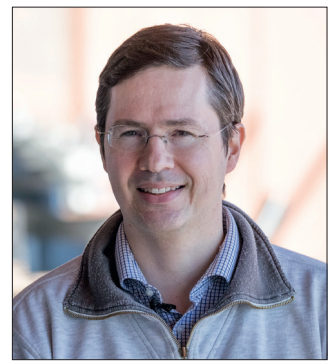

Pierre Kervella

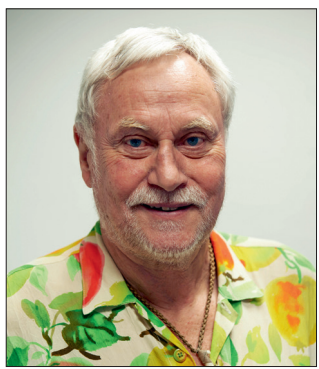

Rolf-Peter Kudritzki

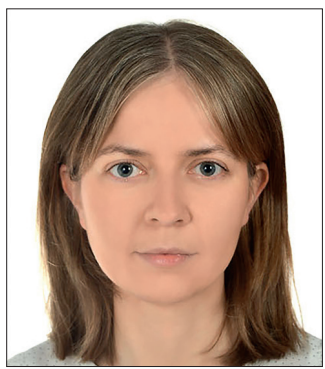

Weronika Narloch

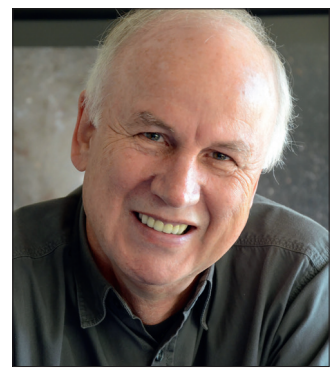

Wolfgang Gieren

# Beyond the Araucaria Project: hikes, social gatherings, and conferences

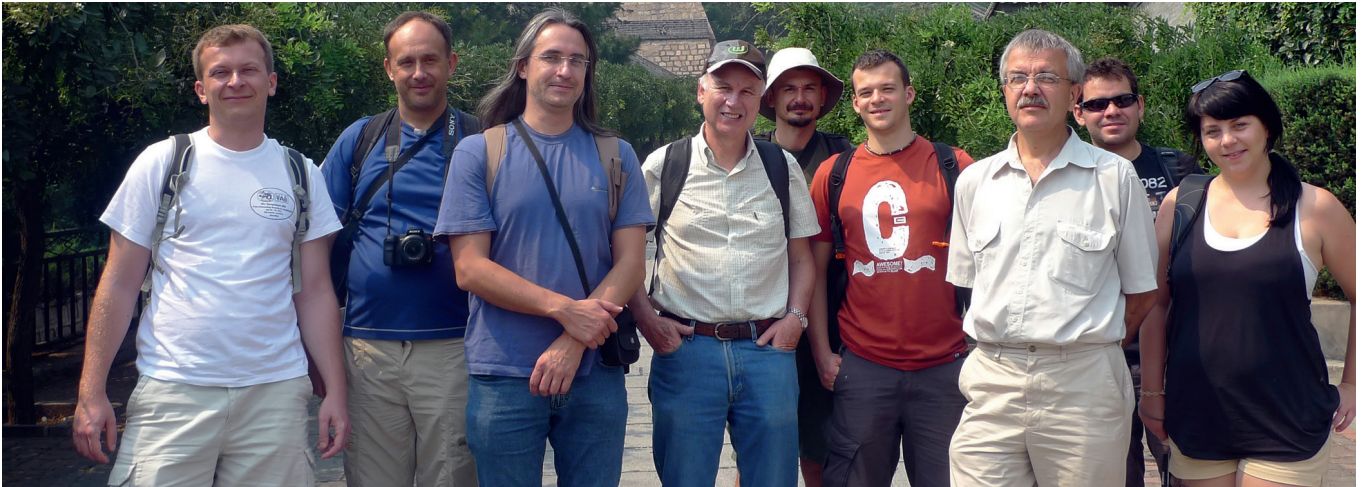

Members and friends of the Araucaria Project at the IAU Symposium "Advancing the physics of the cosmic distance scale" in Beijing, China, 2012.

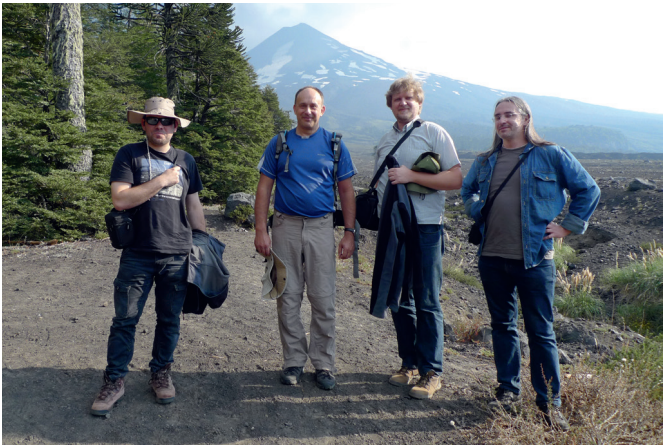

Members of the Araucaria Project on a hike in the Conguillío National Park, Chile, 2015. From the left: Marek, Grzesiek, Zibi, Igor. In the back: Llaima Volcano.

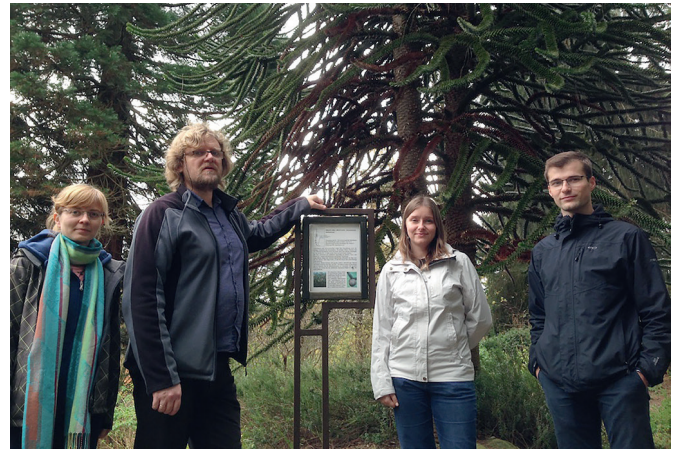

Araucaria-OCA workshop in Bochum, Germany, 2017. From the left: Weronika, Zibi, Araucaria tree, Paulina, Piotrek.

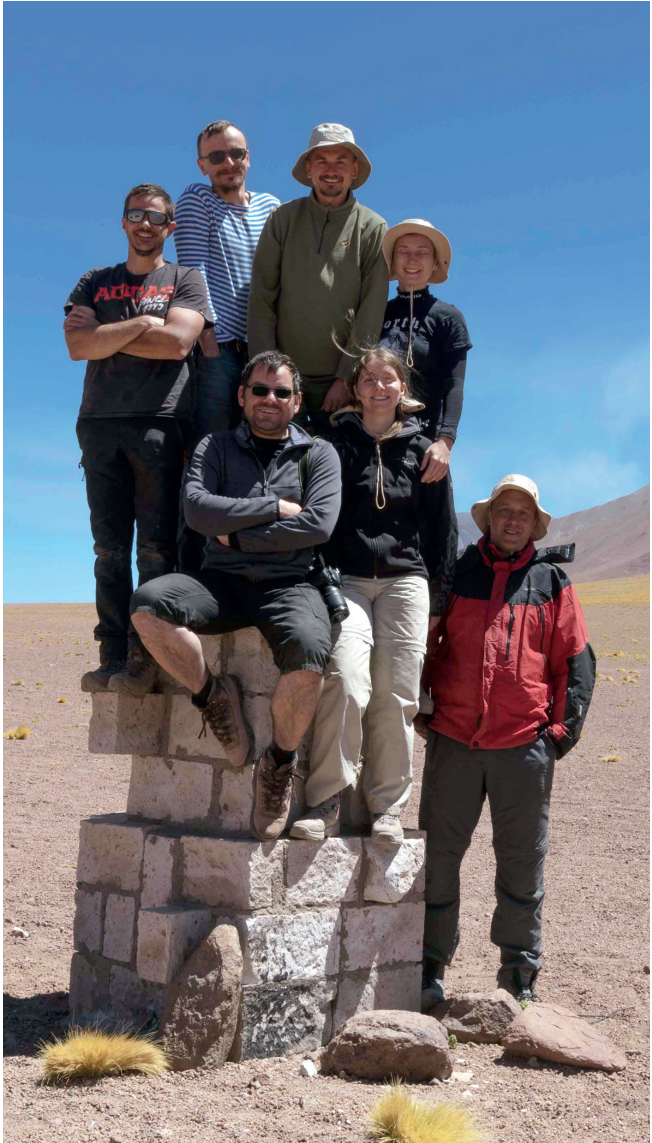

Members and friends of the Araucaria Project at the Tropic of Capricorn, in the Atacama Desert, Chile, 2016.

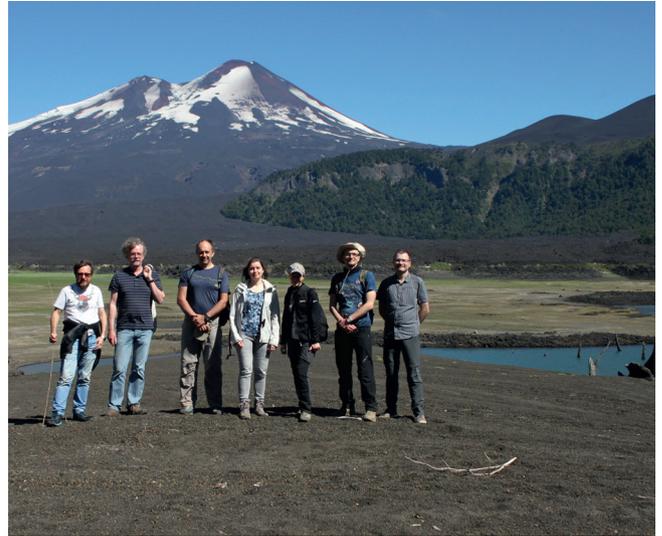

Members and friends of the Araucaria Project on a hike in the Conguillío National Park, Chile, 2019.

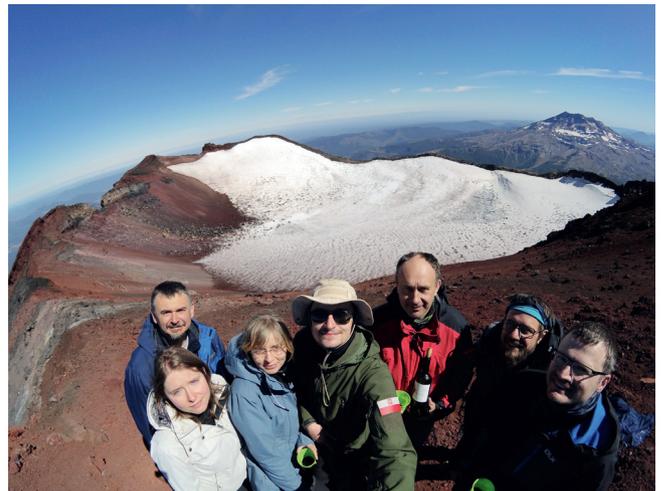

Members and friends of the Araucaria Project on the summit of the Lonquimay Volcano in the Malalcahuello National Reserve, Chile, 2019.

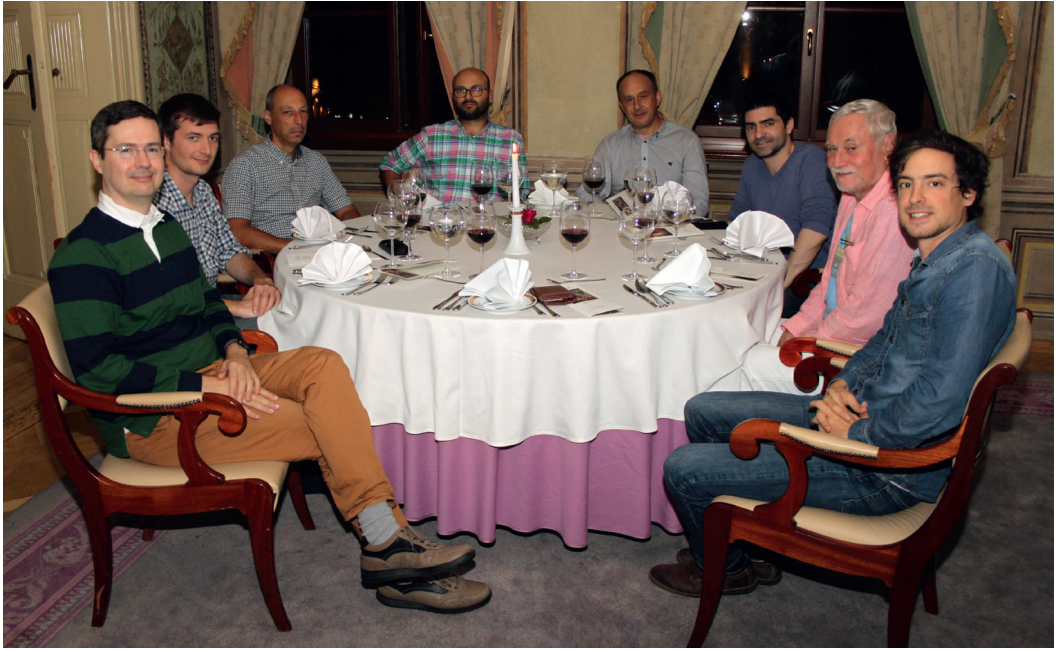

Members and friends at the Araucaria Meeting, during a dinner at the Royal Restaurant Wierzynek, Kraków, Poland, 2017.

Members of the Araucaria Project at the RR Lyrae Conference, during a dinner in the Niepołomice Royal Castle, Poland, 2017.

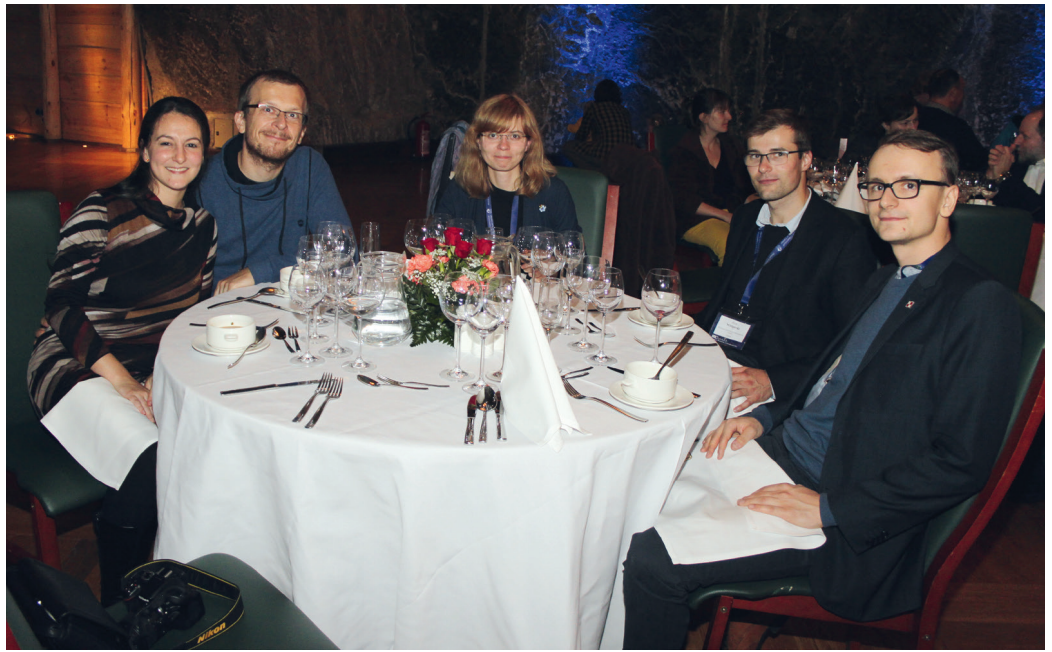

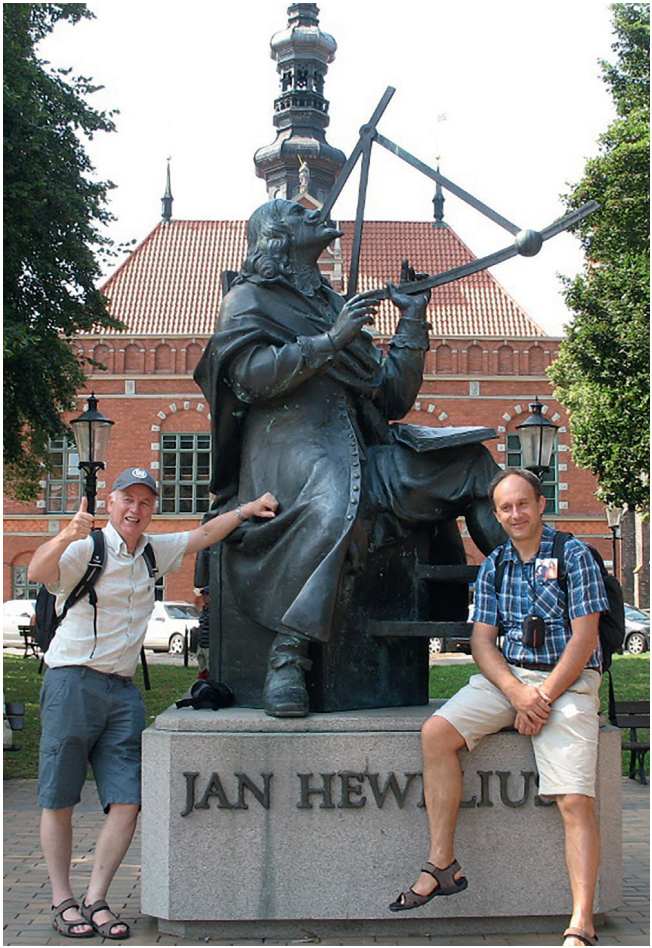

Founders of the Araucaria Project next to the statue of Jan Heweliusz in Gdańsk, Poland, 2013.

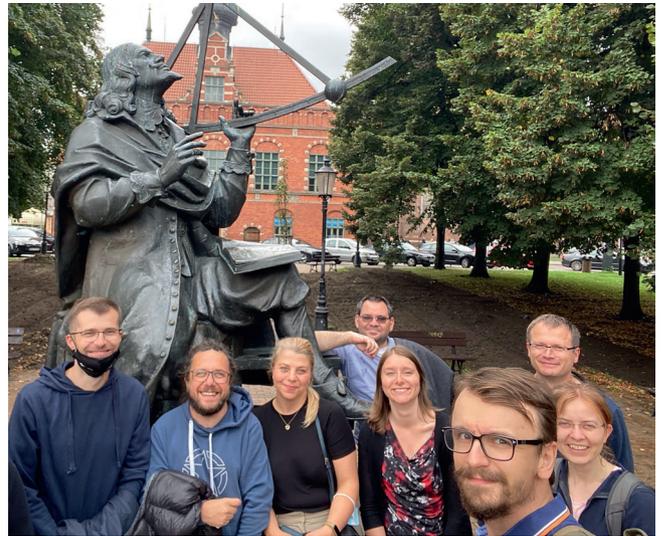

Members of the Araucaria Project casually stroling during the Araucaria Meeting in Gdańsk, Poland, 2021.

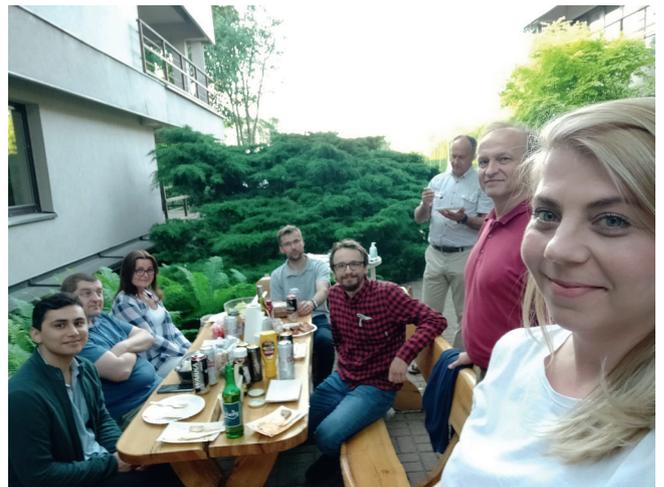

Members of the Araucaria Project barbequing at Nicolaus Copernicus Astronomical Center, Warszawa, Poland, 2021.

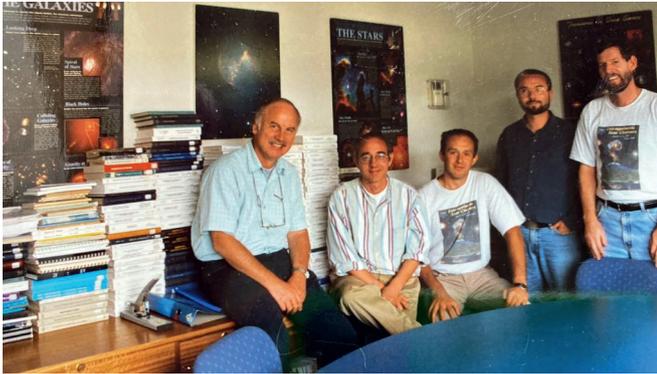

Members of the Araucaria Project in a brand new office at the Universidad de Concepción, Chile, 2002.

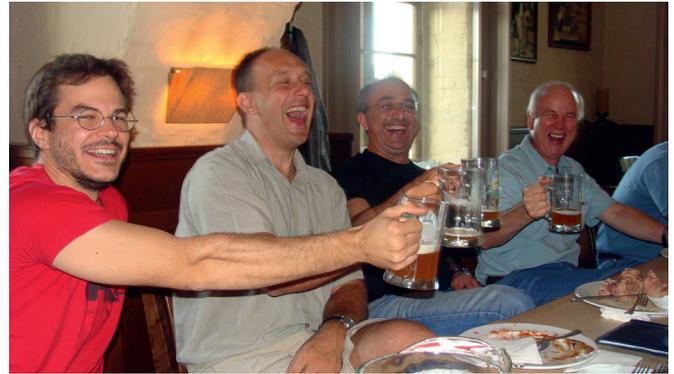

Araucaria members socializing at the meeting in Potsdam, Germany, 2010.

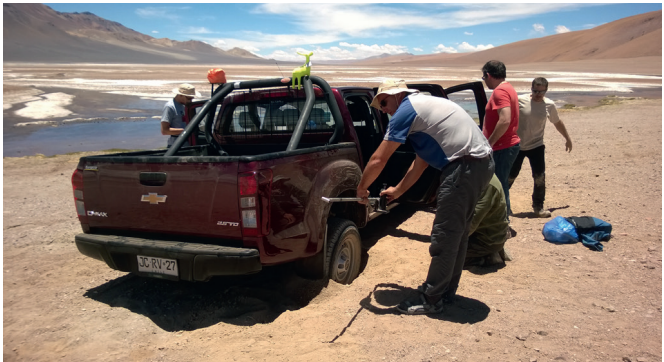

A car bogged down in the sand at the Atacama Desert, on a hike to the Lascar Volcano after the Stellar Pulsation Conference in San Pedro de Atacama, Chile, 2016.

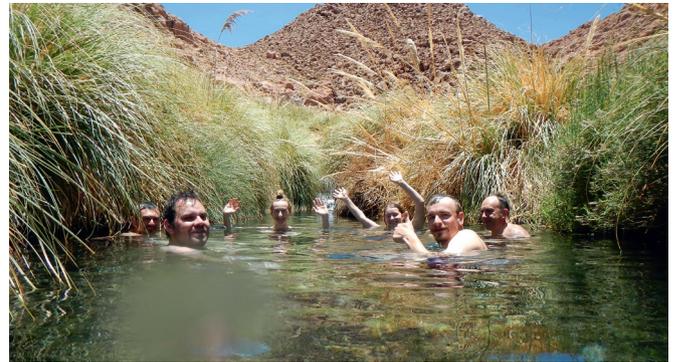

A bath at the Puritana Hot Springs, after the Stellar Pulsation Conference in San Pedro de Atacama, Chile, 2016.

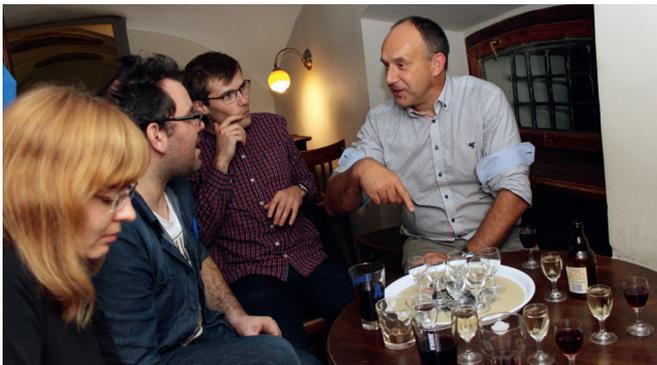

Members of the Araucaria Meeting socializing after the RR Lyrae Conference, Kraków, Poland, 2017.

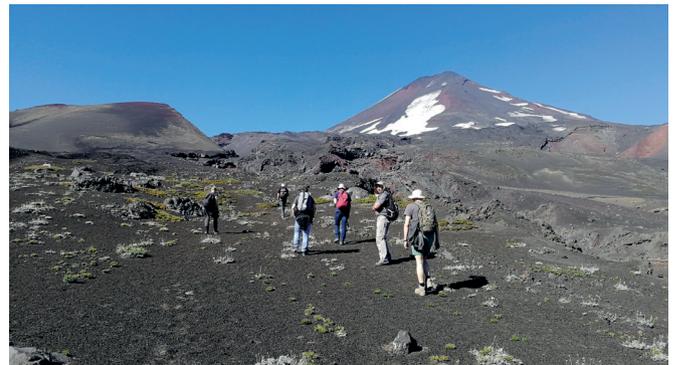

Members and friends of the Araucaria Project on a hike in the Conguillío National Park, Chile, 2019.

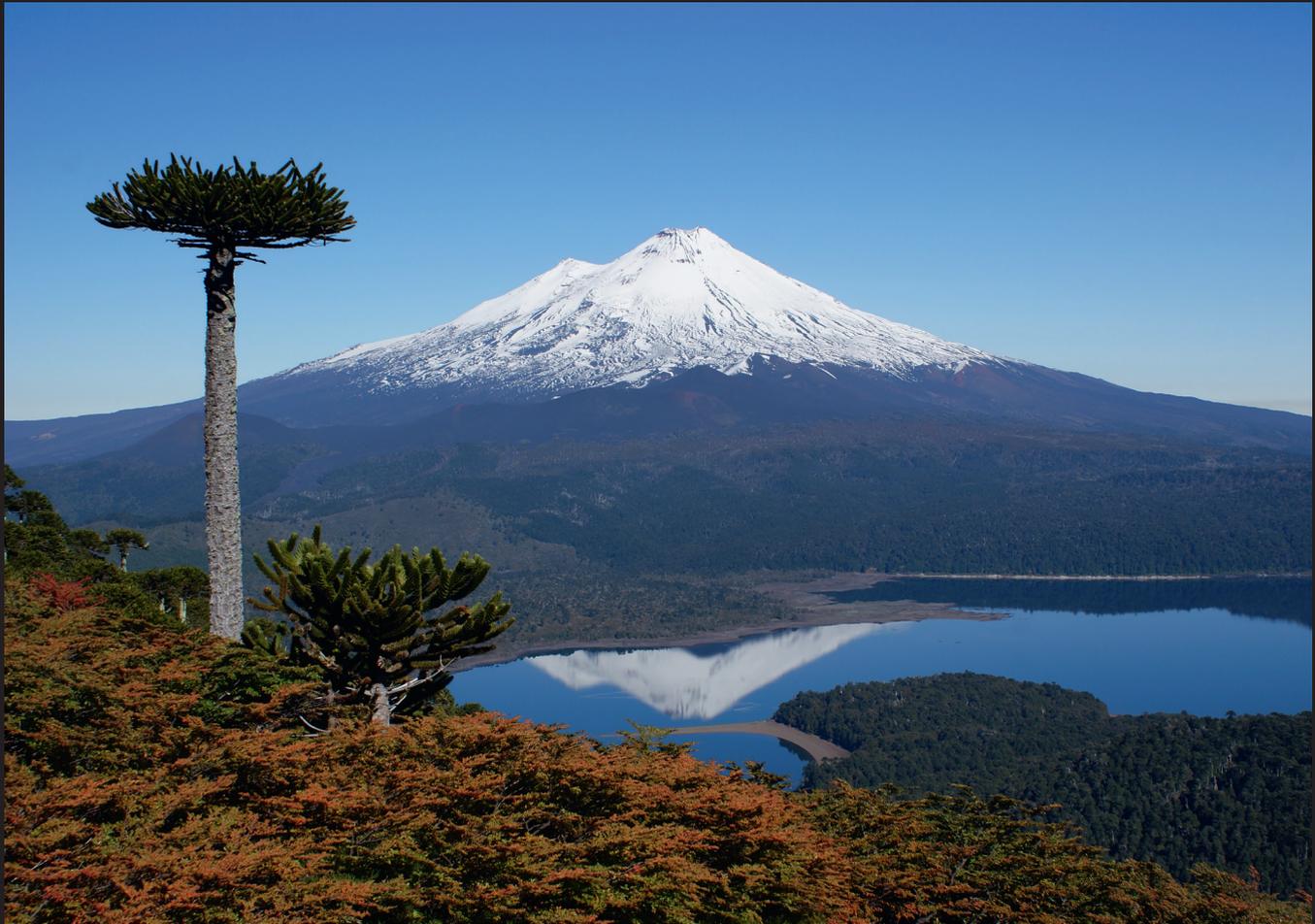

This book is a part of a project that
has received funding from the
European Union's Horizon 2020
research and innovation programme
under grant agreement No 695099.

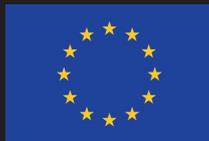

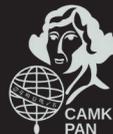

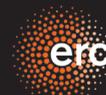

European Research Council
Established by the European Commission

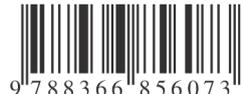

9 788366 856073